\documentclass[aps,rmp,twocolumn,preprintnumbers, amsmath,nofootinbib,floatfix,fleqn,longbibliography]{revtex4-1}
\pdfoutput = 1

\usepackage{color}
\usepackage[dvipsnames]{xcolor}
\usepackage{graphicx}
\usepackage{bm}
 \usepackage[utf8]{inputenc}
 
 \usepackage{float}
 \usepackage{epsfig}% need for subequations

\newcommand{\as}{\alpha_{\mathrm{s}}}
\newcommand{\LA}{\mathrm{A}}
\newcommand{\LD}{\mathrm{D}}
\newcommand{\LB}{\mathrm{B}}

\newcommand{\LF}{\mathrm{F}}

\newcommand{\LL}{\mathrm{L}}

\newcommand{\LN}{\mathrm{N}}
\newcommand{\LO}{\mathrm{O}}
\newcommand{\LP}{\mathrm{P}}
\newcommand{\LR}{\mathrm{R}}

\newcommand{\LT}{\mathrm{T}}

\newcommand{\LZ}{\mathrm{Z}}
\newcommand{\La}{\mathrm{a}}
\newcommand{\Lb}{\mathrm{b}}
\newcommand{\Lc}{\mathrm{c}}
\newcommand{\Ld}{\mathrm{d}}

\newcommand{\Lg}{\mathrm{g}}

\newcommand{\Lp}{\mathrm{p}}

\newcommand{\Ls}{\mathrm{s}}
\newcommand{\Lt}{\mathrm{t}}
\newcommand{\Lu}{\mathrm{u}}

\newcommand{\cD}{\mathcal{D}}
\newcommand{\cE}{\mathcal{E}}

\newcommand{\cL}{\mathcal{L}}
\newcommand{\cN}{\mathcal{N}}

\newcommand{\cT}{\mathcal{T}}

\newcommand{\GeV}{\ \mathrm{GeV}}
\newcommand{\TeV}{\ \mathrm{TeV}}

\definecolor{red}{rgb}{0.8,0,0}
\definecolor{green}{rgb}{0,0.6,0}
\definecolor{blue}{rgb}{0,0,0.8}

\def\mi{{\mathrm i}}
\newcommand{\MSbar}{\overline {\text{MS}}}

\def\ket#1{\big|{#1}\big\rangle}
\def\bra#1{\big\langle{#1}\big|}
\def\brax#1{\big\langle{#1}}   % No "|"
\def\<>#1{\big\langle{#1}\big\rangle}
\def\[]#1{\big[{#1}\big]}

\newbox\charbox
\newbox\slabox
\def\s#1{{      % Feynman slash
        \setbox\charbox=\hbox{$#1$}
        \setbox\slabox=\hbox{$/$}
        \dimen\charbox=\ht\slabox
        \advance\dimen\charbox by -\dp\slabox
        \advance\dimen\charbox by -\ht\charbox
        \advance\dimen\charbox by \dp\charbox
        \divide\dimen\charbox by 2
        \raise-\dimen\charbox\hbox to \wd\charbox{\hss/\hss}
        \llap{$#1$}
}}

\definecolor{red}{rgb}{1,0,0}

\definecolor{darkblue}{rgb}{0.0,0,0.5}
\definecolor{redish}{rgb}{0.675,0,0.2}
\usepackage[unicode=true, bookmarks=false, linkcolor = darkblue, citecolor = redish, breaklinks=false, colorlinks=true, hyperfootnotes=true]{hyperref}

\begin{document}

\date{\textcolor{red}{\today}}

\title{Hadron structure in high-energy collisions}\thanks{Accepted by Reviews of Modern Physics}

\author{Karol Kova\v r\' ik}
\affiliation{
Institut f\"ur Theoretische Physik\\
Westf\"alische Wilhelms-Universit\"at M\"unster\\
Wilhelm-Klemm Stra\ss e 9\\
D-48149 M\"unster,~Germany
}

\author{Pavel\ M.\ Nadolsky}
\affiliation{ Department of Physics, Southern Methodist University, \\ Dallas, TX 75275-0181,~U.S.A.}
\email{nadolsky@smu.edu}

\author{Davison\ E.\ Soper}
\affiliation{Institute of Theoretical Science, University of Oregon,\\ Eugene, OR 97403-5203,~USA}
\email{soper@uoregon.edu}

\preprint{MS-TP-19-09, SMU-HEP-19-05}

\begin{abstract}
Parton distribution functions (PDFs) describe the structure of hadrons as composed of quarks and gluons. They are needed to make predictions for short-distance processes in high-energy collisions and are determined by fitting to cross section data. We review definitions of the PDFs and their relations to high-energy cross sections. We focus on the PDFs in protons, but also discuss PDFs in nuclei. We review in some detail the standard statistical treatment needed to fit the PDFs to data using the Hessian method. We discuss tests that can be used to critically examine whether the needed assumptions are indeed valid. We also present some ideas of what one can do in the case that the tests indicate that the assumptions fail. 
\end{abstract}

\begin{figure}[H]
\centering
\includegraphics[width=0.48\textwidth]{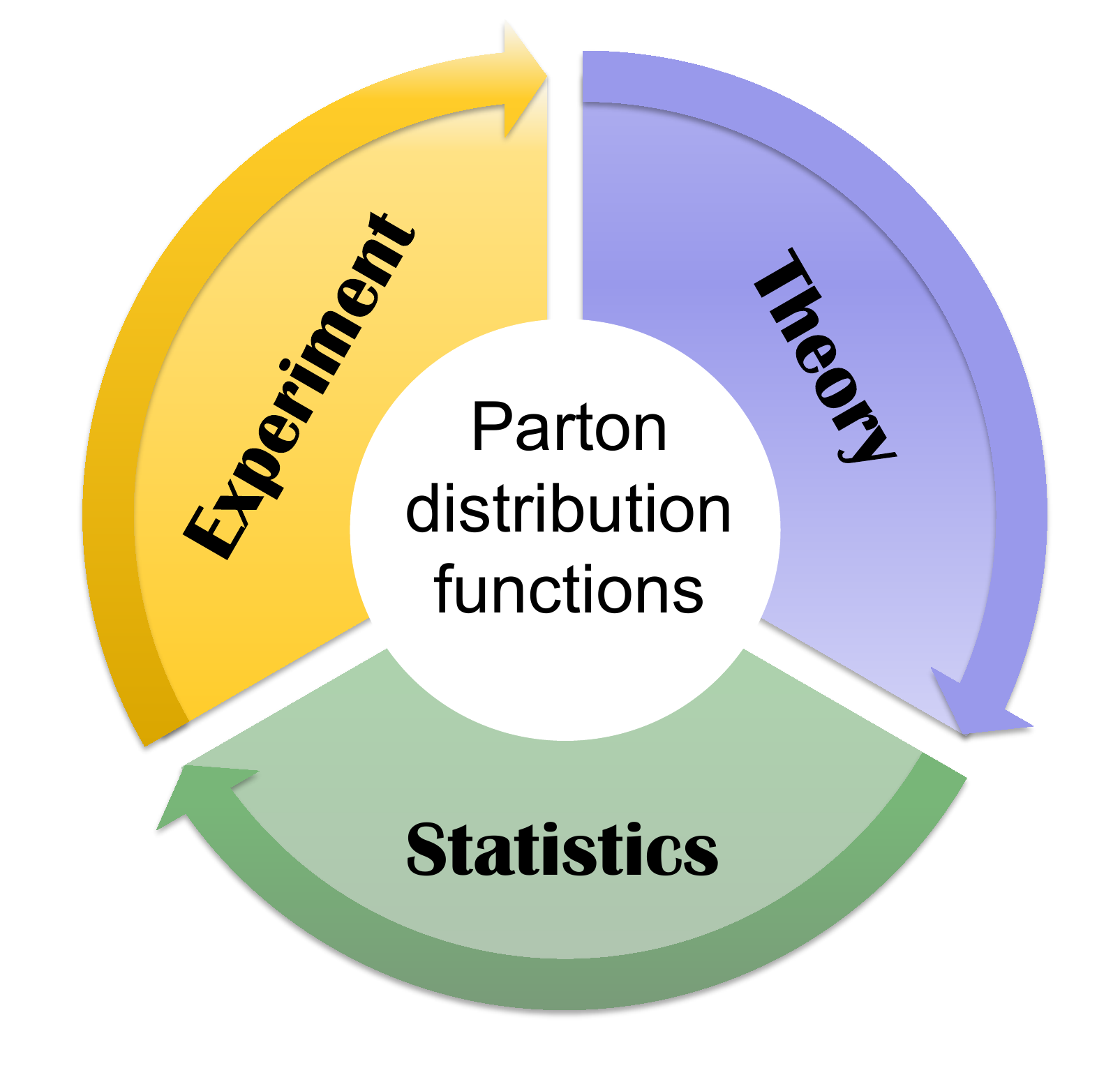}
\end{figure}

\pacs{12.15.Ji, 12.38 Cy, 13.85.Qk}

\keywords{parton distribution functions; large hadron collider}
\maketitle

 \tableofcontents{}

% !TEX root = ../main.tex
\section{Overview}\label{sec:Intro}

%\NOTE{We had mostly parameterize and parameterizaton but often parametrize and parametrization. I changed them all to parameterize and parameterizaton.}

As of 2018, the Large Hadron Collider (LHC) has taken a large data
sample of proton-proton collisions and has been using these data
 to precisely measure the properties of the Higgs boson
and search for physics beyond the Standard Model. The ATLAS and CMS experiments took about 150 fb$^{-1}$ each at the center-of-mass energy of $\sqrt{s}=13\;{\rm TeV}$ and more than 22 fb$^{-1}$ of data at $\sqrt{s}=8\;{\rm TeV}$. In addition, the LHCb
experiment has accumulated more than 9 fb$^{-1}$ of data at various
energies, including data at extreme rapidities. On top of
proton-proton collisions, all experiments at the LHC, and in
particular the ALICE experiment, are taking data in proton-lead and
lead-lead collisions.  

All predictions at the LHC are crucially dependent on  knowledge of the quark and gluon content of the proton.  The probability distributions of the constituents of the nucleon, collectively called {\it partons}, need to be known to make predictions in the theoretical framework of perturbative quantum chromodynamics (PQCD). The concept of a parton originates in a model by Bjorken and Feynman \cite{Bjorken:1968dy, Feynman:1973xc}, where it refers to quasi-independent pointlike constituents observed inside hadrons that undergo deeply inelastic scattering. As the formalism of PQCD developed, partons were shown to be excitations of elementary quantum fields of spins 1/2 and 1 -- quarks, antiquarks, gluons, and even photons -- in a hadron undergoing a hard scattering process. 

Understanding of the structure of the proton is being updated continuously using the wealth of LHC and other world data. Knowledge of the structure of the nucleon needed for a large class of theoretical predictions in PQCD is encoded in the collinear parton distribution functions (PDFs). The PDFs have been determined from data starting in the early 1980s \cite{Duke:1983gd, Eichten:1984eu, Gluck:1980cp}. They are now determined using the method of the {\it global QCD analysis} \cite{Morfin:1990ck, Martin:1987vw, Martin:1988aj, Harriman:1990hi, Owens:1991ej} from experimental measurements at colliders
such as HERA, the Tevatron, and the LHC, and in fixed-target experiments. 
The PDFs are provided in a number of practically useful forms
by several collaborations, including ABM
\cite{Alekhin:2017kpj}, HERAPDF \cite{Abramowicz:2015mha}, CT
\cite{Dulat:2015mca}, CTEQ-JLab \cite{Accardi:2016qay}, MMHT
\cite{Harland-lang:2014zoa}, and NNPDF \cite{Ball:2017nwa}. The modern
PDF parameterizations are provided with families of ``error PDF sets''
\cite{Giele:1998gw,Giele:2001mr,Pumplin:2001ct} that allow the user
to assess the total uncertainty on the PDFs arising from a
variety of experimental and theoretical errors. Methods for
statistical combination of PDF ensembles from various groups exist
\cite{Gao:2013bia,Carrazza:2015aoa,Carrazza:2015hva}, and 
comprehensive guidelines on uses of PDFs at the LHC are published by
the PDF4LHC working group \cite{Butterworth:2015oua}. As of  2020,
close to seven hundred PDF ensembles from various groups are 
distributed as numerical tables in the standard {\it Les Houches Accord
  PDF} (LHAPDF)
format \cite{Buckley:2014ana, Giele:2002hx,
  whalley:2005nh,bourilkov:2006cj} from a public online repository.\footnote{https://lhapdf.hepforge.org.}

The parton distributions in nuclei are equally analyzed by several
collaborations: EPPS \cite{Eskola:2016oht}, nCTEQ
\cite{Kovarik:2015cma}, DSSZ \cite{Deflorian:2011fp}, HKN
\cite{Hirai:2007sx} KA \cite{Khanpour:2016pph} and NNPDF \cite{AbdulKhalek:2019mzd}. They are 
provided either as parametrizations of the nuclear PDFs themselves or
the nuclear correction factor applied to a predefined reference proton
PDF. 

Increasingly precise requirements will be imposed on the
determination of PDFs and their uncertainties during the
high-luminosity (HL) phase of the LHC operation to precisely measure
Higgs boson couplings and electroweak parameters
\cite{Deflorian:2016spz}, and to maximize the HL-LHC reach in a
variety of tests of the Standard Model and new physics searches
\cite{Atlas:2019qfx}.

The purpose of this article is to review select topics related to the theoretical definition, determination, and usage of PDFs in modern
applications.  We will concentrate on  methodological
aspects of the PDF analysis that will be of growing importance in the near-future LHC era. We primarily
focus on theoretical and statistical aspects of the determination of
PDFs in the nucleon and nuclei,  notably, on proper theoretical definitions,
statistical inference of the PDF parameterizations from the
experimental data, and factorization for heavy nuclei.
This work supplements the recent
reviews of phenomenological applications of PDFs available in
\cite{Forte:2013wc,Gao:2017yyd}, as well as extensive comparisons
\cite{Alekhin:2011sk,Watt:2012tq,Butterworth:2015oua,Accardi:2016ndt} of PDFs
from various collaborations and QCD predictions based on these PDFs.
Introductory texts on the
fundamentals of QCD factorization, global PDF analysis, and collider
applications of PDFs are available, e.g.,
in \cite{Collins:2011zzd, Brock:1993sz, Campbell:2017hsr}.

We begin with the parton model and its relation to QCD, the field theory of the strong interactions. 

\subsection{The parton model\label{sec:PartonModel}}

The principal aim of contemporary particle physics is to test the current theory, the Standard Model, and to look for evidence of new physics that is not included in the Standard Model. The Standard Model is a renormalizable quantum field theory with fields for leptons ($\mathrm{e}, \mu, \tau$), their associated neutrinos, quarks (d,u,s,c,b,t), vector bosons ($\gamma, \mathrm{W}^\pm, \mathrm{Z}, \Lg$), and a scalar boson field, the Higgs field. The part of the theory involving the photon, $\mathrm{W}^\pm, \mathrm{Z}$ and Higgs fields involves spontaneous symmetry breaking and is quite subtle. The part of the theory involving the gluon, g, constitutes the theory of the strong interactions, QCD. The gluon couples to quarks, but not to leptons. We cannot offer a review of the Standard Model, but we assume that the reader has some familiarity with it. See, for instance, \cite{Peskin:1995ev, Srednicki:2007qs, Aitchison:2004cs, Schwartz:2013pla}.

One important feature of the Standard Model, and quantum field theory in general, is that the couplings that appear at the vertices of Feynman diagrams representing the theory are best considered to be dependent on a squared momentum scale $\mu^2$. For instance, the QCD coupling constant, $g_\Ls^2/(4\pi) \equiv \as$, becomes $\as(\mu^2)$. The dependence on $\mu^2$ is derived from the theory, even though the value of $\as(\mu^2_0)$ at a particular scale like $\mu^2_0 = M_\LZ^2$ is a free parameter of the theory. Which value of $\mu^2$ is useful in addressing a particular physical problem depends on the typical momentum scale of the problem.

The electroweak coupling, $\alpha(\mu^2)$, increases with $\mu^2$, but for physically relevant momentum scales, $\alpha(\mu^2)$ is small. Thus scattering cross sections for electroweak interactions can be usefully computed as a perturbation series in the small parameter $\alpha(\mu^2)$.

The QCD coupling constant, $\as(\mu^2)$, decreases with $\mu^2$. It is large when $\mu$ is of order 1 GeV or less, and it reduces to $\as(\mu^2) \approx 0.1$ at $\mu = 100 \GeV$. This scale dependence of $\as(\mu^2)$ tells us that perturbation theory should be useful for describing the parts of a physical scattering process with hadrons that involves only very large momentum scales.

However, we run into a complication. How can we use experiments involving protons to investigate the Standard Model? For instance, we know from experiment that we can make Higgs bosons in proton-proton collisions, but how can we understand the Higgs production process quantitatively? The Higgs boson production cross section depends on the internal structure of the initial-state protons. The strong coupling $\as(\mu^2)$ is large for $\mu$ of order the proton mass, $m_\Lp$, or smaller. This means that we cannot expect perturbation theory to be useful for calculating the structure of the proton as a bound state of quarks, antiquarks, and gluons.

To understand the problem and its tentative solution, consider deeply inelastic electron scattering from a proton (DIS). The DIS process is reviewed in Sec.~\ref{sec:DIS}. Suppose an electron scatters from a proton of momentum $P_\LA$ by exchanging a photon with momentum $q^\alpha$.\footnote{We denote Lorentz indices with Latin letters and use the $\{+,-,-,-\}$ sign convention for the Minkowski metric tensor.} We define $Q^2 \equiv - q^2$. Since the 4-momentum $q^\alpha$ is spacelike, we have $Q^2 > 0$.  We demand that $Q^2$ be much larger than $1\mbox{ GeV}^2$. Then there is some hope for a perturbative approach that uses an expansion in powers of $\as(Q^2)$. The lowest-order Feynman diagram, corresponding to $t$-channel electromagnetic scattering $e+ q\!\!\!\!{}^{{}^{\small (-)}}\rightarrow e+q\!\!\!\!{}^{{}^{\small (-)}}$ of an electron on a quark or antiquark, is pretty simple. But how do we relate the initial-state quark field to the proton?

Here is an approach we can follow. Let us use a ``brick-wall'' reference frame in which $\vec q$  lies entirely in the $-z$ direction, $q^0 = 0$, and in which $\vec P_\LA$ lies entirely in the $+z$ direction. Then $|q^z| = Q$, and we can consider the hard quark-photon interaction to be localized in a time interval $\Delta t_\textsc{h} = 1/Q$. In DIS, we also demand that $P_\LA\cdot q$ be large, of order $Q^2$. Then the proton momentum is large, with $P_\LA^z = (P_\LA\cdot q)/Q$. This means that the proton is highly boosted, with a boost factor $e^\omega = (2P_\LA\cdot q)/(m_\Lp Q)$. In the proton rest frame, we can suppose that the time period between successive quark-gluon interactions is of order $\Delta t_s^{\mathrm{rest}} \sim 1/m_\Lp$. In the ``brick-wall'' reference frame, the typical time for these soft internal interactions is $\Delta t_\textsc{s} \sim e^\omega /m_p = 2P_\LA\cdot q/(m_\Lp^2 Q)$ as a consequence of relativistic time dilation. The time interval $\Delta t_\textsc{s}$ is much longer than the hard interaction time scale: $\Delta t_\textsc{s}/\Delta t_\textsc{h} \sim 2P_\LA\cdot q/m_\Lp^2$. This argument implies that the quark-gluon interactions inside the proton are largely frozen during the time interval $\Delta t_\textsc{h}$ while the hard scattering interaction with the virtual photon is taking place. The struck quark is effectively free.

This gives us the parton model of Feynman and Bjorken \cite{Bjorken:1968dy, Feynman:1973xc}.  The word {\em parton} as used now refers to a quark, antiquark, gluon, and sometimes to a photon. Originally, partons meant just the constituents of the proton, whatever those constituents might be. 

It is easy to use the parton model to describe the cross section for DIS. We assume that, inside a fast moving proton, there are partons of various types $a$. A parton can carry a momentum $k^\alpha$ that is a fraction $\xi$ of the momentum of the proton: $k^\alpha = \xi P_\LA^\alpha$. Denote by $f_{a/\Lp}(\xi) d\xi$ the probability to find such a parton with momentum fraction between $\xi$ and $\xi + d\xi$. Let $d\hat\sigma/(dQ^2 d(P_\LA\cdot q))$ be the cross section for the electron to scatter from a free quark with momentum $k = \xi P_\LA$, calculated in lowest-order perturbation theory. Then the cross section for this scattering from a proton should be
\begin{equation}
\label{eq:partonmodel}
\frac{d\sigma}{dQ^2\, d(P_\LA\cdot q)} = \sum_{a}\int\!d\xi\
f_{a/\Lp}(\xi)\,
\frac{d\hat\sigma}{dQ^2\, d(P_\LA\cdot q)}
\;.
\end{equation}
This picture is simple and intuitive. An analogous picture covers processes such as Higgs boson production in high energy proton-proton collisions.

Unfortunately, the parton model as just described does not survive scrutiny when we include gauge field interactions at higher orders of $\as$: one encounters contradictions as soon as one tries to calculate $d\hat\sigma/[dQ^2\, d(P_\LA\cdot q)]$ beyond the leading order in QCD. The basic problem is that QCD is a quantum field theory, in which interactions among the quarks and gluons occur at all time or distance scales, including time scales much smaller than the $\Delta t_\textsc{s}$ scale that one would naively associate with the interior motions inside a highly boosted proton.

Nevertheless, one can turn the parton model picture into a theoretically consistent framework that includes higher-order radiative contributions. One first needs to carefully define what one means by a parton distribution function $f_{a/\Lp}(\xi)$. With a careful definition, the parton distribution functions become $f_{a/\Lp}(\xi, \mu^2)$, with a dependence on a scale $\mu^2$. Then cross sections for DIS and for many processes in hadron-hadron scattering have a property known as {\em factorization}, which means that they satisfy a formula very similar to Eq.~(\ref{eq:partonmodel}).

\subsection{Cross sections and factorization}

In this paper, we will review how the PDFs are systematically defined in the QCD theory and determined by applying statistical inference to experimental observations sensitive to the PDFs. We will also review the general formalism to estimate the uncertainty on the PDFs that results from fitting the experimental data. 

It is most common to define and determine the  the PDFs in the nucleon in the $\overline {\mathrm{MS}}$ factorization scheme discussed in some detail in Sec.~\ref{sec:definition}. Intuitively, these functions, $f_{a/A}(\xi,\mu^2)$, represent the probability to find a parton of type $a$ (a gluon or a particular flavor of quark or antiquark) in a hadron of type $A$, for example a proton, as a function of the fraction $\xi$ of the momentum of the hadron that is carried by the parton. The argument $\mu^2$ in $f$ indicates the momentum scale at which the parton distribution function applies. The $\mu^2$ dependence is given by the
Dokshitser-Gribov-Lipatov-Altarelli-Parisi (DGLAP) evolution equations
\cite{Dokshitzer:1977sg,Gribov:1972ri,Altarelli:1977zs} that
we describe in Sec.~\ref{sec:PDFDGLAP}. With the aid of these evolution
equations, the functions $f_{a/A}(\xi,\mu^2)$ can be determined from
the functions $f_{a/A}(\xi,\mu_0^2)$ at a scale $\mu_0^2$ that is
typically chosen to be around $1 \GeV^2$. The functions
at scale $\mu_0^2$ cannot be calculated  in perturbation theory.

The PDFs at the starting scale $\mu_0^2$ are determined from
experimental data. Consider first a cross section $\sigma[F]$, defined by integrating the completely differential cross section
for any number of final-state particles, multiplied by functions $F$
that describe what is measured in the final state. For instance, if
we observe a single weakly interacting particle, the measurement function $F$
might be simply a product of delta functions that specify the energy
and direction of momentum of the particle. We thus integrate over the momenta of particles that are not measured, giving us an inclusive cross section. The observable $\sigma[F]$ must be ``infrared-safe'', as described later in Sec.~\ref{sec:IRsafety}. For lepton-hadron
scattering, the cross section $\sigma[F]$
is related to parton distributions by
\begin{equation}
\label{eq:factorizationDIS}
\sigma[F] \approx \sum_{a}\int\!d\xi\
f_{a/A}(\xi,\mu^2)\,
\hat \sigma[F]
\;.
\end{equation}
For cross sections at hadron colliders, a parton distribution function is needed for each of two colliding hadrons:
\begin{equation}
\label{eq:factorization00}
\sigma[F] \approx \sum_{a,b}\iint\!d\xi_a\ d\xi_b\
f_{a/A}(\xi_a,\mu^2)\, f_{b/B}(\xi_b,\mu^2)\
\hat \sigma[F].
\end{equation}
We will review this formula in more detail in Sec.~\ref{sec:factorization}.

In order to determine the PDFs at the starting scale $\mu_0^2$, one
selects observables that are sensitive to different combinations of
parton distributions. The parton distributions at scale $\mu_0^2$ are
parameterized by a sufficiently flexible functional form.
The observables are first calculated
using the parton distributions, where the free parameters are given some
initial values and are compared to data. The parameters are then
adjusted until the theoretical predictions describe the data well.  

\subsection{Practical issues in the theory}

Following this very simple strategy requires in reality a very
detailed understanding of many facets of perturbative
QCD.  First, as
the precision of the determination of parton distributions needs to
match the experimental precision, 
fitted observables are calculated at the next-to-next-to-leading order (NNLO) in perturbative QCD for nucleon PDFs. The NNLO accuracy in the global fits corresponds to computing the hard cross sections by including perturbative radiative contributions suppressed by up to two powers of $\alpha_s$. 
Such predictions at NNLO for hard processes suitable for determination of PDFs are increasingly available \cite{Moch:2004xu,Vermaseren:2005qc,Zijlstra:1991qc,vanNeerven:1991nn,Buza:1997mg,Catani:2007vq,Catani:2009sm,Gavin:2010az,Gavin:2012sy,Li:2012wna,Ridder:2015dxa,Gehrmann-DeRidder:2017mvr,Currie:2016bfm,Currie:2017ctp,Campbell:2010ff,Boughezal:2016wmq,Berger:2016inr}. There are also partial results at $\LN^3\LL\LO$, such as \cite{Vermaseren:2005qc,Moch:2007gx,Moch:2008fj,Bierenbaum:2009mv,Ablinger:2010ty,Ablinger:2014nga}.
Theoretical predictions used for nuclear PDFs are still typically at
next-to-leading-order (NLO),  even though first NNLO nuclear PDF analyses exist. 

Incorporating the theoretical calculations into the fit
requires a careful selection of observables that are theoretically
well-defined (infrared-safe) and can be calculated up to the required
order. Due to the nature of higher-order calculations,
the numerical evaluation can be time-consuming. This shortcoming is
usually solved either by using precomputed tables of computationally {\it slow}
point-by-point NNLO corrections applied to {\it fast} NLO
calculations, or increasingly by using fast gridding techniques, such as the ones implemented
in the {\tt fastNLO} \cite{Wobisch:2011ij}, {\tt APPLGRID}
\cite{Carli:2010rw}, {\tt aMCFast} \cite{Bertone:2014zva}, and NNPDF
     {\tt FastKernel} \cite{Forte:2010ta} programs.
     The comparably accurate and fast DGLAP evolution of PDFs
up to NNLO accuracy is implemented in a number of
public codes: {\tt PEGASUS} \cite{Vogt:2004ns},
{\tt HOPPET} \cite{Salam:2008qg}, {\tt QCDNUM}
\cite{Botje:2010ay}, or {\tt APFEL} \cite{Bertone:2013vaa}.

Even after implementing the measured observables and
the corresponding DGLAP evolution for the PDFs at (N)NLO, one still has to
address a number of issues that become important as the PDF analysis
is pushed towards higher precision. On the experimental side,
the NNLO PDFs are increasingly
constrained by high-luminosity measurements, in which the statistical
experimental errors are small, and adequate implementation of many
(sometimes hundreds) of correlated systematic uncertainties is
necessary. Commonly followed procedures for implementation of
systematic uncertainties in the PDF fits are reviewed in Appendix A of Ref.~\cite{Ball:2012wy}. We discuss the treatment of systematic uncertainties in some detail in Sec.~\ref{sec:Statistics}.

From the side of theory, subtle radiative contributions, such as NLO
electroweak or higher-twist contributions, are comparable to NNLO QCD
contributions in some fitted observables. The photon constituents
contribute at a fraction-of-percent level to the total momentum of the
proton. The associated parton distribution for the {\it photon} can be
computed very accurately using the structure functions and nucleon
form factors from lepton-hadron (in)elastic scattering as the input \cite{Manohar:2016nzj,Manohar:2017eqh}.
The resulting {\tt LUXqed} parameterization
of the photon PDF is already implemented in \cite{Bertone:2017bme,Nathvani:2018pys}.
An alternative is to fit a phenomenological parameterization of the photon PDF at the initial scale of evolution together with the rest of the PDFs. Such phenomenological parameterizations constrained just by the global fit  \cite{Ball:2013hta, Schmidt:2015zda,Giuli:2017oii} are less precise than the {\tt LUXqed} form.

Even at NNLO, the residual theoretical uncertainties due to missing
higher-order contributions in $\alpha_s$
may have an impact on PDFs. Such theory uncertainties are partly correlated in a generally
unknown way across experimental data points.
In addition to the traditional estimation of higher-order
contributions by the variation of factorization and renormalization
scales, recently, more elaborate methods for estimation of
higher-order uncertainties have been explored with an eye on
applications in the PDF fits, such as \cite{Olness:2009qd,Cacciari:2011ze,Gao:2011unpublished,Forte:2013mda,Harland-Lang:2018bxd,AbdulKhalek:2019ihb,AbdulKhalek:2019bux}. See the discussion at Eq.~(\ref{eq:Dkdefexact}).
%\begin{figure*}[tb]
%\centering
%\includegraphics[width=0.48\textwidth]{./figs/alphas_scanTct18_pj814a2.pdf}
%\includegraphics[width=0.48\textwidth]{./figs/alphas_scanTct18z_pj814b2.pdf}
%\caption{Dependence of the increase in $\chi^2$ on the strong coupling at the scale of %$M_Z$ for CT18 (left) and CT18Z (right) PDFs at NNLO, 
%shown for all experiments and separately for the DIS, Drell-Yan (DY) pair production, and jet/top pair production data sets. Reproduced from Ref.~\cite{Hou:2019efy}. 
%	}
%\label{fig:lm_alphas}
%\end{figure*}

\subsection{The strong coupling}

Another associated issue is the
treatment of the strong coupling $\alpha_s(\mu^2_R)$. The strong
coupling depends on a scale $\mu^2_R$, called the renormalization
scale. For a review, see any text on QCD, for example
\cite{Collins:2011zzd}. Since $\alpha_s(\mu^2_R)$ obeys a
renormalization group equation, its value at any $\mu^2_R$ can be
determined from its value at a fixed scale $\mu_{R0}^2$. Normally, one sets
$\mu_{R0}$ to be the mass $M_\LZ$ of the $Z$-boson.  All QCD observables 
depend on $\alpha_s(M_\LZ^2)$, and so  do the fitted PDFs. Conventionally, the
world-average value of  $\alpha_s(M_\LZ^2)$ \cite{Tanabashi:2018oca}
is derived from a combination of experimental measurements, with the
tightest constraints imposed by the QCD observables that do not depend on
the PDFs, notably, hadroproduction in electron-positron
collisions, hadronic $\tau$-decays, and quarkonia masses.

Other useful constraints on $\alpha_s$ are imposed by a variety of hadron-scattering
observables (lepton-hadron DIS, jet and
$t\bar{t}$ production, ...) that are simultaneously sensitive to PDFs,
but the constraints of this class
are generally weaker and more susceptible to systematic effects.
As some hadronic observables of the latter class are also included in the global
fit to constrain the PDFs, in principle these observables can 
determine both $\alpha_s$ and the PDFs at the same time.
Consequently several treatments of $\alpha_s$
exist in the current PDF analyses. Most PDF groups publish some global fits
that determine $\alpha_s(M_\LZ^2)$ and PDFs simultaneously. They
typically find that the best-fit $\alpha_s(M_\LZ^2)$ is consistent with in the world average of $\alpha_s(M_\LZ^2)$, but with a considerably larger uncertainty than in the world average. 
ABM fits are representative of this approach \cite{Alekhin:2017kpj}, as well as the dedicated studies performed in the global framework \cite{Ball:2018iqk,Thorne:2018ykq}. 

As an example, the best-fit $\alpha_s(M_\LZ^2)=0.1166\pm 0.0027$ in CT18 at NNLO \cite{Hou:2019efy} is consistent with the world average $\alpha_s(M_\LZ^2)=0.1181 \pm 0.0011$ \cite{Tanabashi:2018oca}, as well as with $\alpha_s(M_\LZ^2)\approx 0.1176$ in \cite{Thorne:2018ykq} and a somewhat higher $\alpha_s(M_\LZ^2)=0.1185 \pm 0.0012$ in \cite{Ball:2018iqk}. The CT18 value results as some trade-off between the DIS experiments (notably, fixed-target DIS), which collectively prefer a somewhat lower $\alpha_s(M_\LZ^2)\approx 0.115$, and jet+top and Drell-Yan experiments, which collectively prefer a higher $\alpha_s(M_\LZ^2)\approx 0.119$ \cite{Hou:2019efy}. The quoted uncertainty on $\alpha_s(M_\LZ^2)$ at 68\% probability level varies among the PDF-fitting groups in 2020 from about 0.001 to 0.0025 depending on the adopted definitions for the uncertainty (with the CT18 uncertainty being the most conservative).  

On the other hand, it is often advantageous to perform the PDF
fits and determine the PDF uncertainty at a fixed world-average value of
$\alpha_s(M_\LZ^2)$, then estimate the $\alpha_s$ uncertainty of the fits by using a few PDF fits with alternative $\alpha_s$ values. If the uncertainties
obey a Gaussian probability distribution, 
it can be rigorously demonstrated that, to compute the total
PDF+$\alpha_s$ uncertainty that includes all correlations, it suffices to
add the resulting PDF and $\alpha_s$ uncertainties in quadrature
\cite{Lai:2010nw}. The empirical probability distributions in the PDF fits
are indeed sufficiently close to being Gaussian, so this prescription for computing the PDF+$\alpha_s$ uncertainty is adopted by the majority of recent fits. 
For example, the PDF4LHC group recommends \cite{Butterworth:2015oua}
calculating the PDF+$\alpha_s$ uncertainty at the 68\% confidence level 
by adding in quadrature the PDF uncertainty computed using
30 (100) PDF4LHC15 error sets for  the world-average
$\alpha_s(M_\LZ^2)=0.1180$, and the $\alpha_s$ uncertainty
computed from two best-fit PDF sets for $\alpha_s(M_\LZ^2)=0.1165$ and
$0.1195$.

\subsection{Heavy-quark masses}

Another issue that needs to be addressed in global fits of PDFs is the
treatment of massive charm and bottom quarks. Mass effects play an
important role in describing, for example, subprocesses with charm (anti)quarks in
DIS. There are several approaches to treating the mass of the quark,
such as the zero-mass variable-flavour-number (ZM-VFN) scheme
\cite{Collins:1978wz,Collins:1986mp} or the 
general-mass variable-flavor-number (GM-VFN) scheme
\cite{Collins:1998rz,Aivazis:1993pi,Kramer:2000hn,Thorne:1997ga,Thorne:2006qt,Forte:2010ta}. In Sec.~\ref{sec:heavyquarks}, we provide a short pedagogical
introduction to the otherwise extensive topic of the treatment of masses
for heavy quarks. For a more thorough review of the heavy-quark schemes we refer the reader to \cite{Butterworth:2015oua,Accardi:2016ndt}. 

\subsection{Special kinematic regions}
\label{sec:specialregions}

Some kinematic regions require special treatment if their respective experimental data are to be included in a global fit of PDFs. 

One such region is where the typical momentum fraction $\xi$ is large,
but the momentum scale of the process is not too large. Then the nonzero mass
of the proton (or in general the target), normally neglected, may
need to be taken into account. These target-mass corrections are
discussed in detail in \cite{Schienbein:2007gr}. In the same region,
deuterium nuclear corrections also have to be considered, when
data in this specific kinematic region are taken on deuterium rather
than proton targets \cite{Accardi:2009br,Accardi:2011fa}. 

The other kinematical region in need of careful treatment is the one where the
momentum transfer $Q^2$ is low, typically below 4 $\mbox{GeV}^2$, at
moderate or large $\xi$. Here power corrections can become
important and can be taken into account, for example as in
\cite{Martin:2003sk,Alekhin:2012ig,Thorne:2014toa}. 

Yet another such region arises in DIS at small $x$ and
$Q$, roughly satisfying $Q^2 < A_{cut}/x^\lambda$ with $A_{cut}\sim
0.5-1.5\mbox{ GeV}^2$, and $\lambda \sim 0.3$ \cite{GolecBiernat:1999qd,Caola:2009iy}. This is
the limit where summation of small-$x$ logarithms will become
necessary, and indeed, a slowdown in perturbative convergence of inclusive DIS cross sections and
resulting small-$x$ PDFs is observed even at NNLO in the affected HERA
region \cite{Abramowicz:2015mha}. The DIS data in this region provide valuable constraints on the small-$x$
behavior of the gluon PDF. The small-$x$ instability in DIS can be cured to a
certain extent by inclusion of power-suppressed (higher-twist) contributions
\cite{Harland-Lang:2016yfn} or, quite effectively in the HERA region
at $Q^2 > 4\mbox{ GeV}^2$,
by using an $x$-dependent factorization scale $\mu^2$ in NNLO
DIS cross sections in some of the PDF sets (CT18X and CT18Z) published in \cite{Hou:2019efy}. Summation of small-$x$ logarithms, matched
to NNLO, has been
successfully implemented in NNPDF \cite{Ball:2017otu}
and {\tt xFitter} analyses \cite{Abdolmaleki:2018jln}. It results
in even better description of the accessible small-$x$ region. Either NNLO+NLLx summation, as in \cite{Ball:2017otu}, or the choice of a special $x$-dependent factorization scale in the {\it fixed-order} NNLO cross section, as in  \cite{Hou:2019efy}, thus leads to a better description of the small-$x$ subsample of the HERA DIS data. The resulting PDFs obtained after these changes tend to have elevated gluon and strangeness components at $x<10^{-2}$ as a result of slower $Q$
dependence of the DIS cross sections at small $x$ than would be predicted at a fixed order with a standard scale $\mu=Q$. LHC predictions based on such modified PDFs, such as CT18Z, may lie outside of the nominal error bands of the PDF set with default choices, such as CT18. 

\subsection{Fitting}

After addressing all necessary features of theory predictions such as the ones spelled out in the previous paragraphs, one compares the theory predictions to the experimental data. The process of fitting the theoretical predictions to data by adjusting the PDFs is the main focus of this review. The reason is that proper determination of PDF uncertainties will be highly important for the analysis of the high-luminosity LHC data, as the PDF uncertainty will soon dominate systematic uncertainties on the theory side in key tests of
electroweak symmetry breaking, including the measurements of Higgs couplings and mass of the charged weak boson \cite{Deflorian:2016spz, Atlas:2019qfx}. The statistical framework of the PDF fits is fundamentally more complex than the one in the electroweak precision fits: while the parametric model of the electroweak fits is uniquely determined by the Standard Model Lagrangian, the parametric model for the parton distribution functions may change within some limits in order to optimize agreement between QCD theory and data.

Consequently, the PDF uncertainty is comprised of four categories of contributions:
\begin{enumerate}
  \item {\it Experimental uncertainties}, including statistical and
 correlated and uncorrelated systematic 
    uncertainties of each experimental data set;
  \item {\it Theoretical uncertainties}, including the absent higher-order and
    power-suppressed radiative contributions, as well as uncertainties in using parton showering programs to correct the data in order to compare to fixed-order perturbative cross sections;
  \item {\it Parameterization uncertainties} associated with the choice
    of the PDF functional form;
  \item {\it Methodological uncertainties}, such as those associated with the
    selection of experimental data sets, fitting procedures, and 
    goodness-of-fit criteria.
\end{enumerate}

\begin{figure*}
\center{\includegraphics[width=0.40\textwidth]{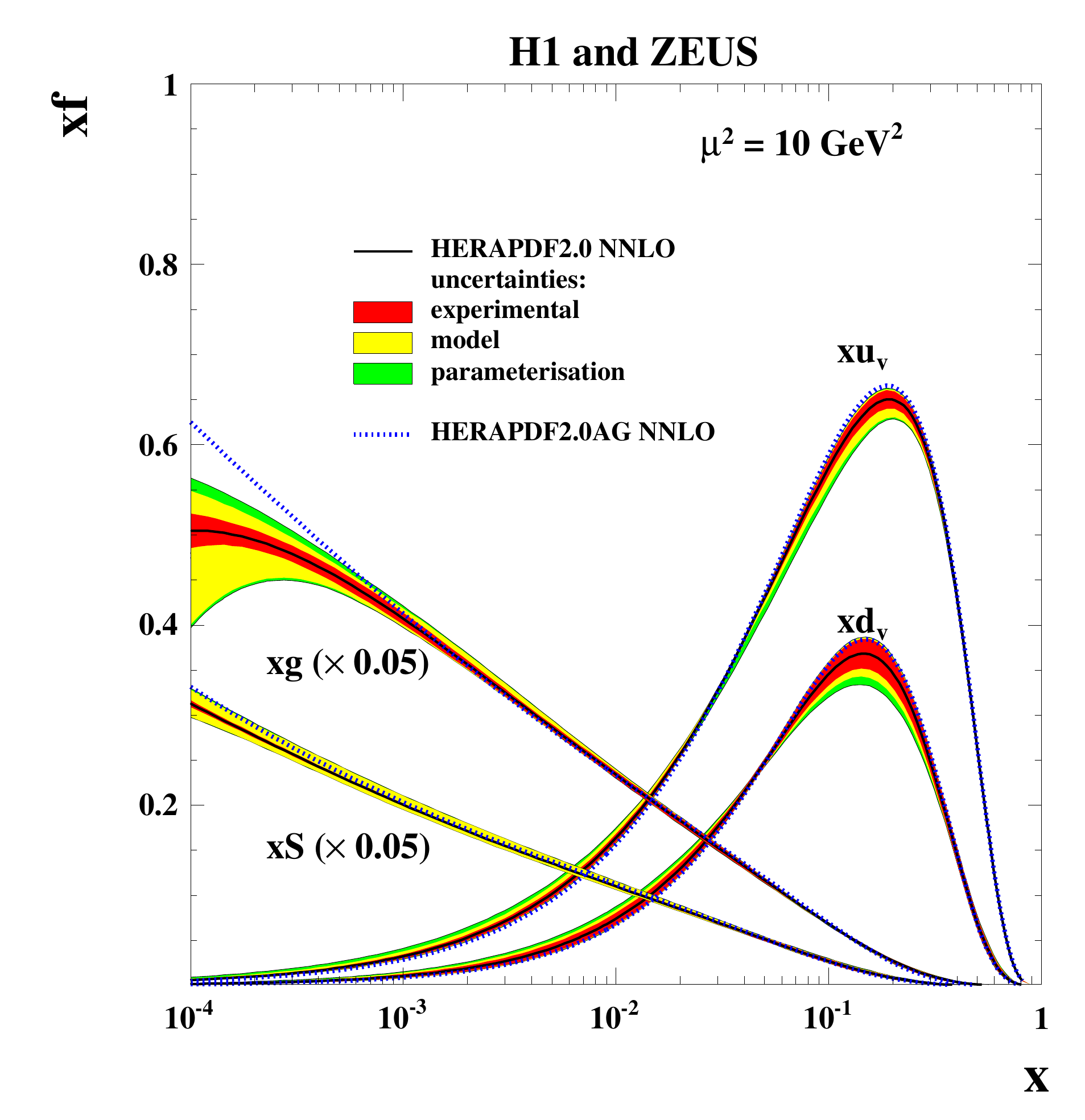}\quad \includegraphics[width=0.57\textwidth]{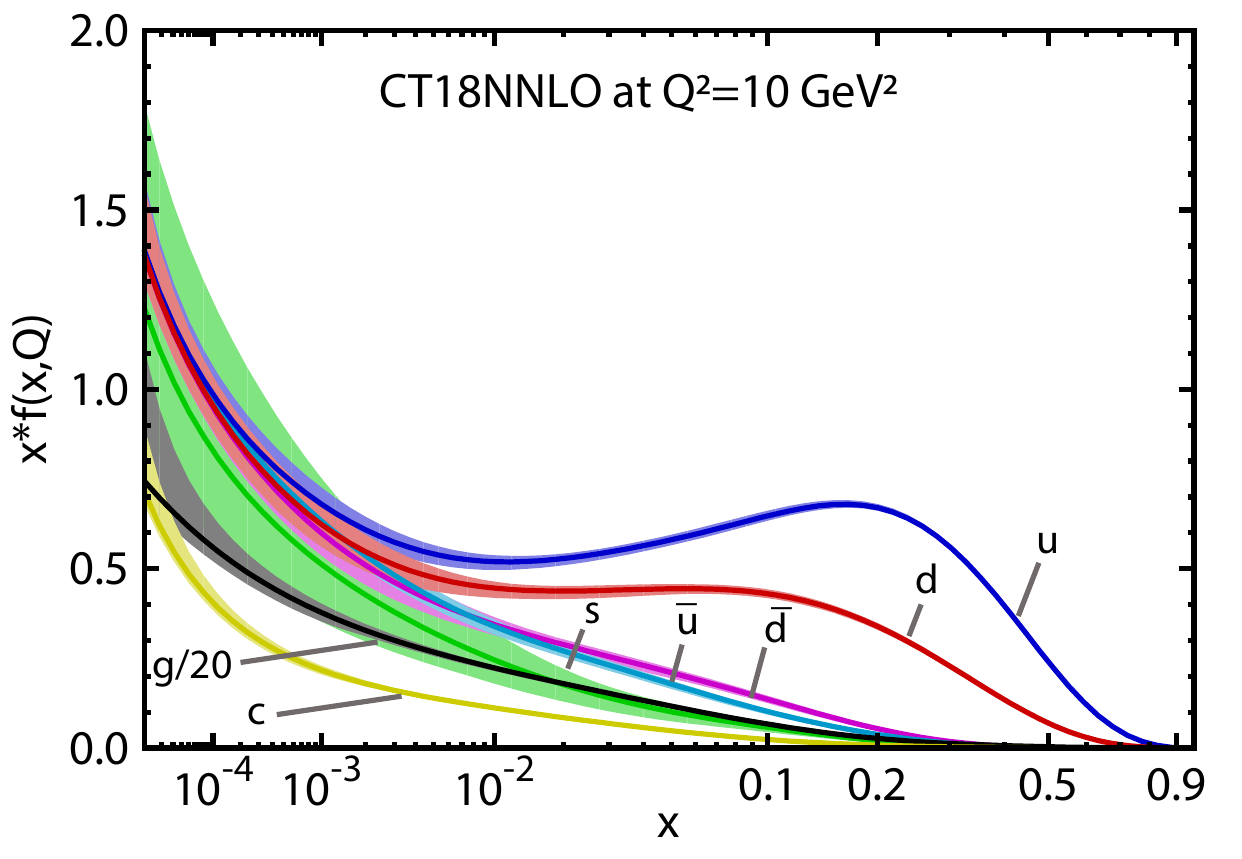}}
\caption{Left: The parton distribution functions $x u_v\equiv x (u-\bar u)$,
  $x d_v\equiv x (d-\bar d)$, $x S\equiv 2 x (\bar u +\bar d + \bar s
  + \bar c)$ and $x g$ of HERAPDF2.0 NNLO at $\mu^2 = 10 \mbox{
    GeV}^2$. The experimental, theoretical model and parameterization
  uncertainties are shown separately. From \cite{Abramowicz:2015mha}.
  Right: the PDF uncertainty bands for CT18
NNLO PDFs \cite{Hou:2019efy} at $\mu^2 = 10 \mbox{ GeV}^2$. }
\label{fig:pdfbands}
\end{figure*}

As an illustration, the left panel of Fig.~\ref{fig:pdfbands}
shows the HERAPDF2.0 parameterizations
determined from the fits exclusively to DIS data.
The PDF uncertainty corresponding to the PDF solutions covering
68\% of the cumulative probability
is comprised of the experimental, theoretical model, and
parameterization components that were estimated for a select fitting methodology \cite{Abramowicz:2015mha}. Other groups may not separate all four components listed above in the total PDF uncertainty. 
In the right panel of
Fig.~\ref{fig:pdfbands}, the CT18 NNLO PDF uncertainty bands
are evaluated for 68\% cumulative probability
according to a two-tier goodness-of-fit
criterion \cite{Lai:2010vv} that accounts both for the agreement with the totality of
fitted data and with individual experimental data sets.  The CT18 analysis includes a variety of data sets on DIS, vector boson, jet, and $t\bar t$ production. While this diversity of data allows one to resolve differences between PDFs of various flavors and probe a broader range of PDF parameterization forms, in practice some incompatibilities (``{\it tensions}'') between constraints on the PDFs from various experiments are introduced and need to be either eliminated or accounted for in the PDF uncertainty estimate.  [The CT18 and HERAPDF2.0 PDFs are fitted to 3690 and 1130 data points,
  respectively. About 100 different parameterization forms have been
  tried in the CT18 analysis, contributing to the spread of the PDF uncertainty.] The width of
the CT18 error bands thus depends on a two-level {\it tolerance} convention
\cite{Lai:2010vv,Pumplin:2002vw} that is
adjusted so as to reflect PDF variations associated with some disagreements between experiments, parameterization and
theoretical uncertainties. We notice, for example, that the
CT18 error bands for some poorly constrained flavors, notably, 
the strangeness PDF $xs(x,\mu^2)$ at small $x$ (green band), may be broader than the
respective HERAPDF2.0 error bands at the same probability level, despite having more experimental data included in the CT18 analysis compared to HERAPDF. The wider error bands reflect, for a large part, the spread in the acceptable PDFs estimated using the CT18 flexible parameterization forms, but also some inflation of the experimental uncertainty to reflect the imperfect agreement among experiments. Sec.~\ref{sec:experimentchisqall} shows how to examine several experiments for their agreement.

To find most likely solutions for PDFs and establish the respective
uncertainties, one must answer a fundamental question: how good, actually, is each PDF fit? We explore this question and advocate using a {\it
  strong} set of goodness-of-fit criteria that go beyond the {\it weak}
criterion based on just the value of the goodness-of-fit function $\chi^2$. 

All PDF fitters employ some version of  minimization of the goodness-of-fit function  
\begin{equation}
	\chi^2(a) = \sum_{ij}(D_i-T_i(a))(D_j-T_j(a))C_{ij}
	\;,
\end{equation}
where $D_i$ are the data values, $T_i(a)$ the corresponding theory
predictions, which depend on free PDF parameters $a$, and $C_{ij}$ is the
covariance matrix. The goodness-of-fit function is used to assess the quality of
the theoretical description of the data and to estimate the
uncertainty in the determination of the fit parameters $a$. The
statistical foundations that motivate the use of the goodness-of-fit functions
are discussed in Sec.~\ref{sec:Statistics}.  While it is most common to find the global minimum of $\chi^2(a)$ numerically, many insights about the global fits can be gleaned analytically in the so called Hessian approach to fitting the PDFs,  first developed in
\cite{Stump:2001gu,Pumplin:2001ct} and refined ever since. We will discuss this
approach in a great detail in order to illustrate various aspects of the fits. It relies on the
observation that the PDFs approximately obey the
multivariate Gaussian probability distribution in the well-constrained kinematic
regions, which in turn allows one to derive the key outcomes of the PDF
analysis in a closed algebraic form. For example, the Hessian method is commonly used at the end of the fit to quantify the uncertainty on the resulting PDFs. There is a powerful alternative
approach, discussed only briefly in this review, to find the best-fit PDFs and determine their uncertainties using stochastic
(Monte-Carlo) sampling of PDFs \cite{Giele:1998gw, Giele:2001mr}
and PDF parameterizations by neural networks \cite{Forte:2002fg}. 
Although our results are demonstrated in the Hessian approximation, they also
elucidate the numerical outcomes of the Monte-Carlo
sampling PDF analyses such as the one used by NNPDF. They also
apply to the approximate techniques
for updating the published PDF ensembles with information from new
data by statistical reweighting of PDF replicas in the 
Monte-Carlo \cite{Giele:1998gw, Ball:2010gb, Ball:2011gg,
  Sato:2013ika} or Hessian \cite{Paukkunen:2014zia, Schmidt:2018hvu,
  Watt:2012tq} representations. 

 Sec.~\ref{sec:goodness} is devoted to the discussion of tests one can
 perform to determine the extent to which the fitting procedure is
 consistent with the statistical hypotheses used in the procedure. 
 This leads to a discussion of the strong goodness-of-fit 
set of criteria.

% !TEX root = main.tex
\section{Review of theory}\label{sec:Theory}

In this section, we provide a brief overview of the theory of PDFs and their relation to cross sections. We start with the definition of PDFs as matrix elements of quantum field operators. Then we discuss the factorization property of QCD, which allows us to relate certain kinds of cross sections to PDFs and perturbatively calculated quantities. Finally, we turn to the treatment of heavy quarks in these relations, although we treat this complex subject only briefly.

%-------------------------------------------------------------------
\subsection{Definition of parton distribution functions
\label{sec:definition}}

In this subsection, we give definitions for PDFs as matrix elements in a proton (or other hadron) of certain operators. 
Instead of simply stating the definitions, we motivate them from basic field theory, following the reasoning in \cite{Collins:1981uw}. For more details, one can consult the book \cite{Collins:2011zzd}.\footnote{Our conventions follow the Particle Data Group \cite{Tanabashi:2018oca} and the book \cite{Collins:2011zzd}. In particular, we choose the sign of the strong coupling $g$ so that the quark-gluon vertex is $-\mi g \gamma^\mu t^a$. This is the opposite from the choice in \cite{Collins:1981uw}.}

\subsubsection{Momenta} \label{sec:momenta}

Consider a proton with momentum $P$ along the +$z$ direction. We define $+$ and $-$ components of vectors using $v^\pm = (v^0 \pm v^3)/\sqrt 2$. Then $P$ has components 
\begin{equation}
(P^+, P^-, \bm P^\perp) = 
\left(P^+,\frac{m_\Lp^2}{2 P^+},\bm 0 \right)
\;.
\end{equation}
It is helpful to think of $P^+$ as being very large, $9.2 \TeV$ for the LHC, but the size of $P^+$ does not matter for the definition of PDFs.

In the exposition below, we frequently have a function of a four-vector $v$. We use two alternative notations: either $f(v)$ or $f(v^+,v^-,\bm v^\perp)$.

We seek to define PDFs, $f_{a/\Lp}(\xi,\mu^2)$, which can be interpreted as giving the probability density for finding a parton of flavor $a$ (a quark, antiquark, or gluon), which carries a fraction $\xi$ of  $P^+$, in the proton with momentum $P$. This function depends on a momentum squared scale $\mu^2$ at which one imagines measuring the presence of the parton.

\subsubsection{Parton distributions in canonical field theory}\label{sec:PDFCanonicalQFT}

To get started with the definition of PDFs, consider an operator $b(\xi P^+, \bm k_\perp,s,c;i)$ that destroys a quark of flavor $i$ having helicity $s$, color $c$, +-momentum $\xi P^+$ and transverse momentum $\bm k_\perp$. This quark then carries a fraction $\xi$ of the $+$-momentum $P^+$ of the proton. The adjoint operator $b^\dagger(\xi P^+, \bm k_\perp,s,c;i)$ then creates a quark with the same quantum numbers. We normalize the creation and destruction operators to have anticommutation relations
\begin{equation}
\begin{split}
[b(\xi' &P^+, \bm k'_\perp,s',c';i)  ,b^\dagger(\xi P^+, \bm k_\perp,s,c;i)]_+
\\
={}& (2 \pi)^3 2\xi P^+ \delta(\xi' P^+ - \xi P^+)
\delta(k'_\perp - k_\perp)
\delta_{s's} \delta_{c'c}
\;.
\end{split}
\end{equation}
Additionally, we suppose that the vacuum state $\ket{0}$ has no quarks in it, so
\begin{equation}
b(\xi P^+, \bm k_\perp,s,c;i)\ket{0} = 0
\;.
\end{equation}

With the quark creation and destruction operator at hand, we can construct the operator that counts the number of quarks in a region of $\xi$ and $\bm k_\perp$:
\begin{align}
\rho(&\xi P^+, \bm k_\perp;i) 
\\& = 
\frac{1}{(2\pi)^3 2\xi} 
\sum_{s,c}b^\dagger(\xi P^+, \bm k_\perp,s,c;i)\,
b(\xi P^+, \bm k_\perp,s,c;i).
\notag
\end{align}
The reader can verify that, if $\ket{\Psi}$ is obtained by applying
quark creation operators to the vacuum, then the integral of $\rho$
over a
momentum-space volume ${\cal V}_3$ counts the number $N({\cal V}_3)$ of quarks in  ${\cal V}_3$:
\begin{equation}
\int_{{\cal V}_3}\!d\xi\,d\bm k_\perp\
\rho(\xi P^+, \bm k_\perp;i) \ket{\Psi} = N({\cal V}_3)\ket{\Psi}
\;.
\end{equation}
We want to define a parton density, the number of partons $f_{i/\Lp}(\xi)$ of flavor $i$ per unit $d \xi$ in a proton. We can take a matrix element of $\rho$ in a proton state to define this:
\begin{equation}
\label{eq:f0}
f_{i/\Lp}^{(0)}(\xi) \brax{P'}\ket{P}
=
\int\! d\bm k_\perp
\bra{P'}\rho(\xi P^+, \bm k_\perp;i)\ket{P}
\;.
\end{equation}
Here, for simplicity, we consider the proton to be spinless, but one can substitute a spin average: $\frac{1}{2}\sum_{s_\Lp} \bra{P',s_\Lp}\cdots\ket{P,s_\Lp}$. As noted at the beginning of this section, we take the proton momentum $P$ to be along the $z$-axis, so that $\bm P_\perp = 0$. However $P^+$ is arbitrary.

We have given $f$ a superscript $(0)$ to indicate that this is a preliminary version of the needed definition.

To make this definition more useful, we can relate the quark creation and destruction operators to the quark field operator $\psi_i(x)$. For this purpose, we use the version of QCD quantized on planes of equal $x^+ = (x^0 + x^3)/\sqrt{2}$ instead of planes of equal time $t = x^0$ \cite{Kogut:1969xa,Bjorken:1970ah}. The fields obey canonical commutation relations on planes of equal $x^+$. To make this work, we use the gauge $A^+(x) = 0$ for the gluon field. With this way of writing the theory, the two components of the four-component Dirac field projected by $P_{\rm dy} = \frac{1}{2}\,\gamma^- \gamma^+$ (such that $ P_{\rm dy}^2 = P_{\rm dy}$) are the independent dynamical fields (`dy') representing quarks. The dynamical part of the quark field at $x^+ = 0$ is related to quark and antiquark creation and destruction operators by
\begin{equation}
\begin{split}
\label{eq:psi}
P_{\rm dy}&\psi_{i,c}(0,x^-,\bm x_\perp) ={} 
\frac{1}{(2\pi)^{3}}\int_0^\infty\!\frac{dk^+}{2k^+} \int d\bm k_\perp 
\\& \sum_s\, \bigg\{
P_{\rm dy}\, u(k,s)\, e^{-i k\cdot x} b(k^+,\bm k_\perp,c,s;i)
\\
&\qquad + 
P_{\rm dy}\, v(k,s)\, e^{+i k\cdot x} d^\dagger(k^+,\bm k_\perp,c,s;i)
\bigg\}
\;.
\end{split}
\end{equation}
Here $k\cdot x = k^+ x^- - \bm k_\perp \cdot \bm x_\perp$, and $d^\dagger$ is an antiquark creation operator, analogous to the quark creation operator $b^\dagger$. The field $\psi$ carries a flavor index $i$. It also carries a color index $c$, which we normally suppress. The spinors $u$ and $v$ are the usual solutions of the free Dirac equation, normalized to $\overline u(k,s)\gamma^+ u(k,s) = 2 k^+$ and $\overline v(k,s)\gamma^+ v(k,s) = 2 k^+$. Then one easily finds that $P_{\rm dy}\, u(k,s)$ and $P_{\rm dy}\,v(k,s)$ depend only on the $+$ component of $k$.

When we combine Eqs.~(\ref{eq:f0}) and (\ref{eq:psi}), we obtain, quite directly,
\begin{equation}
\begin{split}
f_{i/\Lp}^{(0)}&(\xi) \brax{P'}\ket{P} =  
\frac{P^+}{2\pi}\int\!dy^-\, e^{-\mi \xi P^+ y^-}
\int\!dx^- \ d\bm x_\perp
\\&\times
\bra{P'}
\bar\psi_i(0,x^-+y^-,\bm x_\perp) \gamma^+ \psi_i(0,x^-,\bm x_\perp)
\ket{P}
\;.
\end{split}\label{eq:fiLp}
\end{equation}
We can eliminate the factor $\brax{P'}\ket{P}$ by using translation invariance to write
\begin{equation}
\begin{split}
\bra{P'} 
\bar\psi_i(0,&x^- + y^-,\bm x_\perp) \gamma^+ \psi_i(0,x^-,\bm x_\perp)
\ket{P}
\\
={}& e^{\mi[(P' - P)\cdot x^- - (\bm P_\perp' - \bm P_\perp)\cdot \bm x_\perp]}
\\&\times
\bra{P'}
\bar\psi_i(0,y^-,\bm 0) \gamma^+ \psi_i(0,0,\bm 0)
\ket{P}
\;.
\end{split}
\end{equation}
Then we can perform the $x^-$ and $\bm x_\perp$ integrations to give delta functions that set $P'$ to $P$. We normalize our proton state vectors to
\begin{equation}
\brax{P'}\ket{P} = 
(2 \pi)^3 2 P^+ \delta(P^{\prime +} - P^+)\,
\delta(\bm P'_\perp - \bm P_\perp)
\;.
\end{equation}
Then the delta functions from $\brax{P'}\ket{P}$ in Eq.~(\ref{eq:fiLp}) cancel. We set $P'$ to $P$ to get
\begin{align}
\label{eq:f0f}
& f_{i/\Lp}^{(0)}(\xi)  
\\&\quad =
\frac{1}{4\pi} \int\!dy^-\, e^{-\mi \xi P^+ y^-}
\bra{P}
\bar\psi_i(0,y^-,\bm 0) \gamma^+ \psi_i(0)
\ket{P}
\notag
\;.
\end{align}

We have presented this result in some detail to emphasize that the PDF
for quarks is simply the proton matrix element of the number density
operator for quarks as obtained in canonically quantized field
theory. 

\subsubsection{Gauge invariance \label{sec:WilsonLine}}

Next, without changing $f_{i/\Lp}^{(0)}(\xi)$, we can rewrite the definition in a way that makes it gauge-invariant. The canonical field theory that our
derivation has relied on makes use of the lightlike axial gauge $A^+(x) = 0$ for the gluon field. In an arbitrary gauge, we merely insert a Wilson line factor,
\begin{equation}
\label{eq:WilsonLine}
W(y^-,0) = {\cal P} \exp\left(
-\mi g \int_0^{y^-}\!\!d\bar y^-  A^+(0,\bar y^-,\bm 0)_a\, t_a
\right)
.
\end{equation}
This is a matrix in the color indices carried by the quark fields;
$t_a$ is the $SU(3)_c$ generator matrix in the $\bm 3$ representation. The $\cal P$ indicates path ordering of the operators and matrices, with more positive $y^-$ values to the left. The revised definition is
\begin{equation}
\begin{split}
\label{eq:f0final}
f_{i/\Lp}^{(0)}(\xi)  
={}&
\frac{1}{4\pi} \int\!dy^-\, e^{-\mi \xi P^+ y^-}
\\&\times
\bra{P}
\bar\psi_i(0,y^-,\bm 0) \gamma^+ W(y^-,0)\, \psi_i(0)
\ket{P}
\;.
\end{split}
\end{equation}
The factor $W$ is just 1 if we use $A^+(x) = 0$ gauge. If we change
the gauge by a unitary transformation $U(x)$, we replace
\begin{equation}
\begin{split}
\psi_i(0) \to{}& U(0)\, \psi_i(0)
\;,
\\
\bar\psi_i(0,y^-,\bm 0) \to{}& \bar \psi_i(0,y^-,\bm 0)\, U(0,y^-,\bm 0)^{-1}
\;,
\\
W(y-,0) \to{}&  U(0,y^-,\bm 0)\, W(y-,0)\, U(0)^{-1}
\;.
\end{split}
\end{equation}
Thus, when we include the operator $W$, the right-hand side of the equation is invariant under a change of gauge.

If we use a covariant (Bethe-Salpeter) wave function for the proton state, we can use Eq.~(\ref{eq:f0final}) for perturbative calculations. The field $\psi_i(0)$ absorbs a quark line from the wave function. Similarly, $\bar\psi_i(0,y^-,\bm 0)$ creates a quark line that goes into the conjugate wave function. These quark lines can emit and absorb gluons. The factor $W(y^-,0)$ is conveniently written as $W(y^-,\infty)$ times $W(\infty,0)$. The operators $W$ contains gluon fields that create and absorb gluons. In a simple intuitive picture, we don't just destroy a quark at position 0, leaving its color with nowhere to go. Rather we scatter it, so that it moves to infinity along a fixed lightlike line in the minus direction, carrying its color with it. Then its color comes back to $(0,y^-,\bm 0)$ to provide the color for the quark that we create.

\subsubsection{Renormalization}\label{sec:PDFRenormalization}

The function $f_{i/\Lp}^{(0)}(\xi)$ has so far been defined using
``bare'' fields, a bare coupling, bare parton masses, and a bare
operator product of fields in a canonical formulation of the field
theory. This will not do. Even the simplest one-loop calculation
reveals that the bare $\alpha_s$, quark masses, and
$f_{i/\Lp}^{(0)}(\xi)$ contain ultraviolet (UV) divergences. Thus we need
to renormalize everything. The standard way to do this is to apply $\MSbar$ renormalization with scale $\mu^2$. For this, we need to choose a number $N_f$ of active flavors.\footnote{ For instance, if we are following the
  $N_f=5$ convention, then neither $\alpha_s$ nor PDFs include
  contributions from top quarks. Then top-quark virtual loops can still
  occur within the Feynman diagrams, but they are treated using the CWZ 
  prescription \cite{Collins:1978wz}, in which the UV divergencies that they
  introduce are subtracted at zero incoming momenta and do not
  affect scale dependence of $\alpha_s$ or PDFs. Depending on the circumstances, 
  one uses different numbers of active flavors, as we will discuss in 
  Sec.~\ref{sec:heavyquarks}.}
The $\MSbar$-renormalized entities acquire dependence on $\mu^2$, which can be chosen so as to improve perturbative convergence for the short-distance cross section
$\widehat{\sigma}$ in Eq.~(\ref{eq:factorization00}). Physically, quark and gluon interactions at distance scales smaller than $1/\mu^2$ are not resolved in these objects.

This gives us our final definition for the quark distribution \cite{Collins:1981uw},
\begin{align}
\label{eq:fquarkfinal}
f_{i/\Lp}(\xi,\mu^2)  ={}& 
\frac{1}{4\pi} \int\!dy^-\, e^{-\mi \xi P^+ y^-}
\\&\times
\bra{P}
\bar\psi_i(0,y^-,\bm 0) \gamma^+ W(y^-,0)\, \psi_i(0)
\ket{P}
\notag
\;,
\end{align}
where $W(y^-,0)$ is given by Eq.~(\ref{eq:WilsonLine}). We understand now that the formulas refer to fields and couplings and field products that are renormalized with the $\MSbar$ prescription for all active quarks and gluons.

For antiquarks, the analogous definition is
\begin{align}
\label{eq:fantiquarkfinal}
f_{\bar i/\Lp}(\xi,\mu^2)  ={}& 
\frac{1}{4\pi} \int\!dy^-\, e^{-\mi \xi P^+ y^-}
\\&\times
\bra{P}
{\rm Tr}\big[\gamma^+\psi_i(0,y^- \!,\bm 0)W(y^-,0)\bar\psi_i(0)\big] 
\ket{P}
, \notag
\end{align}
where the color generator matrices in $W(y^-,0)$ are in the $\bar{\bm 3}$ representation of SU(3).

We should understand that no approximations are made in Eq.~(\ref{eq:fquarkfinal}). In particular, we do not treat the quarks as being massless. We cannot calculate $f_{i/\Lp}(\xi,\mu^2)$ at any finite order of perturbation theory, but we could, in principle, calculate it using lattice gauge theory. In such a calculation, we would use our best estimates for the parameters in the QCD lagrangian, including the strong coupling and the quark masses.

In fact, PDFs can be calculated using lattice gauge theory
\cite{Lin:2017snn}, but the accuracy of such calculations is still
limited. One can obtain much better accuracy by fitting the parton
distributions to data, as described in this review. However, the
definition of the parton distributions is not affected by the
approximations that we make in the fitting procedure. For instance the
calculated cross sections used in the fit can be leading order (LO),
next-to-leading order (NLO), or next-to-next-to-leading order
(NNLO). The resulting fits are often referred to as LO, NLO, or
NNLO. However, it is the fits that carry these designations. The
functions $f_{i/\Lp}(\xi,\mu^2)$ that we are trying to estimate are
non-perturbative objects whose definitions are independent of the
fitting method. 

\subsubsection{Gluons}\label{sec:PDFGluons}

In the previous subsections, we have defined the PDFs for quarks and antiquarks by beginning with the number operator for quarks or antiquarks in unrenormalized canonical field theory using null-plane quantization in $A^+ = 0$ gauge. The starting definition is then generalized to be gauge invariant and use $\MSbar$ renormalized operators. One can follow the same sort of logic for the gluon field. We simply state the result \cite{Collins:1981uw}:
\begin{align}
\label{eq:fgluon}
f_{g/\Lp}(\xi,\mu^2)  ={}& 
\frac{1}{2\pi\,\xi P^+} \int\!dy^-\, e^{-\mi \xi P^+ y^-}
\\&\times
\bra{P}
G(0,y^-,\bm 0)^{+\nu}  W(y^-,0)\, G(0)_\nu^{\ +}
\ket{P}
\;,\notag
\end{align}
where $G_{\mu\nu}$ is the gluon field operator,
\begin{equation}
G_{\mu\nu}^a = \partial_\mu A_\nu^a - \partial_\nu A_\mu^a
- g f^{abc} A_\mu^b A_\nu^c
\;,
\end{equation}
and the Wilson line operator $W(y^-,0)$ is given by Eq.~(\ref{eq:WilsonLine}), now using SU(3) generator matrices $(t_a)_{bc}=-i\ f_{abc}$ in the adjoint representation.

\subsubsection{Evolution equation}\label{sec:PDFDGLAP}
Let us take a closer look at the renormalization of the PDFs. 
The $\MSbar$ renormalization of the strong coupling $\alpha_s\equiv g^2/(4\pi)$ and
the fields $\psi_i(x)$ and $A^\mu(x)_a$ in $n=4-2\epsilon$ dimensions
proceeds in the usual way by subtracting $1/\epsilon$ poles and
some finite terms from two-point subgraphs,
three-point subgraphs, and four-gluon subgraphs with loops containing gluons and the $N_f$ active quarks. Another sort of pole arises in matrix elements for PDFs from operator products like $\bar\psi_i(0,y^-,\bm 0) \gamma^+
\psi_i(0)$ in Eq.~(\ref{eq:fquarkfinal}). Consider a graph in which a gluon is emitted from a
propagator representing the quark that is destroyed by $\psi_i(0)$,
then absorbed by a propagator representing the quark
created by $\bar\psi_i(0,y^-,\bm 0)$. This gluon line creates a loop
subgraph that is UV-divergent in four dimensions.

We subtract the divergence using the $\MSbar$ prescription, which
creates dependence of $f_{a/\Lp}(\xi,\mu^2_F)$ on the {\it factorization
  scale} $\mu^2_F$, possibly
different from the renormalization scale $\mu^2_R$. In much of our discussion, we will assume that $\mu_R^2$ and $\mu_F^2$ are the same and denote both as $\mu^2$.

By examining the structure of the UV divergences, one finds that the functions
$f_{a/\Lp}(\xi,\mu_\LF^2)$ obey DGLAP evolution equations, 
\begin{equation}
\begin{split}
\label{eq:pdfevolution}
&\frac{d}{d\log \mu^2_\LF}\,f_{a/\Lp}(\xi,\mu^2_\LF)
\\& \qquad
= \sum_{\hat a} \int_\xi^1\!\frac{dz}{z}\ P_{a \hat a}(z,\as(\mu^2_\LF))\,
f_{\hat a/\Lp}(\xi/z,\mu^2_\LF)
\;.
\end{split}
\end{equation}
The functions $f_{a/\Lp}(\xi,\mu^2_\LF)$ are nonperturbative, but,
since the dependence on $\mu^2_\LF$ arises from the UV divergences of graphs for $f_{a/\Lp}(\xi,\mu^2_\LF)$, the evolution kernels $P_{a b}(z,\as(\mu^2_\LF))$ are perturbatively calculable as expansions in powers of $\as(\mu^2_\LF)$:
\begin{align}
\label{eq:pdfevolutionseries}
P_{a \hat a}(z,\as(\mu^2_\LF))
={}& \frac{\as(\mu^2_\LF)}{2\pi}\,P_{a \hat a}^{(1)}(z)
+ \left[\frac{\as(\mu^2_\LF)}{2\pi}\right]^2\,P_{a \hat a}^{(2)}(z)
\notag
\\&
+ \cdots \;.
\end{align}

The exact evolution kernels $P_{a b}$ have been known up to three loops
(NNLO) since 2004 \cite{Moch:2004pa,Vogt:2004mw},
with active efforts now underway
on computing the four-loop terms  $P_{a b}^{(4)}$, cf. \cite{Ueda:2018xms}.
It is significant that the functions $P_{a \hat a}^{(n)}(z)$ do not depend on quark masses. In graphs for the PDFs, there are masses in quark propagators, $(\s k + m)/(k^2 - m^2 + \mi\epsilon)$. However the ultraviolet poles of these graphs are determined by the behavior of the propagators for $k \to \infty$. In this limit, the masses do not contribute. This is an advantage of using the $\MSbar$ scheme for renormalizing the PDFs.

\subsection{Infrared safety and factorization}\label{sec:IRsafetyandFactorization}

PDFs are used to compute cross sections in collisions of a lepton with a hadron and
collisions between two hadrons. We concentrate in this subsection on
hadron-hadron collisions, since these are currently the subject of
investigation at the Large Hadron Collider (LHC). Lepton-hadron
collisions, as in deeply inelastic scattering, are simpler. 

In one sense, the use of PDFs to describe proton-proton collisions is very simple. Suppose that we are interested in the cross section $d^2\sigma/(dp_\LT dy)$, to produce a jet with transverse momentum $p_\LT$ and rapidity $y$ plus anything else in the collision of a hadron of type $A$ and a hadron of type $B$. Or, suppose that we are interested in the cross section $d\sigma/ dy$, to produce an on-shell Higgs boson with rapidity $y$ plus anything else. We can consider many cases at once by saying that we are interested in a cross section $\sigma[F]$ to measure a general observable $F$ that is ``infrared-safe'' in the sense that will be explained in subsection \ref{sec:IRsafety}. Then the PDFs relate $\sigma[F]$ to a calculated cross section $\hat \sigma_{ab}[F]$ for the collision of two partons. In its briefest form, the relation is
\begin{equation}
\label{eq:factorization0}
\sigma[F] \approx \sum_{a,b}\iint\!d\xi_a\ d\xi_b\
f_{a/A}(\xi_a,\mu^2)\, f_{b/B}(\xi_b,\mu^2)\
\hat \sigma_{ab}[F].
\end{equation}
Here we sum over the possible flavors $a$ and $b$ of partons that we might find in the respective hadrons. We integrate over the momentum fractions $\xi_a$ and $\xi_b$ of these partons. Then we multiply by $\hat\sigma_{ab}[F]$, which plays the role of a cross section for the collision of these partons to produce the final state that we are looking for. 

We say that Eq.~(\ref{eq:factorization0}) expresses ``factorization.''\footnote{The word ``factorization'' is applied to many formulas in which a physical quantity is expressed as a convolution of a product of factors. Eq.~(\ref{eq:factorization0}) is sometimes called inclusive collinear factorization.}
Factorization seems simple, but it is not. First, it works only when the observable to be measured, $\sigma[F]$, has a certain property, ``infrared safety.'' Second, factorization is approximate, and we need to understand what is left out. Third, its validity is not self-evident, as one finds when trying to calculate $\hat \sigma_{ab}[F]$ beyond the leading order and encountering infinities if the calculation is not carefully formulated. Fourth, while $\hat\sigma_{ab}[F]$ plays the role of a cross section to produce a certain final state in the collision of two partons, actually its calculation beyond the leading order in perturbative QCD involves subtractions, as discussed in subsection \ref{sec:IRsafety} below.

\subsubsection{Kinematics}

Consider a hard scattering process in the collisions of two high energy hadrons, A and B. The hadrons carry momenta $P_\LA$ and $P_\LB$. The hadron energies are high enough that we can simplify the equations describing the collision kinematics by treating the colliding hadrons as being massless. Then with a suitable choice of reference frame, the hadron momenta are
\begin{equation}
\begin{split}
P_\LA ={}& 
\left(P_\LA^+,0,\bm 0 \right)
\;,
\\
P_\LB ={}& 
\left(0, P_\LB^-,\bm 0 \right)
\;.
\end{split}
\end{equation}
We then imagine a parton level process in which a parton from hadron A, with flavor $a$ and momentum $\xi_\La P_\LA$ collides with a parton from hadron B, with flavor $b$ and momentum $\xi_\Lb P_\LB$. This collision produces $m$ partons with flavors $f_i$ and momenta $p_i$. Each final state parton has rapidity $y_i$ and transverse momentum $\bm p_{i,\perp}$, so that the components of its momentum are
\begin{equation}
p_i = (e^{y_i} \sqrt{(\bm p_{i,\perp}^2 + m_i^2)/2}, 
e^{-y_i} \sqrt{(\bm p_{i,\perp}^2 + m_i^2)/2}, \bm p_{i,\perp})
\;.
\end{equation}
Then momentum conservation gives us
\begin{equation}
\begin{split}
\label{eq:momentumconservation}
\sum_{i = 2}^m \bm p_{i,\perp} ={}& - \bm p_{1,\perp}
\;,
\\
\sum_{i = 1}^m e^{y_i} \sqrt{(\bm p_{i,\perp}^2 + m_i^2)/2} ={}& 
\xi_\La P_\LA^+
\;,
\\
\sum_{i = 1}^m e^{-y_i} \sqrt{(\bm p_{i,\perp}^2 + m_i^2)/2} ={}& 
\xi_\Lb P_\LB^-
\;.
\end{split}
\end{equation}

\subsubsection{Infrared safety}
\label{sec:IRsafety}

In order for the factorization to work, the observable should be infrared-safe. The basic physical idea for this was introduced in \cite{Sterman:1977wj}. We will follow the development in
\cite{Kunszt:1992tn} and  define what we mean by measuring the cross section for an observable $F$ and what it means for $F$ to be infrared safe. To keep the discussion simple, we temporarily assume that all of the partons involved are light quarks and gluons, which we consider to be massless, and that the observable does not distinguish the flavors of the partons. For this simple case, we express the  parton-level cross section for an observable $F$ using the definition
\begin{equation}
\begin{split}
\label{eq:sigmaF}
\hat\sigma_{ab}[F] ={}& 
\frac{1}{2!}\int\! dy_1\,dy_2\,d\bm p_{2,\perp}
\frac{d\hat\sigma_2}
{dy_1\,dy_2\,d\bm p_{2,\perp}}
F_2(p_1,p_2)
\\&+
\frac{1}{3!}
\int\! dy_1\,dy_2\,dy_3\,d\bm p_{2,\perp}\,d\bm p_{3,\perp}
\\&\quad \times
\frac{d\hat\sigma_3}
{dy_1\,dy_2\,dy_3\,d\bm p_{2,\perp}\,d\bm p_{3,\perp}}
F_3(p_1,p_2,p_3)
\\&+
\cdots
\;.
\end{split}
\end{equation}
Here we start with the cross section to produce $m$ partons with momenta $\{p_1, \dots, p_m\}$. We multiply the cross section by a function $F_m(p_1,\dots,p_m)$ that specifies the measurement that we want to make on the final-state partons. These functions are taken to be symmetric under interchange of their arguments. Accordingly, we divide by the number $m!$ of permutations of the parton labels. We integrate over the momenta of the final-state partons. The transverse momentum of parton 1 and the needed momentum fractions for the incoming partons are determined by Eq.~(\ref{eq:momentumconservation}). Finally, we sum over the number $m$ of final-state partons. 

An example may be useful. If we want to evaluate the cross section for the sum of the absolute values of the transverse momenta of the partons to be bigger than 100 GeV, then we would choose a step function $F_m = \theta(\sum_{i = 1}^m |\bm p_{i,\perp}| > 100 \GeV)$. More generally, the $F_m$ for jet cross sections are made of step functions and delta functions constructed according to the jet algorithm used.

Infrared safety is a property of the functions $F_m$ that relates each function $F_{m+1}(p_1,\dots,p_m,p_{m+1})$ to the function $F_{m}(p_1,\dots,p_m)$ with one fewer parton. There are two requirements needed for $F$ to be infrared safe.

First, consider the limit in which partons $m + 1$ and $m$ become collinear:
\begin{equation}
\begin{split}
\label{eq:collinearlimit}
&p_{m+1} \to z\tilde p_m
\;,
\\
&p_{m} \to{} (1-z)\tilde p_m
\;.
\end{split}
\end{equation}
Here $\tilde p_m$ is a lightlike momentum and $0 \leq z \leq
1$.  
Therefore, $p_m + p_{m+1} \to \tilde p_m$.
We can concentrate on just partons with labels $m+1$ and $m$ because the functions $F$ are assumed to be symmetric under interchange of the parton labels. In order for $F$ to be infrared safe, we demand that
\begin{equation}
\label{eq:IRsafe1}
F_{m+1}(p_1,\dots,p_{m-1},p_m,p_{m+1})
\to F_m(p_1,\dots,p_{m-1},\tilde p_m)
\end{equation}
in the collinear limit (\ref{eq:collinearlimit}). 

Second, consider also the limit in which parton $m+1$ becomes collinear to one of the beams: 
\begin{equation}
\label{eq:collinearA}
p_{m+1} \to \lambda  P_\LA
\end{equation}
or
\begin{equation}
\label{eq:collinearB}
p_{m+1} \to \lambda  P_\LB
\;.
\end{equation}
Here $\lambda \ge 0$. When $\lambda = 0$, parton $m+1$ is simply becoming infinitely soft. In order for $F$ to be infrared safe, we demand that
\begin{equation}
\label{eq:IRsafe2}
F_{m+1}(p_1,\dots,p_m,p_{m+1})
\to F_m(p_1,\dots,p_{m})
\end{equation}
in either limit (\ref{eq:collinearA}) or (\ref{eq:collinearB}).

Briefly, then, when the partonic masses are negligible, infrared safety means that the result of the measurement is not sensitive to whether or not one parton splits into two almost collinear partons, and it is not sensitive to any partons that have very small momenta transverse to the beam directions.

Sometimes an observable $F$ with this property is referred to as infrared and collinear safe (IRC-safe) instead of just infrared safe (IR-safe). The meaning is the same.

Cross sections to produce jets are infrared safe as long as we use a suitable algorithm, encoded in the functions $F_m$, to define jets. However, the cross section to produce a jet containing a specified number of charged particles would not be infrared safe: a collinear splitting of a gluon to a quark plus an antiquark increases the number of charged particles by two.

In Eq.~(\ref{eq:sigmaF}), we have cross sections to produce specified partons. This is the formula that we use to construct $\hat\sigma_{ab}[F]$ on the right-hand side of Eq.~(\ref{eq:factorization0}). On the left-hand side of Eq.~(\ref{eq:factorization0}), we measure the cross section $\sigma[F]$ experimentally by using the same functions $F_m$ applied to hadrons instead of partons. Relating a hadron cross section to a parton cross section is evidently an approximation, so that there is an error in Eq.~(\ref{eq:factorization0}) no matter how many orders of perturbation theory we use in the calculation. The physical idea is that the infrared safe measurement involves a large scale $Q$, which is incorporated into the functions $F_m$. Processes that involve scales $M$ with $M \ll Q$ do not substantially affect the measurement. In particular, combining partons to form hadrons involves scales $M \sim 0.3 \GeV$. Thus turning partons into hadrons does not substantially affect a measurement with a much larger scale $Q$.

While we can best understand the idea of IR safety by sticking to massless partons and flavor-independent measurement functions, it is important to keep in mind that a quark does have a mass, however small or large; QCD factorization need not assume that the partons are massless. For quark masses $m_i$ that are small compared to the scale $Q$ of the measurement that we have in mind, we should amend the definition of infrared safety of the functions $F_m$ to include taking the limit $m_i \to 0$. Sometimes a quark mass is comparable to the scale $Q$. For instance, this typically applies to top quarks. It also applies to, say, charm quarks when we take $Q$ to be only a couple of GeV. In such cases, we need not consider limits $m_i \to 0$ for ``heavy'' quarks and can operate with parton distribution functions only for the ``light'' quarks. See Sec.~\ref{sec:heavyquarks} below.

\subsubsection{Factorization}
\label{sec:factorization}

With the needed preparation accomplished, we can now state how the PDFs are used to calculate the cross section for whatever observable $F$ we want -- as long as $F$ is infrared safe. For this condition to apply, the observable $F$ must be sufficiently inclusive. The formula we use was stated in Eq.~(\ref{eq:factorization0}), and we restate it here in a slightly more detailed form:
\begin{equation}
\begin{split}
\label{eq:factorization}
\sigma[F] ={}& \sum_{a,b}\int\!d\xi_a \int\!d\xi_b\
f_{a/A}(\xi_a,\mu_\LF^2)\, f_{b/B}(\xi_b,\mu_\LF^2)
\\&\quad\times
\hat \sigma_{a,b,\xi_a,\xi_b,\mu_\LF^2}[F]
+ {\cal O}\!\left({M}/{Q}\right)
\;.
\end{split}
\end{equation}
Our convention in Eq.~(\ref{eq:factorization}) is to use the name $\mu_\LF$, the ``factorization scale,'' for the scale parameter in the parton distribution functions. The physical cross section cannot depend on $\mu_\LF$, so $\hat \sigma$ must also depend on $\mu_\LF$. Note that in Eq.~(\ref{eq:factorization}), we have not yet expanded in powers of $\as$.

The intuitive basis for Eq.~(\ref{eq:factorization}) is very simple. The factor $f_{a/A}(\xi_a,\mu_\LF^2)\, d\xi_\La$ plays the role of the probability to find a parton of flavor $a$ in a hadron of flavor $A$. For the other hadron, the corresponding probability is  $f_{b/B}(\xi_b,\mu_\LF^2)\, d\xi_\Lb$. Then $\hat \sigma[F]$ plays the role of a cross section to obtain the observable $F$ from the scattering of these partons, as given in Eq.~(\ref{eq:sigmaF}). Naturally, this parton level cross section depends on the parton variables $a,b,\xi_a,\xi_b$, as indicated by the subscript notation. Here the differential cross sections to produce
$m$ final state partons contain delta functions that relate the momentum fractions $\xi_\La$ and $\xi_\Lb$ to the final state parton momenta, according to Eq.~(\ref{eq:momentumconservation}).

A similar formula applies for lepton-hadron scattering, the process
providing important constraints on the PDF parameterizations (See Sec.~\ref{sec:DIS}).
Then there is only one hadron in the initial state, so the formula is simpler:
\begin{equation}
\label{eq:DISfactorization}
\sigma[F] = \sum_{a}\int\!d\xi\,
f_{a/A}(\xi,\mu_\LF^2)\
\hat \sigma_{a,\xi,\mu_\LF^2}[F]
+ {\cal O}\!\left({M}/{Q}\right)
\;.
\end{equation}

The cross section $\hat \sigma[F]$ in Eq.~(\ref{eq:factorization}) (or
Eq.~(\ref{eq:DISfactorization})) has a perturbative expansion in
powers of $\as(\mu_\LR^2)$, where the renormalization scale
$\mu_\LR^2$ can be chosen independently from the factorization scale $\mu_\LF^2$. That is,
\begin{equation}
\begin{split}
\label{eq:hatsigmaexpansion}
\hat \sigma_{a,b,\xi_a,\xi_b,\mu_\LF^2}[F]
={}& \left[\frac{\as(\mu_\LR^2)}{2\pi}\right]^B
\hat \sigma^{(B)}_{a,b,\xi_a,\xi_b,\mu_\LF^2,\mu_\LR^2}[F]
\\&
+
\left[\frac{\as(\mu_\LR^2)}{2\pi}\right]^{B+1}\!
\hat \sigma^{({B+1})}_{a,b,\xi_a,\xi_b,\mu_\LF^2,\mu_\LR^2}[F]
\\&
+
\cdots
\;.
\end{split}
\end{equation}
Here $B$ is the integer that tells us how many powers of $\as$ appear in the Born level cross section: {\it e.g.} 0 for Z boson production, 2 for two jet production. Perturbative calculations can be at lowest order (LO), corresponding to one term in the expansion, next-to-lowest order (NLO) with two terms, sometimes NNLO, and, in general, $\LN^k \LL\LO$.

When $\hat\sigma$ is expanded in powers of $\as$, as in Eq.~(\ref{eq:hatsigmaexpansion}), our convention is to use the name $\mu_\LR$, the ``renormalization scale,'' for the scale parameter in $\as$. The physical cross section cannot depend on $\mu_\LR$, so the functions $\hat \sigma^{B+k}$ in Eq.~(\ref{eq:hatsigmaexpansion}) must also depend on $\mu_\LR$. (Other conventions for the precise meaning of $\mu_\LF$ and $\mu_\LR$ are possible.)

One useful property of Eqs.~(\ref{eq:DISfactorization}) and (\ref{eq:hatsigmaexpansion}) is that the dependence of the calculated cross section on $\mu_\LF^2$ and $\mu_\LR^2$ diminishes as we go to higher orders. Indeed, the cross section in nature, $\sigma[F]$, does not depend on $\mu_\LF^2$ and $\mu_\LR^2$. Thus if we calculate to order $\as^{B+k}$, the derivative of the calculated cross section with respect to $\mu_\LF^2$ and $\mu_\LR^2$ will be of order $\as^{B+k+1}$. Because of this property, one often uses the effect of varying $\mu_\LF$ or $\mu_\LR$ by a fixed factor ({\it e.g.} 2 or 1/2) to provide an estimate of the error caused by calculating only to a finite perturbative order.

Note that, in order for terms of order $\as^{B+k}$ in $\hat \sigma$ to match the terms for order $\as^{k}$ in the evolution equation for the parton distributions, we need to include at least the terms up to $\as^{k}P^{(k)}_{a\hat a}(z)$ in Eqs.~(\ref{eq:pdfevolution}) and (\ref{eq:pdfevolutionseries}) giving the evolution of the PDFs. Since 
the lowest-order term in the evolution kernel is $P^{(1)}_{a\hat a}(z)$, including terms up to $P^{(k)}_{a\hat a}(z)$ is referred to as $\LN^{k-1} \LL\LO$ evolution, while we say that including terms up to $\as^{B+k}$ in the partonic cross section gives an $\LN^{k} \LL\LO$ calculation. Thus, for example, if we have an NNLO cross section calculation, we will not obtain the proper matching for the $\mu_\LF$ scale dependence unless we have at least NLO evolution for the PDFs. Normally, one uses yet higher-order evolution for the PDFs because this evolution determines the accuracy with which we know the PDFs at a high scale, given experimental inputs at much lower scales.

The error terms ${\cal O}\!\left({M}/{Q}\right)$ in
Eqs.~(\ref{eq:factorization}) and (\ref{eq:DISfactorization})
arise from power-suppressed contributions that are beyond the accuracy of the factorized representation \cite{Collins:2011zzd, Collins:1998rz, Collins:1982wa, Collins:1985ue, Bodwin:1984hc, Collins:1988ig}. No matter how many terms are included in $\hat \sigma$, there are contributions that are left out. These terms are suppressed by a power of $M \sim 1 \GeV$ divided by a large scale parameter $Q$ that characterizes the hard scattering process to be measured. Here we choose $1 \GeV$ as a nominal scale for hadronic bound state physics. This value is somewhat smaller than the scale $\mu$ at which $\as(\mu^2) = 1/2$, somewhat larger than the proton mass, and about four times larger than the inverse of the radius of a proton.

When data are included in a PDF fit but, for these data, $M/Q$ is not small enough to be completely negligible, it may be useful to include a nonperturbative model for the power-suppressed corrections. The model can have one or more parameters that can be fit to the data. For instance, one can add a contribution to the cross section that is proportional to $\kappa^2/Q^2$ where $\kappa$ is a parameter to be fit. This possibility is especially relevant for deeply inelastic scattering (Sec.~\ref{sec:DIS}), for which one can make use of a theoretical expansion known as the operator product expansion \cite{Wilson:1969zs} to suggest the form of the first power-suppressed contribution. See such fits, for example, in \cite{Alekhin:2012ig,Alekhin:2017kpj}.
 
The power-suppressed contributions in Eq.~(\ref{eq:factorization}) arise from the approximations needed to  obtain the result. For instance, if a loop momentum $l$ flows through the wave function of quarks in a proton, we have to neglect $l$ compared to the hard momenta, say the transverse momentum of an observed jet. We can illustrate the issue of power corrections and, at the same time, see something about the interplay of ultraviolet and infrared singularities by looking at a very simple model for an integral that might occur in a calculation.

Consider a model integral for a correction to the Born cross section for a process involving only one initial state hadron. In this model, the Born cross section is
\begin{equation}
I_\mathrm{Born} = \int_0^\infty \!dl^2\, |\psi(l^2)|^2 \times 1
\;,
\end{equation}
where $l^2$ represents the transverse momentum of a quark that is restricted to be of order $1 \GeV^2$ by a hadronic wave function $|\psi(l^2)|^2$. The Born level PDF is $\int_0^\infty dl^2 |\psi(l^2)|^2$. The Born hard scattering function is simply 1.  At order $\as$, we should have an $\as$ correction to the Born PDF times the Born hard scattering function, plus the Born PDF times an $\as$ correction to the hard scattering function. 

In our model, the order $\as$ correction to the cross section is
\begin{equation}
\label{eq:integral1}
I = \int_0^\infty \!dl^2\,|\psi(l^2)|^2\, \as
\int_0^\infty\! dk^2\ \left[\frac{\mu_\LF^2}{k^2}\right]^{\epsilon}
\frac{1}{k^2 + l^2}\frac{Q^2}{k^2 + Q^2}
\;.
\end{equation}
Here $Q^2$ is the ``hard scale,'' with $l^2 \ll Q^2$. We supply a factor $\left[{\mu_\LF^2}/{k^2}\right]^{\epsilon}$ that mimics dimensional regularization in $n=4-2\epsilon$ dimensions. This integral is, however convergent in the infrared (IR) and in the ultraviolet (UV), so we could simply set $\epsilon \to 0$ after its computation.

For $k^2 \gg l^2$, we could approximate $1/(k^2 + l^2) \to 1/k^2$. This would define a part of the integral $I$ that represents the probability distribution $|\psi(l^2)|^2$ times an order $\as^1$ contribution to the hard scattering function,
\begin{equation}
\begin{split}
\label{eq:IUV}
I_\mathrm{UV} ={}& \int_0^\infty \!dl^2\,|\psi(l^2)|^2
\\&\times
\as \cdot \left( \int_0^\infty\! dk^2\ \left[\frac{\mu_\LF^2}{k^2}\right]^{\epsilon}
\frac{1}{k^2}\frac{Q^2}{k^2 + Q^2}
+ \frac{1}{\epsilon}\right)
\;.
\end{split}
\end{equation}
The integral is IR divergent. We have subtracted its IR pole, proportional to $-1/\epsilon$.

For $k^2 \ll Q^2$, we could approximate $Q^2/(k^2 + Q^2) \to 1$. The integral would then define an $\alpha_s$ correction to the PDF times the Born hard scattering function,
\begin{equation}
\begin{split}
\label{eq:IIR}
I_\mathrm{IR} ={}& 
\int_0^\infty \!dl^2\,|\psi(l^2)|^2
\\&\times
 \as \cdot \left( 
\int_0^\infty\! dk^2\ \left[\frac{\mu_\LF^2}{k^2}\right]^{\epsilon}
\frac{1}{k^2 + l^2}
- \frac{1}{\epsilon}
\right) \cdot 1
\;.
\end{split}
\end{equation}
The PDF integral is UV divergent, so we have ``renormalized'' it by subtracting its UV pole, proportional to $+1/\epsilon$.

We can also use
\begin{equation}
I_0 = \int_0^\infty \!dl^2\,|\psi(l^2)|^2\,
\as 
\int_0^\infty\! dk^2\ \left[\frac{\mu_\LF^2}{k^2}\right]^{\epsilon}
\frac{1}{k^2}
= 0
\;.
\end{equation}
The $k$-integral equals zero because it equals $1/\epsilon - 1/\epsilon$. 

A simple calculation using the integrands of our integrals gives
\begin{equation}
\begin{split}
I ={}& I_\mathrm{UV} + I_\mathrm{IR} - I_0 - \int_0^\infty \!dl^2\,|\psi(l^2)|^2
\\&\times
\frac{l^2}{Q^2}\,\as
\int_0^\infty\! dk^2\ \left[\frac{\mu_\LF^2}{k^2}\right]^{\epsilon}
\frac{1}{k^2 + l^2}\frac{Q^2}{k^2 + Q^2}.
\end{split}
\end{equation}
Thus our integral can be decomposed into the UV integral (including an IR subtraction), the IR integral (including its UV renormalization subtraction), and a remainder that is suppressed by a factor $l^2/Q^2$, where $l^2$ is of order $1 \GeV^2$. This is the power-suppressed correction.

Notice that dimensional regularization with factorization scale $\mu_\LF^2$ followed by subtracting $1/\epsilon$ poles serves two functions.  The integral $I_\textrm{UV}$ in Eq.~(\ref{eq:IUV}) is a model for how one calculates the hard scattering function at NLO. This integral has an IR divergence, which is removed by subtracting its pole. This subtraction matches the subtraction necessary to renormalize UV divergence in the correction to the model parton distribution function in Eq.~(\ref{eq:IIR}).

This model calculation is somewhat misleading because it suggests that PDFs have a useful perturbative expansion. They do not. To compute a hard cross section in real QCD, one can replace the distribution of partons in a proton by the distribution of partons in a parton, which is singular but perturbatively calculable. See \cite{Collins:2011zzd} for details.
 
Not much is known about the general form of the power corrections for hadron-hadron collisions.\footnote{See, however, \cite{Qiu:1990xxa, Qiu:1990xy}.} It is important that they are there, but, if $Q$ is of order hundreds of GeV, then the power corrections are completely negligible. However, if $Q$ is of order 5 GeV, then we ought not to claim 1\% accuracy in the calculation of $\sigma[F]$, no matter how many orders of perturbation theory we use.

Eq.~(\ref{eq:factorization}), representing  inclusive collinear factorization, is the basis of every prediction for hard processes at hadron colliders like the LHC, including both Standard Model processes and processes that might produce new heavy particles. So far as we know, it is a theorem for infrared-safe QCD observables dependent on energy-momentum variables that are of the same order of magnitude.  There are other formulas in QCD that go under the name of ``factorization'' and typically apply either to amplitudes rather than cross sections or to observables dependent on several momentum variables of disparate orders of magnitude. These include $k_\LT$ factorization and soft-collinear-effective-theory (SCET) factorization. These other forms of factorization may fail, thus requiring a more complex analysis, or at least are more subject to doubt than Eq.~(\ref{eq:factorization}). See, for example, \cite{Collins:2007nk, Catani:2011st,Forshaw:2012bi, Rothstein:2016bsq, Schwartz:2018obd}.

Early attempts to establish  inclusive collinear factorization \cite{Amati:1978by, Ellis:1978ty} were instructive, but incomplete. Later proofs of Eq.~(\ref{eq:factorization}) \cite{Collins:1982wa, Collins:1985ue, Bodwin:1984hc, Collins:1988ig,Collins:1998rz} are far from simple. They could perhaps benefit from more scrutiny than they have received. One issue is that the published proofs have considered only the Drell-Yan process, not more complex processes like jet production. A more serious issue is that there is no known general method that can deal with the boundaries between integration regions in the Feynman diagrams. On the other hand, any breakdown in the  inclusive collinear factorization in Eq.~(\ref{eq:factorization}) could lead to infinities in calculations of $\hat \sigma$, and no problems have been observed so far even in  $\mathrm{N}^3 \mathrm{LO}$ calculations \cite{Anastasiou:2015yha}.

\subsection{Treatment of heavy quarks}%%%%%%%%%%%%%%%%
\label{sec:heavyquarks}

In order to accurately describe data at energies from one to thousands of GeV, modern global PDF fits not only change the number $N_f$ of active flavors depending on the scales $\mu_{\LR,\LF}^2$, but also retain relevant quark mass dependence in the hard-scattering cross sections. The most comprehensive approach to do this is to work in one of the general-mass variable-flavor-number (GM-VFN) factorization schemes \cite{Aivazis:1993pi, Buza:1996wv, Chuvakin:1999nx, Kramer:2003jw, Kniehl:2004fy, Kniehl:2005mk, Thorne:1997ga, Thorne:2006qt, Forte:2010ta}. The massive fixed-flavor number schemes \cite{Gluck:1980cp, Gluck:1987uk} are also applicable under the right circumstances and may result in simpler predictions. Such computations are a complex subject that we cannot cover in any depth in the space available. We will illustrate some of the key ideas by mostly following \cite{Kramer:2000hn}. 

Consider the perturbative calculation of an infrared-safe cross
section $\sigma$ with scale $Q^2$ when $Q^2 \gg m_i^2$, where $i$
denotes any of the
$\Lu$, $\Ld$, $\Ls$, $\Lc$, and $\Lb$ quarks.
We can greatly simplify this calculation by neglecting masses of the
five quarks. But what if $Q$ is high enough,
and we expect that top quarks contribute either in the final state
or in the virtual loop corrections? We rarely can set $m_\Lt=0$,
as $m_\Lt \approx 174 \GeV$ is so large that we
seldom have $Q^2 \gg m_\Lt^2$ even at the LHC. 

There is a simple answer: in the range of energies comparable to $m_t$, we can use the $\MSbar$ scheme with five
active quark flavors $\Lu$, $\Ld$, $\Ls$, $\Lc$, and $\Lb$. Top quarks are included
in the relevant Feynman graphs, but, in accord with the CWZ
prescription, we use the zero-momentum subtraction,
instead of $\MSbar$ subtraction, for
the UV renormalization of loop subgraphs with top-quark lines. Then, terms involving $m_t$ appear only in the hard cross section $\widehat \sigma$ of Eq.~(\ref{eq:factorization}). 

One consequence of this is that the evolution equation for $\as(\mu^2_\LR)$ uses the 5-flavor beta-function. Radiative contributions to the top quark mass also need to be renormalized, see the later part of the subsection. The 5-flavor scheme introduces nonzero PDFs $f_{i/\Lp}(\xi,\mu^2_\LF)$ for $i \in \{\Lg, \Lu, \bar \Lu, \Ld, \bar \Ld, \dots, \Lb, \bar \Lb\}$. 
There are no parton distributions for $i = \Lt$ or $i = \bar \Lt$ in this scheme. The PDFs $f_{i/\Lp}(\xi,\mu^2_\LF)$ are defined as in the previous sections and evolve using 5-flavor DGLAP kernels. Another (not obvious) result is that top-quark contributions
are negligible in the limit $Q^2 \ll m_\Lt^2$: that is, top quarks decouple when the momentum scale $Q$ of the problem is much smaller than top-quark mass.

In a general case, we can distinguish among several versions of an $N_f$-flavor scheme. In the zero-mass (ZM) scheme, only the Feynman graphs with massless active quarks, $q_i$ with $i\leq N_f$, are included in $\widehat \sigma$ in the factorized hadronic cross section (\ref{eq:factorization}). In the fixed-flavor number (FFN) scheme, the massless active quark contributions for $i\leq N_f$ are included as in the ZM scheme, and also the Feynman graphs with massive inactive (anti)quarks, $q_i$ with $i>N_f$, are included only in the hard cross section $\widehat \sigma$. Finally, the most complete general-mass (GM) scheme retains non-negligible quark mass terms from both active and inactive quarks in all parts of the hadronic cross section. 

When discussing the region $Q^2 \gg m_\Lb^2$, we can use the 5-flavor FFN scheme and set masses of $\Lu, \Ld, \Ls, \Lc, \Lb$ quarks to zero. Consider now the region $Q^2 \sim m_\Lb^2$ and $Q^2 \gg m_\Lc^2$. Here, we can use the 4-flavor FFN scheme. There are parton distributions for flavors $\Lu, \Ld, \Ls, \Lc$, and these quarks are considered massless, while there are no parton distributions for $\Lb$ and $\bar{\mathrm{b}}$. In fact, the 4-flavor scheme is also fine for the subregion of $Q^2 \gg m_\Lb^2$ in which $\as \log(Q^2/m_\Lb^2) \ll 1$, where we do not need to sum logs of $Q^2/m_\Lb^2$.\footnote{One can also use a 3-flavor FFN scheme. For instance, this scheme is used for fitting $\Lc$- and $\Lb$-quark production in deeply inelastic scattering in \cite{Alekhin:2017kpj}. It introduces PDFs for $\Lu$, $\Ld$, and $\Ls$ quarks, which are treated as massless. The $\Lc$ and $\Lb$ quarks appear, with their masses, in $\hat \sigma$.}

We now have two possible FFN schemes for calculating a cross section at $m_\Lb^2/Q^2 \ll 1$ but $\as \log(Q^2/m_\Lb^2) \ll 1$: the 5-flavor scheme with  $\as^{(5)}(\mu^2)$ and $f_{i/\Lp}^{(5)}(\xi,\mu^2)$; and the 4-flavor scheme with $\as^{(4)}(\mu^2)$ and $f_{i/\Lp}^{(4)}(\xi,\mu^2)$. (Here, we set $\mu_\LR^2=\mu_\LF^2=\mu^2$.) The physical predictions must be the same, order by order in $\as$, in either scheme. This condition gives us matching relations between $\alpha_s$ and PDFs in the 4-flavor and 5-flavor schemes. At lowest order in $\alpha_s$, these relations are very simple. We should not use $\as^{(5)}(\mu^2)$ and $f_{i/\Lp}^{(5)}(\xi,\mu^2)$ for calculating physical cross sections unless $\mu^2 \gg m_\Lb^2$, but if we simply use their analytic forms for $\mu^2 = m_\Lb^2$, we have 
\begin{equation}
\begin{split}
\label{eq:4to5flavors}
& \as^{(5)}(m_\Lb^2) = \as^{(4)}(m_\Lb^2)
\;,
\\
& f_{i/\Lp}^{(5)}(\xi,m_\Lb^2) = f_{i/\Lp}^{(4)}(\xi,m_\Lb^2)
\quad \mbox{ for }
i \notin \{\Lb, \bar \Lb\}
\;,
\\
& f_{\Lb/\Lp}^{(5)}(\xi,m_\Lb^2) = f_{\bar{\Lb}/\Lp}^{(5)}(\xi,m_\Lb^2) = 0
\;.
\end{split}
\end{equation}
At higher orders of $\alpha_s$, $f_{\Lb/\Lp}^{(5)}(\xi,m_\Lb^2)\neq 0$, and the matching conditions are different and depend on whether $m_b$ is an $\MSbar$ or pole mass \cite{Ablinger:2017err,Bierenbaum:2007qe,Buza:1995ie}. Then, to obtain the $f_{i/\Lp}^{(5)}(\xi,\mu^2)$ for $\mu^2 > m_\Lb^2$, we solve the $N_f=5$ evolution equation with a boundary condition (\ref{eq:4to5flavors}) at $\mu^2 = m_\Lb^2$. One can also use a different scale, $\mu_\Lb^2 \ne m_\Lb^2$, as well as alternative presriptions, for the matching \cite{Bertone:2017ehk,Bertone:2017djs,Kusina:2013slm}.

We can derive analogous matching relations between the 3-flavor and
4-flavor schemes at $\mu=m_\Lc\approx 1.3\mbox{ GeV}$. The full range
$\mu^2 \geq m_c^2$ is then described by a sequence of the schemes
with $N_f=3,4,$ and $5$ that together comprise a VFN scheme. 

The scheme described so far is conceptually simple, but involves awkward switches between different values of $N_f$ at yet unspecified energy values. For instance, suppose we wish to switch between the $N_f = 3$ FFN calculation and the $N_f = 4$ calculation at a switching value $\mu = Q\equiv \mu_{4}$ somewhere above $m_\Lc$. At the corresponding  value of $Q=\mu_4$, the calculated cross section will be discontinuous, which is detrimental in a PDF fit. If $\mu_4^2$ is too close to $m_\Lc^2$, the $N_f=4$ FFN cross section will miss important terms proportional to $m_\Lc^2$. If $\mu_4^2$ is too far above $m_\Lc^2$, the missing higher-order collinear logs $\as \log(Q^2/m_\Lc^2)$ make the $N_f=3$ FFN cross section unreliable in the upper range of $Q$. 

To avoid that, many global fits use some version of a general-mass variable flavor number (GM-VFN) scheme, see references above. These schemes allow for $N_f$ switching exactly at the corresponding quark mass, $\mu_{i}=m_{i}$, and they achieve a smooth interpolation between the switching points. In such a scheme, one retains numerically {\it nonneglible} masses for {\it any} quark type
in $\hat \sigma$. For instance, the ZM VFN and GM VFN schemes differ only
in the treatment of the terms of order ${\cal O}(m_i^2/Q^2)$ in the
short-distance cross sections. 
They have the same mass dependence of PDFs. All that we are doing is including the essential ${\cal O}(m_i^2/Q^2)$ terms in $\hat \sigma$.

The logic that we just outlined is closely followed by
the simplified Aivazis-Collins-Olness-Tung (S-ACOT)
scheme \cite{Aivazis:1993pi,Kramer:2000hn}.
It is proved to all $\alpha_s$ orders in
Ref.~\cite{Collins:1998rz}
and applied in DIS up to NNLO \cite{Guzzi:2011ew} for use in CTEQ-TEA fits.
In the ACOT family of schemes, the flavor number in
$\alpha_s$, masses, and PDFs is varied according to the
CWZ prescription. Other GM-VFN schemes are
perturbatively equivalent to the (S-)ACOT scheme. 
The reader can find
comparisons between the GM-VFN approaches in \cite{Gao:2017yyd, Guzzi:2011ew, Alekhin:2009ni, Binoth:2010nha}. 

We have discussed variable flavor number schemes with up to five active flavors. Also, at a 100 TeV collider, introduction of a top-quark PDF may be warranted, see Sec. 3.4 in \cite{Mangano:2016jyj}. 
 Groups doing global fitting often also present PDFs determined in fixed-flavor number schemes.

{\bf Heavy-quark masses.} All nonzero quark masses in the calculation require renormalization. Either the heavy-quark $\MSbar$ masses or pole masses provided by the Review of Particle Physics \cite{Tanabashi:2018oca} are used as input parameters when fitting the PDFs. Their values can be even extracted from PDF fits \cite{Gizhko:2017fiu,Alekhin:2017kpj,Ball:2016neh,Gao:2013wwa}. The $\MSbar$ mass for charm quark is better defined in pQCD and more precisely constrained by the world data. On the other hand, some pQCD calculations in the global fits use the pole mass as the input. 
Perturbative relations to convert the $\MSbar$ mass into the pole mass, or back, are known to a high order in pQCD \cite{Chetyrkin:2000yt}.

{\bf Fitted charm.} The PDF for charm quarks is of special interest. Consider first the most standard treatment. Suppose, hypothetically, that we have data for a cross section at a scale a scale $Q$ around $m_\Lc$, and that, in this cross section, the hard interaction produces a charm quark  and antiquark in the final state. The 3-flavor scheme would be useful for predicting this cross section, containing nonperturbative parameterizations of $f_{i/\Lp}^{(3)}(\xi,\mu_0^2)$ for $i \in \{\Lg, \Lu, \bar \Lu, \Ld, \bar \Ld, \Ls, \bar \Ls\}$. There is no PDF for charm quarks. If we now change to a 4-flavor scheme, we have PDFs $f_{i/\Lp}^{(4)}(\xi,\mu^2)$ for $i \in \{\Lg, \Lu, \bar \Lu, \Ld, \bar \Ld, \Ls, \bar \Ls, \Lc, \bar \Lc\}$. These would be obtained by matching to the 3-flavor PDFs at scale $\mu^2 = m_\Lc^2$. For the charm quark, the matching gives $f_{\Lc/\Lp}^{(4)}(\xi,m_\Lc^2) = 0$ (at leading order). Matching gives us what may be called {\em perturbative charm} because the charm PDF for $\mu^2 > m_\Lc^2$ is produced solely by the perturbative matching and DGLAP evolution.

Note that the leading-order $3\rightarrow 4$  matching condition $f_{\Lc/\Lp}^{(4)}(\xi,m_\Lc^2) = 0$ can be traced to the assumption that the power-suppressed corrections of order  ${M}^2/{m_c^2}$ to a hypothetical cross section at $Q^2 \sim m_\Lc^2$ would be negligible. If we think that this is perhaps not the case, we could allow $f_{\Lc/\Lp}^{(4)}(\xi,m_\Lc^2)$ to be non-zero. Then, solving the DGLAP evolution equation with the boundary condition at $\mu^2 = m_\Lc^2$ would result in a possibly non-zero value of $f_{\Lc/\Lp}^{(4)}(\xi,m_\Lc^2)$, which in turn will influence the prediction of experimental results at scales $Q^2$ larger than $m_\Lc^2$, even $Q^2 \gg m_\Lc^2$. If we put these experimental results into the PDF fitting program, we can fit $f_{\Lc/\Lp}^{(4)}(\xi,m_\Lc^2)$. This gives us {\em fitted charm}.

The logic of fitted charm just outlined implies that, if we {\em do} consider experiments at $Q^2 \sim m_\Lc^2$, we should allow for fitted non-perturbative charm quark contributions to the predictions for these experiments. If we use a 4-flavor scheme with $m_\Lc \ne 0$ as described below, these contributions can come from $f_{\Lc/\Lp}^{(4)}(\xi,\mu^2)$ for $\mu^2 \sim m_\Lc^2$. 
%If we use a 3-flavor scheme, these contributions can come from power-suppressed contributions of order ${M}^2/{m_c^2}$.

Representative Feynman diagrams that contribute to the fitted component of the charm PDF
can be viewed in Fig.~3 of \cite{Hou:2017khm}. These diagrams
introduce contributions of order ${\cal O}\!\left({M}^2/{m_c^2}\right)$ to the charm PDF. The available data can accommodate or even mildly prefer contributions $f_{\Lc/\Lp}^{(4)}(\xi,m_\Lc^2)$ carrying up to about 1\% of the net proton's momentum
\cite{Ball:2016neh,Jimenez-Delgado:2014zga,Hou:2017khm}.

%-------------------------------------------------------------------
\subsection{Deeply Inelastic Scattering}\label{sec:DIS}

We conclude the theory overview by a brief discussion of
 deeply inelastic lepton scattering (DIS), $\ell(k) + A(p_A) \to \ell'(k') + X$,
a very important class of processes for the determination of
PDFs  \cite{Devenish:2004}. Here $\ell$ and $\ell'$ can be either electrons, neutrinos, or
muons with specified momenta $k$ and $k'$. Hadron $A$ is a proton, nucleus,
or pion with momentum $p_A$; and
$X$ stands for unobserved particles. The interaction between the
leptons and hadron proceeds by an exchange of a virtual $\gamma^*$, $Z$,
or $W$ boson with momentum $q = k - k'$.

Not only were the measurements in DIS
historically influential in the development of QCD, while diverse DIS
data from HERA and fixed targets still serve as the backbone for global fits, 
but also
projections \cite{Khalek:2018mdn, Hobbs:2019gob}
show that DIS data will continue to provide essential 
constraints on the PDFs in the high-luminosity LHC era.

It is conventional to define three Lorentz-invariant variables,
\begin{equation}
\begin{split}
\label{eq:DISvariables}
Q^2 ={}& - q^2
\;,\
x_{\it bj} = \frac{Q^2}{2 P_\LA \cdot q}
\;,\ 
y = \frac{P_\LA\cdot q}{P_\LA \cdot k}
\;.
\end{split}
\end{equation}
``Deeply inelastic'' means that $Q^2 \gg 1 \GeV^2$, and $x_{\it bj}$ is
not too small or too close to 1. The use of the variable $x_{\it bj}$ was first suggested by James Bjorken, who proposed that the cross section would have simple properties in the deeply inelastic limit \cite{Bjorken:1968dy}.

If only the final-state lepton is observed,
one often writes the spin-independent
cross section as a linear combination of three ``structure
functions,'' $F_1(x_{\it bj},Q^2)$, $F_2(x_{\it bj},Q^2)$, and
$F_3(x_{\it bj},Q^2)$. To determine the $F_i$'s, one needs data from
two c.m.~energies, $\sqrt s$. Otherwise, an approximate assumption is
needed to extract the $F_i$. The structure function $F_3$ is nonzero
only if the current violates parity. 

The structure functions
can be written in terms of PDFs in the form
\begin{equation}
\begin{split}
\label{eq:factorizationDIFi}
F_i(x_{\it bj},Q^2) &= \sum_{a}\int\!d\xi\
f_{a/A}(\xi,\mu^2)\,
\hat F_{i,a}\left(\frac{x_{\it bj}}{\xi},\ \frac{\mu^2}{Q^2}\right) \\
 & + {\cal O}(M/Q), 
\end{split}
\end{equation}
where $M \sim 1 \mbox{ GeV}$.
Usually, one chooses $\mu^2 = Q^2$. This is really just another form
of Eq.~(\ref{eq:DISfactorization}). It offers one nice result.
In $eA$ scattering with a virtual
photon exchange from the electron, at lowest order in $\alpha_s$,
$\hat F_2$ contains a delta function that sets $\xi \to x_{\it bj}$:
\begin{equation}
F_2(x_{\it bj},Q^2) \approx \sum_{i} \mathcal{Q}_i^2\, x_{\it bj}\,f_i(x_{\it bj},Q^2)
\;,
\label{eq:F2LO}
\end{equation}
where $\mathcal{Q}_i$ is the electric charge (in units of $e$) of partons of type
$i$: $Q_{\Lu, \Lc, \Lt}=2/3$, and $Q_{\Ld,\Ls,\Lb}=-1/3$. Because of the charge
factors $Q_i^2$, the neutral-current DIS via the photon exchange is
about four times more sensitive to up-type (anti)quark PDFs than to
down-type ones. It is more difficult for DIS to constrain the
$d$-quark and especially $s$-quark PDFs, so the uncertainties on these
flavors tend to be higher than for $u$ and $c$ PDFs, as we already
saw in Fig.~\ref{fig:pdfbands}. We caution, however, that the simple result in Eq.~(\ref{eq:F2LO}) does not hold beyond the lowest order, and that the structure functions are not to be confused with PDFs.

Note that we have used $\xi$ for the momentum fraction argument of PDFs and $x_{\it bj}$ for the kinematic variable defined in Eq.~(\ref{eq:DISvariables}). In the particle physics literature, one often sees the notation $x$ used for the PDF momentum fraction argument $\xi$. In fact, we will sometimes use $x$ with this meaning later in this review. We always use $x_{\it bj}$ and not simply $x$ for the kinematic variable defined in Eq.~(\ref{eq:DISvariables}).

%-------------------------------------------------------------------

% !TEX root = main.tex
\section{Statistical inference in fitting the parton distributions}
\label{sec:Statistics}

In this section, we derive the key statistical results relevant for
the extraction of the PDFs from the
experimental data. We use the simplest possible framework, in which
experimental errors can be approximated as having a Gaussian
distribution, and the theoretical predictions are approximated as
linear functions of the parameters used to describe the parton
distribution functions. This framework is sometimes called the Hessian
method \cite{Pumplin:2001ct}, since a certain matrix called the
Hessian matrix plays a prominent role. The Hessian approach is
motivated by the observation that many essential features of the PDF
fits are captured by assuming an approximately Gaussian behavior of
the underlying probability distribution of the experimental data. The PDF
functional forms can be determined (``inferred'') from the experimental
data by applying Bayes' theorem \cite{Alekhin:1996za} as
is briefly summarized in Sec.~\ref{sec:Bayes}. The
Hessian approximation provides a simplified solution to the problem of
Bayesian inference when the PDFs are well-constrained, and the deviations
from the most likely solution for PDFs are relatively small. 
The only non-standard feature
that we add is the inclusion of a set of parameters $R_k$ that
represent the possibility that the theory, with an ideal choice of
parameters for the PDFs, is not perfect and
may not fit data exactly. 

\subsection{Bayes' theorem}
\label{sec:Bayes}

Fitting parton distributions to data involves accounting for the statistical and systematic errors in the data. Thus we will need a statistical analysis. For this, we use a Bayesian framework in this paper. The alternative is a frequentist framework, but we find that the Bayesian framework is simple to understand and makes us more aware of assumptions that are otherwise left obscure.

We begin with Bayes' theorem. At its base, this is a simple matter of counting. Consider a population in which each individual can have one or both of two characteristics, $T_1$ and $D$. For a concrete example, the population might consist of people in California. $T_1$ might be ``has a certain genetic marker'', while $D$ might be ``tests positive for this genetic marker.'' Denote by $P(T_1)$ the probability that an individual has characteristic $T_1$. That is, $P(T_1)$ is the number of individuals with characteristic $T_1$ divided by the total number of individuals. Similarly, let $P(D)$ be the probability that an individual has characteristic $D$. Denote by $P(T_1|D)$ the conditional probability that an individual that is known to have characteristic $D$ has characteristic $T_1$. That is, $P(T_1|D)$ is the number of individuals with both characteristics $T_1$ and $D$ divided by the number of individuals with characteristic $D$. Similarly, let $P(D|T_1)$ the conditional probability that an individual that is known to have characteristic $T_1$ has characteristic $D$. With these definitions, we have Bayes' theorem:
\begin{equation}
\label{eq:Bayes0}
P(T_1|D)\, P(D) = P(D|T_1)\, P(T_1)
\;.
\end{equation}
By evaluating the {\it likelihood} $P(D|T_1)$ according to an explicit prescription presented in the next section and by having estimates for $P(D)$ and $P(T_1)$, we could {\it infer} the ``posterior probability'' $P(T_1|D)$ by rearranging the factors in Eq.~(\ref{eq:Bayes0}):
\begin{equation}
\label{eq:Bayes00}
P(T_1|D) = \frac{P(D|T_1)\, P(T_1)}{P(D)}
\;.
\end{equation}

Now consider another characteristic $T_2$, for instance ``does not have the genetic marker.'' Then Bayes' theorem gives us
\begin{equation}
\label{eq:Bayes1}
\frac{P(T_1|D)}{P(T_2|D)} = \frac{P(D|T_1)}{P(D|T_2)}\,\frac{P(T_1)}{P(T_2)}
\;.
\end{equation}
Note that $P(D)$ cancels in this equation. If we know the quantities on the right-hand side of Eq.~(\ref{eq:Bayes1}), this tells us the relative probabilities of finding characteristics $T_1$ and $T_2$ among individuals that are known to have characteristic $D$. 

Eq.~(\ref{eq:Bayes1}) is directly applicable and useful in cases in which there is a large number of individuals, and we know the probabilities on the right-hand side of the equation. For instance, a physician of a patient who tests positive for a rare health condition will want to use Eq.~(\ref{eq:Bayes1}) to help decide on prescribing a specific treatment to address this condition. 

We will, however, use Eq.~(\ref{eq:Bayes1}) in a more subtle case. Suppose that there is only one individual. Suppose further that you have a subjective belief, based on your prior experience, that the probability that this individual has property $T_i$ is $P(T_i)$ for $i \in \{1,2\}$.  This ``prior probability'' could be formed based on your past observations.

Now suppose that the individual is observed to have property $D$. Assume that you know how to compute the probability $P(D|T_i)$ for an individual to have property $D$ if the individual has property $T_i$. You can turn this knowledge around with the help of Bayes theorem (\ref{eq:Bayes00}) to calculate the probability  $P(T_i|D)$ for having property $T_i$ on the condition that the individual was observed to have property $D$. You can also use Eq.~(\ref{eq:Bayes1}) to compute the ``posterior probability ratio'' $P(T_1|D)/P(T_2|D)$, by multiplying the prior probability ratio $P(T_1)/P(T_2)$, reflecting your knowledge before the measurement, by the ratio $P(D|T_1)/P(D|T_2)$ of the likelihoods calculated for each property $T_i$. Then, you declare an updated prior probability ratio: 
\begin{equation}
\left(\frac{P(T_1)}{P(T_2)}\right)_{\rm new} \equiv \frac{P(T_1|D)}{P(T_2|D)}= 
\frac{P(D|T_1)}{P(D|T_2)}\,\left(\frac{P(T_1)}{P(T_2)}\right)_{\rm old}.
\label{eq:Bayes2}
\end{equation}

In our physics application, $T_1$ and $T_2$ are possible theoretical models describing a physical system, or perhaps one model of the system with two choices of parameters. Even when we cannot interpret $P(T_i)$ by counting the number of instances of the system in which $T_i$ holds, we often have some idea about which theory is more likely, and we express this belief as the prior probabilities $P(T_i)$. The prior probabilities are often based on previous experiments, but are partly subjective. Thus your prior probabilities may not be the same as mine. Now we make a measurement and observe that the system has a property $D$. With these new data in mind, we update the probability ratio for the two theories according to Eq.~(\ref{eq:Bayes2}).

Your new estimate is still partly subjective and may not agree with mine. However, as data accumulates, it frequently happens that the subjective nature of our probabilities ceases to really matter. Suppose, for example, that your prior probability ratio is $P(T_1)/P(T_2) = 10$, and mine is $P(T_1)/P(T_2) = 0.1$. Then we do not agree which theory is more likely. However, suppose that, after lots of data become available, the likelihood ratio for the data is $P(D|T_1)/P(D|T_2) = 10^6$. Then your posterior probability ratio is $P(T_1|D)/P(T_2|D) = 10^7$, while mine is $P(T_1|D)/P(T_2|D) = 10^5$.  Either ratio is very large and strongly disfavors theory $T_2$ in comparison to theory $T_1$. At this point, I just agree that you were right, and we stop discussing the matter.

Notice that our subjective probability estimates change by the same factor, equal to the likelihood ratio $P(D|T_1)/P(D|T_2)$. We agree on this factor because we agree on the data $D$ as well as on the calculation. For the agreement to be possible, the theories must be {\it natural}, in the sense that the theory predictions $T$  and, by extension, the probabilities depending on them, are smooth functions  of theoretical parameters. The technical reason is that estimation of correlated uncertainties requires inversion of a large Hessian matrix $H_{\alpha\beta}$ constructed from derivatives $\partial T_k/\partial a_\alpha$ of model predictions $T_k$ with respect to parameters $a_\alpha$. This inversion is numerically stable only if $\partial T_{k}/\partial a_\alpha$ are well-behaved. Naturalness is thus required for reliable estimates of the probabilities and  derived quantities, including the model-discriminating ratios $P(D|T_1)/P(D|T_2)$ and uncertainties on theoretical predictions.

\subsection{Gaussian model for the distribution of data}

We wish to fit parameters $a_\alpha$, $\alpha \in \{1,\dots,N_P\}$, to data. The data are given by values $D_k$, $k \in \{1,\dots,N_D\}$. Each $D_k$ represents the number of counts divided by an integrated luminosity in a certain bin of measured momenta in a certain experiment that is to be included in the fit. In the Gaussian approximation, we suppose that, after accounting for experimental errors, the data have the form
\begin{equation}
\label{eq:Dkdef}
D_k = \langle D_k \rangle + \sigma_k\, \Delta_k 
+ \sigma_k\sum_J \beta_{kJ} \bar\lambda_J
\;.
\end{equation}
Here $\langle D_k \rangle$ is the value that the datum $D_k$ would have if there were no experimental uncertainties from either counting statistics or systematic effects such as detector calibration. The value $\sigma_k$ is the statistical uncertainty (one standard deviation quoted in the data tables with the ``$\pm$'' sign) associated with $D_k$. The variable $\Delta_k$ represents fluctuations in $D_k$ from counting statistics or other sources. In accord with our Gaussian approximation, the $\Delta_k$ are normalized to be independent Gaussian random variables with mean 0 and variance 1. We will use the notation $\cN(0,1)$ for this distribution. In a typical particle experiment, there can be experimental systematic uncertainties, for example, associated with some imprecision in measurements of luminosity or particle energies. In general, the systematic uncertainties can be decomposed into two types: random fluctuations that are fully uncorrelated point-by-point and contribute to $\sigma_k\Delta_k$ in Eq.~(\ref{eq:Dkdef}); and fully correlated variations between the experimental data points, 
contributing to the last term in Eq.~(\ref{eq:Dkdef}). Then, we would interpret $\sigma_k$ as the full uncorrelated uncertainty, composed from the uncorrelated statistical and systematic uncertainties added in quadrature. 

Let the number of sources of correlated systematic uncertainties be $N_\lambda$. We represent the correlated systematic uncertainties by a correlation matrix $\sigma_k \beta_{kJ}$. This is a matrix with indices $k,J$. There is no sum over $k$. Then in Eq.~(\ref{eq:Dkdef}) there is a sum  over an index $J$ that labels the sources of correlated systematic uncertainties. The contribution to $D_k$ from a systematic source $J$ is written as  $\sigma_k\beta_{kJ} \bar\lambda_J$, where $\bar\lambda_J$ is an independent random variable with an $\cN(0,1)$ distribution.\footnote{In general, both statistical and systematic uncertainties can be asymmetric and quoted as such by the experimental groups. In practice, neglecting the asymmetry has been often acceptable in the PDF analyses.} 

An example may be helpful. Each $D_k$ is obtained by dividing a number of counts by a measured value $\cL_0$ of the integrated luminosity $\cL$ for the corresponding experiment. There is some uncertainty in the luminosity measurement, which we express by writing $\cL = \cL_0 + \cL_0\sigma_\cL \bar\lambda_{J_\cL}$, where $\sigma_\cL$ is an estimated fractional uncertainty in the luminosity, ${J_\cL}$ is the index we choose for this source of uncertainty, and $\bar\lambda_{J_\cL}$ is the corresponding Gaussian random variable. Then in Eq.~(\ref{eq:Dkdef}),
\begin{equation}
\sigma_k \beta_{k{J_\cL}} = \langle D_k \rangle \sigma_\cL
\;.
\end{equation}

We will refer to the random variables $\bar\lambda_{J}$ as correlated systematic error variables. They are analogous to the variables $\Delta_k$, whose fluctuations embody the uncorrelated errors. In one way of analyzing the data, we do not try to determine the values of the $\bar\lambda_{J}$. In another analysis, introduced in Sec~\ref{sec:altchisq}, we introduce {\it nuisance parameters} $\lambda_{J}$ that are intended to estimate the values of the $\bar\lambda_{J}$. 

The distribution of data $D$ is given by noting that, if $f$ is a function of the $\Delta_k$ and $\bar \lambda_J$, its expectation value is
\begin{equation}
\begin{split}
\label{eq:expectation1}
\langle f \rangle ={}& 
(2\pi)^{- (N_D+N_\lambda)/2}\int d^{N_D} \Delta\ 
\int d^{N_\lambda} \bar\lambda\
f(\Delta,\bar \lambda)
\\&\times
\exp\!\left(- \frac{1}{2}\sum_k \Delta_k^2 - \frac{1}{2}\sum_J \bar\lambda_J^2
\right)
\;.
\end{split}
\end{equation}
Using Eq.~(\ref{eq:Dkdef}) in Eq.~(\ref{eq:expectation1}) and observing that $\langle \Delta_k \rangle = 0$ and $\langle \bar \lambda_J \rangle = 0$, we see that the expectation value of $D_k$ is the value $\langle D_k \rangle$ that appears in Eq.~(\ref{eq:Dkdef}). Then, using Eq.~(\ref{eq:Dkdef}) in Eq.~(\ref{eq:expectation1}) again, and noting that $\langle \Delta_i \Delta_j \rangle = \delta_{ij}$, $\langle \bar \lambda_J \bar\lambda_L \rangle = \delta_{JL}$, and $\langle \Delta_i \bar\lambda_J \rangle = 0$, we find that
\begin{equation}
\begin{split}
\label{eq:Dkexpectations}
\langle (D_i - \langle D_i \rangle)(D_j - \langle D_j \rangle) \rangle ={}& 
C^{-1}_{ij}
\;,
\end{split}
\end{equation}
where
\begin{equation}
\label{eq:Cinversedef}
C^{-1}_{ij} = \sigma_i\sigma_j \left\{ \delta_{ij} 
+ \sum_J \beta_{i J} \beta_{j J} \right\}
\;.
\end{equation}
The matrix $C^{-1}$ is the inverse of a matrix $C$ that is called the covariance matrix. When experimental systematic errors $\beta_{iJ}$ are present, $C$ is generally not a diagonal matrix.\footnote{We may note that, if we start with the matrix $C^{-1}$, instead of Eq.~(\ref{eq:Dkdef}), we can diagonalize $C^{-1}$ to obtain a non-unique representation in the form of Eq.~(\ref{eq:Cinversedef}), which is consistent with Eq.~(\ref{eq:Dkdef}).} We can use this simple result to show that, if $f$ is a function of the data $D_k$, then its expectation value is
\begin{align}
\label{eq:expectation2}
\langle f\rangle ={}& 
\frac{\sqrt{\det C}}{(2\pi)^{N_D/2} }
\int d^{N_D} D\ 
f(D)
\\&\times
\exp\!\left(- \frac{1}{2}\sum_{ij}
(D_i - \langle D_i \rangle)(D_j - \langle D_j \rangle)\,C_{ij}
\right)
. \notag
\end{align}
This gives the the probability density for the data $D_k$, with the variables $\bar\lambda_J$ integrated out.

The relation between Eq.~(\ref{eq:Dkexpectations}) and Eq.~(\ref{eq:expectation2}) is an important general result that we can use whenever variables $y_i$ are distributed in a generalized Gaussian fashion, that is, with a probability density proportional to $\exp(-\sum y_i y_j M_{ij})$ for a matrix $M$. To prove it, we use Eq.~(\ref{eq:expectation2}) to compute $\langle (D_i - \langle D_i \rangle)(D_j - \langle D_j \rangle) \rangle$. We change variables in (\ref{eq:expectation2}) to $x_i = (\sqrt C)_{ij} (D_j - \langle D_j \rangle )$.\footnote{We can define the matrix $\sqrt C$ because $C$ is a real symmetric matrix with all positive eigenvalues.} We use $d^{N_D}D = (1/\sqrt{\det C})\, d^{N_D} x$. Then the exponent is $-\frac{1}{2}\sum_{i}x_i^2$. We then obtain $\langle x_k x_l \rangle = \delta_{kl}$, which gives us the result in Eq.~(\ref{eq:Dkexpectations}).

\subsection{Determining theory parameters using $\chi^2$}
\label{sec:theorypredictions}

We now introduce theoretical predictions $T_k(a)$ for the data $D_k$. The prediction depends on some parameters $a$, most notably the parameters of the PDFs. We have defined $\langle D_k \rangle$ to be the value of $D_k$ if the experimental errors are negligible. For a given choice of the parameters $a$, the theoretical prediction may not be correct, but if the prediction is perfect then $T_k(a) = \langle D_k \rangle$. 

If we substitute $\langle D_k \rangle \to T_k(a)$ in Eq.~(\ref{eq:expectation2}), we see that the probability to obtain the experimental results $D$ if the theory represented by $T(a)$ is correct is
\begin{equation}
    \label{eq:PDTa}
P(D|T(a)) = d\mu(D)\, \exp\left(- \frac{1}{2}\chi^2(D,a)\right)\ ,
\end{equation}
where
\begin{equation}
\label{eq:chisqDa}
\chi^2(D,a) \equiv 
\sum_{ij}
(D_i - T_i(a))(D_j - T_j(a))\, C_{ij}
\;,
\end{equation}
and the data space measure is
\begin{equation}
\label{eq:muD}
d\mu(D) \equiv (2\pi)^{- N_D/2}\sqrt{\det C}\
d^{N_D} D
\;.
\end{equation}
Consider two choices, $a_1$ and $a_2$, for the parameters. Suppose that before seeing the experimental results $D$, we judge the probability that theory $T(a_1)$ is correct to be $P(T(a_1))$, and we judge the probability that theory $T(a_2)$ is correct to be $P(T(a_2))$. Perhaps these prior probabilities are based on previous experiments, or perhaps they are based on some sort of dynamical model of parton behavior. Whatever our prior belief was, it should be modified after we know the experimental results. Let the new probabilities based on the experimental results $D$ be $P(T(a_1)|D)$ and $P(T(a_2)|D)$, respectively. Then Bayes' theorem (\ref{eq:Bayes1}) gives us
\begin{equation*}
\frac{P(T(a_1)|D)}{P(T(a_2)|D)}
=
\frac{P(D|T(a_1))}{P(D|T(a_2))}\
\frac{P(T(a_1))}{P(T(a_2))}
\;.
\end{equation*}
The information from experiment is contained in the likelihood ratio 
\begin{equation}
\label{eq:likelihoodR}
\frac{P(D|T(a_1))}{P(D|T(a_2))} = 
\exp\!\left(- \frac{
\chi^2(D,a_1)-\chi^2(D,a_2)}{2}
\right)
.
\end{equation}
Thus $\chi^2(D,a)$ is the function that we need for parameter estimation: differences in $\chi^2$ give us the likelihood ratio that tells us how to adjust our judgments of which parameter choices are favored, using Bayes' theorem (\ref{eq:Bayes1}). The maximum of the likelihood $P(D|T(a))$ with respect to $a$ is achieved at the global minimum of $\chi^2(D,a)$.

We can use $\chi^2(D,a)$ as a measure of goodness of fit: the theory matches the experimental results well when the differences $D_i - T_i(a)$ are small, meaning that $\chi^2(D,a)$ is small. However, in our opinion, there are reasons to be critical of $\chi^2(D,a)$ as the only measure of goodness of fit when the number $N_D$ of data is large. We will return to this question in Sec.~\ref{sec:goodness}.

\subsection{Another definition of $\chi^2$}
\label{sec:altchisq}

Given a sample of data $D_k$, we can measure how well the theory matches the data by the parameter $\chi^2(D,a)$ defined in Eq.~(\ref{eq:chisqDa}). We can rewrite Eq.~(\ref{eq:chisqDa}) as
\begin{equation}
\begin{split}
\chi^2(D,a) ={}& 
\sum_{ij}
(D_i - \langle D_i \rangle)(D_j - \langle D_j \rangle)\, C_{ij}
\\&
+ 2 \sum_{ij} 
(D_i - \langle D_i \rangle)(\langle D_j \rangle - T_j(a))\, C_{ij}
\\ &
+ \sum_{ij}
(\langle D_i \rangle 
- T_i(a))(\langle D_j \rangle - T_j(a))\, C_{ij}
,
\end{split}
\end{equation}
with the expectation value
\begin{equation}
\langle\chi^2(D,a)\rangle =
N_D + \chi^2 \left(\langle D\rangle,\ a \right),
\end{equation}
which we derived using Eqs.~(\ref{eq:chisqDa}) and (\ref{eq:Dkexpectations}). This is minimized when $T_i(a) = \langle D_i \rangle $.

The function $\chi^2(D,a)$ is derived from the differential probability $P(D|T(a))$ to obtain the data $D$ if the theory $T(a)$ is correct, cf. Eq.~(\ref{eq:PDTa}). We can define another version of $\chi^2$, based on the differential probability $P(D|T(a),\lambda)$ to obtain the data $D$ if the theory $T(a)$ is correct, and if the random systematic error variables $\bar \lambda$ take the values $\bar \lambda_J = \lambda_J$. Using Eqs.~(\ref{eq:expectation1}) and (\ref{eq:Dkdef}) with $\langle D_k \rangle$ replaced by $T_k(a)$, this probability is
\begin{align}
\label{eq:PDTalambda1}
P(D|&T(a),\lambda) 
\notag
\\&=
(2\pi)^{- (N_D+N_\lambda)/2}\int d^{N_D} \Delta\ 
\int d^{N_\lambda} \bar\lambda
\notag
\\&\times
\exp\!\left(- \frac{1}{2}\sum_k \Delta_k^2 - \frac{1}{2}\sum_J \bar\lambda_J^2
\right)
\\ &\times
\prod_k \delta\!\left(D_k - \left[
T_k(a) + \sigma_k\Delta_k + \sigma_k\sum_J \beta_{kJ}\bar\lambda_J
\right]\right)
\notag
\\&\times
\prod_J \delta\!\left(\bar\lambda_J - \lambda_J\right)
\;.\notag
\end{align}
Performing the integrations, we have
\begin{equation}
\begin{split}
\label{eq:PDTalambda2}
P(D|T(a),\lambda) ={}&
(2\pi)^{- (N_D+N_\lambda)/2}\left[\prod_k\frac{1}{\sigma_k}\right]
\\&\times
\exp\!\left(- \frac{1}{2}\chi^2(D,a,\lambda)
\right)
\;.
\end{split}
\end{equation}
where
\begin{equation}
\begin{split}
\label{eq:chisqDalambda}
\chi^2(D,a,\lambda) ={}&
\sum_k
\left[
\frac{D_k - T_k(a)}{\sigma_k} - \sum_I \beta_{kI}\lambda_I
\right]^2
\\&
+\sum_J \lambda_J^2
\;.
\end{split}
\end{equation}
Then, as in Eq.~(\ref{eq:likelihoodR}), the information from experiment needed to apply Bayes' Theorem to the determination of $a$ and $\lambda$ is contained in the likelihood ratio 
\begin{equation}
\label{eq:likelihoodRalambda}
\frac{P(D|T(a_1),\lambda_1)}{P(D|T(a_2),\lambda_2)}  =
\exp\!\left(- \frac{
\small{\chi^2(D,a_1,\lambda_1) -
\chi^2(D,a_2,\lambda_2)}
}{2}
\right)
,
\end{equation}
Note that the parameters $\lambda_J$ are not necessarily equal to the true systematic error variables $\bar\lambda_J$. Rather, the $\lambda_J$ are parameters that one can fit to the data $D_k$. The best fit values of $\lambda_J$ then approximate the true $\bar\lambda_J$. The $\lambda_J$ are called {\em nuisance parameters}. We maximize the likelihood of obtaining the observed data with parton parameters $a$ and nuisance parameters $\lambda$ by minimizing $\chi^2(D,a,\lambda)$ with respect to $a$ and $\lambda$.

The function $\chi^2(D,a,\lambda)$ is useful if we want to use data to estimate not only the true values of the parton parameters, $\bar a$, but also the systematic error parameters $\bar\lambda$. Since we are normally not so interested in the $\bar\lambda$ values, the function $\chi^2(D,a,\lambda)$ may seem less important than the function $\chi^2(D,a)$. However, the function $\chi^2(D,a,\lambda)$ has the advantage that it does not involve the covariance matrix $C_{ij}$. 

It is significant that if we fit values of $\lambda$ by minimizing $\chi^2(D,a,\lambda)$, we can obtain $\chi^2(D,a)$. With the manipulations outlined in Appendix \ref{sec:appendix}, we can write $\chi^2(D,a,\lambda)$ in an instructive form:
\begin{equation}
\begin{split}
\label{eq:altchisqDalam}
\chi^2(D,a,\lambda) ={}&
\sum_{ij} 
(D_i - T_i(a))(D_j - T_j(a))\, C_{ij}
\\&
+ \sum_{I J} 
\lambda'_I \lambda'_J B_{IJ}
\;,
\end{split}
\end{equation}
where $C$ is the covariance matrix defined in Eq.~(\ref{eq:Cinversedef}),
$B$ is a matrix with elements
\begin{equation}
\label{eq:BIJbis2}
B_{IJ} \equiv \delta_{IJ} 
+\sum_k \beta_{kI}\beta_{kJ}
\;,
\end{equation}
and $\lambda'$ is a shifted version of $\lambda$,
\begin{equation}
\label{eq:lambdashift2}
\lambda'_I = \lambda_I - \sum_J \sum_k \frac{(D_k - T_k(a))}{\sigma_k}\,
\beta_{kJ} B^{-1}_{JI}
\;.
\end{equation}
The minimum of $\chi^2(D,a,\lambda)$ with respect to $\lambda$ occurs when $\lambda' = 0$, corresponding to $\lambda =\lambda^\textrm{fit}$. Thus 
\begin{equation}
\min_\lambda \chi^2(D,a,\lambda) \equiv
\chi^2(D,a,\lambda^\textrm{fit})=\chi^2(D,a)
\;.
\end{equation}
The PDF-fitting groups use either form of $\chi^2$. See, for example, the review of various conventions for $\chi^2$ in Appendix A of \cite{Ball:2012wy}. 

In $\chi^2(D,a)$, the experimental systematic errors are encoded in the matrix $C$. This form is used, e.g., by the NNPDF analyses. In $\chi^2(D,a,\lambda)$, we have the systematic errors expressed explicitly using parameters $\lambda$. This is the convention adapted by CTEQ analyses, starting with CTEQ6 \cite{Pumplin:2002vw}. There is then an extra term $\sum_J\lambda_J^2$ in $\chi^2(D,a,\lambda)$. We are instructed to fit the parameters $\lambda$ to the data by minimizing $\chi^2(D,a,\lambda)$. In the following sections, we will see how to fit the theory parameters $a$ by minimizing $\chi^2(D,a)$ with respect to the parameters $a$. This is then equivalent to minimizing $\chi^2(D,a,\lambda)$ with respect to $a$ and $\lambda$. 

If we use $\chi^2(D,a)$, then we do not need to be concerned with the systematic error parameters $\bar\lambda$. With $\chi^2(D,a)$, we have a matrix $C_{ij}$. The fact that this matrix is not diagonal indicates that the errors are correlated. The presence of $C_{ij}$ makes the formulas a little complicated, but there are no real conceptual complications: $C_{ij}$ acts as a metric tensor on the space of the data, so that one could think of $u_i C_{ij} v_j$ as simply $u\cdot v$.

The minimum of $\chi^2(D,a,\lambda)$ occurs at values $\lambda^\textrm{fit}_J$ of the nuisance parameters. What is the relation between the $\lambda^\textrm{fit}_J$ and the systematic error variables $\bar\lambda_J$? The correlated error variables $\bar\lambda_J$ influence the data $D_k$, but the uncorrelated error variables $\Delta_k$ also influence the $D_k$ and, furthermore, we do not know the exact parton parameters $a_\alpha$, so one cannot expect to be able to recover the $\bar\lambda_J$ exactly from the data. However, we will see below that the $\lambda^\textrm{fit}_J$ approximate the $\bar\lambda_J$ when there are many data $D_k$ and the parameters $\beta_{kJ}$ that give the influence of the $\bar\lambda_J$ on the data are not too small. Specifically, we will see that the $\lambda^\textrm{fit}_J$ approximate well the $\bar\lambda_J$ when the matrix elements $B_{IJ}$ are large. This happens when the sum over the data index $k$ in Eq.~(\ref{eq:BIJbis2}) includes many terms, and the parameters $\beta_{kI}$ are not too small.

The analysis is simple. We begin with $\chi^2(D,a,\lambda)$ in Eq.~(\ref{eq:chisqDalambda}) and substitute
\begin{equation}
D_k - T_k(a) =  \sigma_k\, \Delta_k 
+ \sigma_k\sum_J \beta_{kJ} \bar\lambda_J
- (T_k(a) - \langle D_k \rangle)
\end{equation}
from Eq.~(\ref{eq:Dkdef}). This gives
\begin{align}
\label{eq:chisqDalambda2}
 \chi^2(D,a,\lambda) = 
\sum_k
\Biggl[
\Delta_k 
& + \frac{T_k(a) - \langle D_k \rangle}{\sigma_k}  \\
& - \sum_J \beta_{kJ}(\lambda_J - \bar\lambda_J)
\Biggr]^2
+\sum_J \lambda_J^2 \notag
\;.
\end{align}
The partial derivatives vanish at the best fit $\lambda_I=\lambda^\textrm{fit}_I$,
\begin{equation}
\begin{split}
0 = -\frac{1}{2}\left.\frac{\partial \chi^2}{\partial \lambda_I}\right|_{\lambda=\lambda^\textrm{fit}} = & 
\sum_k
\left[
\Delta_k 
+ \frac{T_k(a) - \langle D_k \rangle}{\sigma_k}
\right]\beta_{kI}
\\&
- \bar\lambda_I
- \sum_J B_{IJ}
(\lambda^\textrm{fit}_J - \bar\lambda_J)
\;,
\end{split}
\end{equation}
so
\begin{align}
\label{eq:lambdafit}
&\lambda^\textrm{fit}_I - \bar\lambda_I
\\&\quad
= \sum_J B^{-1}_{IJ}
\left\{
\sum_k
\beta_{kJ}
\left[
\Delta_k 
+ \frac{T_k(a) - \langle D_k \rangle}{\sigma_k}
\right]
- \bar\lambda_J
\right\}
.\notag
\end{align}
On the right-hand side of this equation, the $\bar \lambda_J$ are of order 1, the quantities $T_k(a) - \langle D_k\rangle$ should be small if we use values of $a$ fit to the data, and, since the $\Delta_k$ are independent random variables with $\cN(0,1)$ distributions, $\sum_k \beta_{kJ} \Delta_k$ should have fluctuations of order a typical $\beta_{kJ}$ coefficient times the square root of the number of contributing indices $k$. Thus the quantity in braces is not large. However, the matrix elements $B^{-1}_{IJ}$ are small. Thus we expect the $\lambda^\textrm{fit}_I - \bar\lambda_I$ to be small.

If we use $\chi^2(D,a,\lambda)$, then the treatment of the systematic error parameters $\lambda$ is similar to the treatment of the theory parameters $a$. We obtain values $\lambda^\textrm{fit} \approx \bar\lambda$. The values $\bar\lambda$ are, by assumption, distributed according to the $\cN(0,1)$ distribution. Thus the values $\lambda^\textrm{fit}$ should be approximately distributed with this distribution. In Sec.~\ref{sec:nuisance}, we use this to test the assumptions that we have used.

We can also consider the limit in which $\sum_k \beta_{kI} \beta_{kJ}$ is small rather than large for some value of $I$ and all values of $J$. Then we have $B_{IJ} \approx \delta_{IJ}$. This gives us
\begin{equation}
\label{eq:lambdafitmod}
\lambda^\textrm{fit}_I
\approx 
\sum_k
\beta_{kI}
\left[
\Delta_k 
+ \frac{T_k(a) - \langle D_k \rangle}{\sigma_k}
\right]
.
\end{equation}
Then $\lambda^\textrm{fit}_I$ is not approximately $\bar \lambda_I$ but is instead quite small. The test in Sec.~\ref{sec:nuisance} suggests that this happens for some of the nuisance parameters.

\subsection{Dependence on the theory parameters \label{sec:deptheopar}}

In section \ref{sec:theorypredictions}, we introduced theory predictions $T_k(a)$ for the data $D_k$. The $T_k(a)$ depend on a number $N_P$ of parameters $a_\alpha$. 
Now we suppose that, in the neighborhood of the global $\chi^2$ minimum, the functions $T_k(a)$ are approximately linear in the parameters. In order to keep the notation as simple as possible, we define the origin of the parameter space so that the neighborhood of the global minimum is a region near $a = 0$. That is, if we were fitting a function $x^{A_1} (1-x)^{A_2}$ for parameters $\{A_1, A_2\}$ and preliminary fits gave $A_1 \approx -1.3$, $A_2 \approx 4.5$, we would define new parameters by $A_1 = -1.3 + a_1$ and $A_2 = 4.5 + a_2$. Then we would be interested in small values of $\{a_1,a_2\}$. Then we assume that, for the purpose of examining the fitting procedure, a linear approximation is adequate:
\begin{equation}
\label{eq:theoryT1}
T_k(a) = T_k(0) + T_{k\alpha} a_\alpha
\;.
\end{equation}
Here and in what follows, we use the Einstein summation convention for parameter indices $\alpha$, $\beta$, \dots. 

Finding the best-fit PDFs does not typically rely on the approximation (\ref{eq:theoryT1}) of linear dependence of $T_k(a)$ on the PDF parameters $a$. However, the PDF error analysis using the Hessian method does rely on this approximation (\ref{eq:theoryT1}). We will adopt it in this review. Using Eq.~(\ref{eq:theoryT1}), $\chi^2$ is a quadratic function of the parameters $a$. In fact, within the range of $a$ important for the fit, this quadratic dependence is a pretty reasonable approximation, as one may note, for example in Fig.~\ref{fig:Einconsistent} in Sec.~\ref{sec:consistency}. One should be careful to check whether this approximation is adequate in an actual fit, see Sec.~\ref{sec:goodness}.

When the statistical and systematic experimental errors that contribute to the data in Eq.~(\ref{eq:Dkdef}) vanish, the data $D_k$ equal their expectation values $\langle D_k \rangle$. We suppose that there are ideal values $\bar a$ of the parton parameters, related to the expectation values $\langle D_k \rangle$ of the data by
\begin{equation}
\label{eq:theoryT2}
T_k(\bar a) =
\langle D_k \rangle + R_k
\;.
\end{equation}
We make the definition of the ideal parameters $\bar a$ more precise
in Eq.~(\ref{eq:theoryT4}) below. In Eq.~(\ref{eq:theoryT2}), we have
included constants $R_k$ that represent imperfections in the theory,
such that even when we use parameters $\bar a$, the theory does not
match $\langle D_k \rangle$ exactly. Of course, one commonly assumes
that the $R_k$ are zero, but in this review we want to at least
consider the possibility that something goes wrong. 
For example, the imperfections represented by $R_k$ could arise because
we omitted higher-order
contributions, there is beyond-the-standard-model physics in the data
but not in the theory, or the parameterization that we use for the PDFs
cannot match the true PDFs exactly. 

Another way to include imperfections in the theory would be to incorporate theory errors into the analysis. In Eq.~(\ref{eq:Dkdef}), we can set the expectation value $\langle D_k\rangle$ of the data in bin $k$ to the cross section in that bin calculated exactly in the Standard Model. We call this exact cross section $\overline T_k(\bar a)$, so that Eq.~(\ref{eq:Dkdef}) becomes
\begin{equation}
\label{eq:Dkdefexact}
D_k = \overline T_k(\bar a) + \sigma_k\, \Delta_k 
+ \sigma_k\sum_{J\in \cE} \beta_{kJ} \bar\lambda_J
\;,
\end{equation}
where $\cE$ is the set of experimental systematic errors. Now, we do not have the exact prediction $\overline T_k(\bar a)$ available. All we have is the cross section $T_k(\bar a)$ calculated, say, at NNLO (but with true PDF and related parameters $\bar a$). In the linear approximation that we use, we can parameterize our ignorance in the form
\begin{equation}
\label{eq:newtheory}
\overline T_k(\bar a) =
T_k(\bar a)
+ \sigma_k\sum_{J\in \cT} \beta_{kJ} \bar\lambda_J
\;,
\end{equation}
where $\cT$ is a set of sources of theory errors, and the parameters $\bar\lambda_J$ are random variables that are chosen from some distribution such as $\cN(0,1)$. Then the term $\sigma_k\sum_{J \in \cT}  \beta_{kJ} \bar\lambda_J$ is our  estimate of theory contributions to the true cross section $\overline T_k(\bar a)$ that are not included in our NNLO calculation $T_k(\bar a)$. The parameters $\beta_{kJ}$  represent our estimate of the dependence of $\overline T_k(\bar a)$ on the $J$-th source of theoretical uncertainty.
Of course, in reality only one value, $\lambda_J^\textrm{true}$, of each $\bar\lambda_J$ will be realized in an exact calculation. The $R_k$ in Eq.~(\ref{eq:theoryT2}) are then 
\begin{equation}
R_k =  -\sigma_k\sum_{J\in \cT} \beta_{kJ} \lambda_J^\textrm{true}
\;.
\end{equation}
When we combine Eqs.~(\ref{eq:Dkdefexact}) and (\ref{eq:newtheory}), we obtain
\begin{equation}
\label{eq:Dkfullerrors}
D_k = T_k(\bar a) + \sigma_k\, \Delta_k 
+ \sigma_k\sum_{J\in \cE \cup \cT} \beta_{kJ} \bar\lambda_J
\;,
\end{equation}
where now both experimental systematic errors and our estimated theory errors are included. The representations of theory errors are discussed in \cite{AbdulKhalek:2019bux,Olness:2009qd,Cacciari:2011ze,Gao:2011unpublished,Forte:2013mda,Harland-Lang:2018bxd}. Until recently, PDF fits typically omitted theory errors. In this review, we do not include theory errors and instead represent imperfections in the theory by the $R_k$ in Eq.~(\ref{eq:theoryT2}).

There is a certain freedom in the definition of $\bar a$ and $R_k$ in Eq.~(\ref{eq:theoryT2}). We can use this freedom to simplify the later analysis. Suppose that we say that Eq.~(\ref{eq:theoryT2}) applies for ideal parameters $\bar a^{(0)}$ and imperfection parameters $R^{(0)}_k$:
\begin{equation}
\label{eq:theoryT21}
T_k(\bar a^{(0)}) =
\langle D_k \rangle + R^{(0)}_k
\;.
\end{equation}
Let $\bar a^{(0)} = \bar a + \delta a$, where the $\delta a_\alpha$ are small parameters that we are free to choose. Then
\begin{equation}
\label{eq:theoryT22}
T_k(\bar a) + T_{k\alpha}\delta a_\alpha =
\langle D_k \rangle + R^{(0)}_k
\;.
\end{equation}
This gives us Eq.~(\ref{eq:theoryT2}) with $R_k = R^{(0)}_k - T_{k\alpha}\delta a_\alpha$. What should the $\delta a_\alpha$ be? In the analysis below, the vector $\sum_{kj}R_k C_{kj} T_{j\beta}$ plays an important role. This vector equals
\begin{align}
\sum_{kj}R_k C_{kj} T_{j\beta} ={}&
\sum_{kj}R^{(0)}_k C_{kj} T_{j\beta}
- H_{\alpha\beta} \delta a_\alpha
\;,
\end{align}
where
\begin{equation}
\label{eq:Halphabeta}
H_{\alpha\beta} \equiv \sum_{kj} T_{k\alpha} C_{kj} T_{j\beta}
\end{equation}
is the Hessian matrix, which will play a major role in the subsequent analysis. We choose
\begin{equation}
\label{eq:specialchoice}
\delta a_\alpha = H^{-1}_{\alpha\beta}
\sum_{kj}R^{(0)}_k C_{kj} T_{j\beta}
\;,
\end{equation}
so that
\begin{equation}
\label{eq:Rkproperty}
\sum_{kj}R_k C_{kj} T_{j\beta} =
0
\;.
\end{equation}
This is a useful property, as we will see.

We define $\chi^2(D,a)$ by Eq.~(\ref{eq:chisqDa}), so that
\begin{equation}
\begin{split}
\label{eq:chisqDa2}
\chi^2(D,a) ={}&
\sum_{ij}
\left[
D_i - T_i(0) - T_{i\alpha}\, a_\alpha 
\right]
\\&\qquad\times
\left[
D_j - T_j(0) - T_{j\beta}\, a_\beta 
\right]
C_{ij}
\;.
\end{split}
\end{equation}
The minimum of $\chi^2(D,a)$ is at parameters such that
\begin{equation}
0 = -\frac{1}{2}\,\frac{\partial\chi^2}{\partial a_\beta}
= \sum_{ij}
\left[
D_i - T_i(0) - T_{i\alpha}\, a_\alpha
\right]
C_{ij} T_{j\beta}
\;.
\end{equation}
This is
\begin{equation}
\label{eq:HatoD}
H_{\beta\alpha} a_\alpha = \cD_\beta
\;,
\end{equation}
where 
\begin{equation}
\begin{split}
\label{eq:cDbeta}
\cD_\beta \equiv {}& 
\sum_{ij} 
(D_i - T_i(0))
C_{ij} T_{j \beta}.
\end{split}
\end{equation}
Thus the fit parameters are
\begin{equation}
\label{eq:bestfit}
a^\textrm{fit}_\alpha = H^{-1}_{\alpha\beta} \cD_\beta
\;.
\end{equation}

What is the expectation value of $a^\textrm{fit}_\alpha$? To answer this question, we need a result for $\bar a_\alpha$. Using Eq.~(\ref{eq:theoryT2}), we have
\begin{equation}
\label{eq:Tabar}
T_{i\alpha} \bar a_\alpha = \langle D_i \rangle - T_i(0) + R_i
\;.
\end{equation}
Thus
\begin{equation}
\label{eq:theoryT3}
\sum_{ij} C_{ji}T_{j \beta}T_{i\alpha} \bar a_\alpha =
\sum_{ij} C_{ji}T_{j \beta} \left(\langle D_i \rangle - T_i(0) + R_i \right)
\;.
\end{equation}
Using the definition (\ref{eq:Halphabeta}) of the Hessian matrix and the property (\ref{eq:Rkproperty}) of the $R_k$,  this is
\begin{equation}
\label{eq:theoryT33}
H_{\beta\alpha} \bar a_\alpha =
\sum_{ij} \left( \langle D_i \rangle - T_i(0)\right)
C_{ij}T_{j \beta}
\;.
\end{equation}
Thus
\begin{equation}
\label{eq:theoryT4}
\bar a_\alpha 
= H^{-1}_{\alpha\beta}
\sum_{ij} (\langle D_i \rangle - T_i(0)) C_{ij}T_{j \beta}
\;.
\end{equation}
Comparing to Eq.~(\ref{eq:bestfit}) gives us
\begin{equation}
\label{eq:theoryT44}
a^\textrm{fit}_\alpha - \bar a_\alpha
= H^{-1}_{\alpha\beta}
\sum_{ij} 
\left(
D_i - \langle D_i \rangle
\right)
C_{ij}T_{j \beta}
\;.
\end{equation}
The result is that the expectation value of $a_\textrm{fit}$ is $\bar a$:
\begin{equation}
\label{eq:expectationa}
\langle a^\textrm{fit}_\alpha - \bar a_\alpha \rangle
= 0
\;.
\end{equation}

This also gives the correlations of the parameters $a_\alpha$ with the data $D_i$ and with each other. From Eqs.~(\ref{eq:theoryT44}) and (\ref{eq:Dkexpectations}), we have
\begin{equation}
\big\langle 
\big( D_k - \langle D_k\rangle \big)\, 
\big(
a^\textrm{fit}_\alpha 
- \bar a_\alpha
\big)
\big\rangle
=
\sum_{ij} 
H^{-1}_{\alpha\beta}
C^{-1}_{ki}
C_{ij}
T_{j \beta}
\;.
\end{equation}
Thus
\begin{equation}
\label{eq:daexpectation}
\big\langle 
\big( D_k - \langle D_k\rangle \big)\, 
\big(
a^\textrm{fit}_\alpha 
- \bar a_\alpha
\big)
\big\rangle
= H^{-1}_{\alpha\beta}T_{k \beta}
\;.
\end{equation}
For $\langle (a^\textrm{fit}_\alpha - \bar a_\alpha) ( a^\textrm{fit}_\beta - \bar a_\beta)\rangle$, Eq.~(\ref{eq:theoryT44}) gives
\begin{equation}
\begin{split}
\big\langle 
\big(
a^\textrm{fit}_\alpha 
- \bar a_\alpha
\big) 
&
\big(
a^\textrm{fit}_\beta 
- \bar a_\beta
\big)
\big\rangle
\\
={}& 
H^{-1}_{\alpha \sigma}H^{-1}_{\beta \tau} 
\sum_{ijkl} C_{ik}T_{k\sigma}
C_{jl}T_{l\tau}
\\&\qquad\times
\langle (D_i - \langle D_i \rangle) (D_j - \langle D_j \rangle)\rangle
\\
={}& H^{-1}_{\alpha \sigma}H^{-1}_{\beta \tau} 
\sum_{ijkl} C_{ik}T_{k\sigma}
C_{jl}T_{l\tau}
C^{-1}_{ij}
\\
={}& H^{-1}_{\alpha \sigma}H^{-1}_{\beta \tau} 
\sum_{kl} C_{lk}T_{k\sigma}T_{l\tau}
\\
={}& H^{-1}_{\alpha \sigma}H^{-1}_{\beta \tau} 
H_{\sigma \tau}
\;.
\end{split}
\end{equation}
Thus
\begin{equation}
\label{eq:expectationaa}
\big\langle 
\big(
a^\textrm{fit}_\alpha 
- \bar a_\alpha
\big) 
\big(
a^\textrm{fit}_\beta 
- \bar a_\beta
\big)
\big\rangle
 = H^{-1}_{\alpha\beta}
\;.
\end{equation}
Compare this to Eq.~(\ref{eq:Dkexpectations}), $\langle (D_i - \langle D_i \rangle)(D_j - \langle D_j \rangle) \rangle = C^{-1}_{ij}$. Thus we see the importance of the Hessian matrix: its inverse is the correlation matrix for the fitted parton parameters.

\subsection{Distribution of the parameters}
\label{sec:parameterdist}

Since the data are Gaussian distributed, and the fit parameters are linearly related to the data, the fit parameters will be Gaussian distributed. The expectation values (\ref{eq:expectationaa}) give us the distribution of the best fit parameters, analogously to what we found in Eq.~(\ref{eq:expectation2}): given a function $f(a_\textrm{fit} - \bar a)$, we have
\begin{align}
\label{eq:adistribution}
\langle f \rangle
={}& \sqrt{\det{H}}\,(2\pi)^{- N_P/2}
\int d^{N_P}(a_\textrm{fit} - \bar a)\
f(a_\textrm{fit} - \bar a)
\notag
\\&
\times
\exp\!\left(- \frac{1}{2}H_{\alpha\beta}\, 
(a^\textrm{fit}_\alpha - \bar a_\alpha) 
(a^\textrm{fit}_\beta - \bar a_\beta)\right)
.
\end{align}
It is good to be clear about where this comes from. We consider an ensemble of repetitions of the experiments. As the data fluctuate, the values of $a_\textrm{fit}$ fluctuate according to the distribution (\ref{eq:adistribution}). However, we can turn this around. Given the data $D_k$, we find the corresponding best-fit parameters $a_\textrm{fit}$. We do not know $\bar a$. But if we repeat the experiments many times, the values of the difference $(a_\textrm{fit} - \bar a)$ will fluctuate around zero according to Eq.~(\ref{eq:adistribution}). Then Eq.~(\ref{eq:adistribution}) gives us a measure of the error in estimating $\bar a$ by $a_\textrm{fit}$.

Eq.~(\ref{eq:adistribution}) applies in the $N_P$-dimensional space of fit parameters. It is instructive to consider how this works in a particular coordinate system.  We let $\{e^{(1)}, e^{(2)},\dots, e^{(N_P)}\}$ be a set of basis vectors for the parameter space. We take these basis vectors to be orthogonal and normalized using $H$ as the metric tensor\footnote{ We do not distinguish between upper and lower indices $\alpha, \beta, \dots$. If we did, parton parameters would have upper indices, $a^\alpha$, and the metric tensor $H$ would have lower indices, $H_{\alpha \beta}$. Then $e^{(n)}$ and $H^{-1}$ would have upper indices,  $e_{(n)}^\alpha$ and $(H^{-1})^{\alpha \beta}$.}: 
\begin{equation}
\label{eq:enproducts}
e^{(n)}_\alpha H_{\alpha\beta}\, e^{(m)}_\beta = \delta_{mn}
\;.
\end{equation}
The corresponding completeness relation is
\begin{equation}
\label{eq:encompleteness}
\sum_n e^{(n)}_\alpha  e^{(n)}_\beta = H^{-1}_{\alpha\beta}
\;.
\end{equation}
(To prove this, we define $\sum_n e^{(n)}_\alpha  e^{(n)}_\beta = A_{\alpha\beta}$ and use Eq.~(\ref{eq:enproducts}) to show that $A_{\alpha\beta}H_{\beta\gamma}e^{(n)}_\gamma = e^{(n)}_\alpha$.) 

It is often useful to choose the basis vectors to be the eigenvectors of $H$:  $H_{\alpha \beta} e^{(n)}_\beta = h_n\, e^{(n)}_\alpha$. However, any choice of basis vectors obeying Eq.~(\ref{eq:enproducts}) will do. For instance, $e^{(1)}$ could be a vector normalized to $e^{(1)}_\alpha H_{\alpha\beta}\, e^{(1)}_\beta  = 1$, pointing in a direction that is of particular interest. Then the other $e^{(n)}$ could be chosen to satisfy Eq.~(\ref{eq:enproducts}). We will use this construction in Sec.~\ref{sec:crosssection}.

Using the basis vectors $e^{(n)}$, we can expand a general vector of parameters $a$ about $a_\textrm{fit}$ in the form
\begin{equation}
\label{eq:at}
a_\alpha(t)
=
a^\textrm{fit}_\alpha
+
\sum_n t_n\,e^{(n)}_\alpha
\;.
\end{equation}
Here the argument $t$ in $a(t)$ denotes $\{t_1,\dots,t_{N_P}\}$. How does $\chi^2$ depend on the parameters $t_n$? To find out, we evaluate $\chi^2(D,a(t))$. Using Eqs.~(\ref{eq:Halphabeta}), (\ref{eq:Rkproperty}), (\ref{eq:chisqDa2}), (\ref{eq:Tabar}), and (\ref{eq:theoryT44}), we obtain for a general choice of $a$,
\begin{align}
\label{eq:chisq3}
\chi^2(D,a) ={}&
\sum_{ij}
\left[
D_i - \langle D_i \rangle  - R_i
\right]
\left[
D_j - \langle D_j \rangle  - R_j
\right]
C_{ij}
\notag
\\&
- 2 H_{\alpha\beta}(a^\textrm{fit}_\alpha - \bar a_\alpha)
(a_\beta - \bar a_\beta)
\\&
+ H_{\alpha\beta}
(a_\alpha - \bar a_\alpha)
(a_\beta - \bar a_\beta)
\;. \notag
\end{align}
Here $a_\textrm{fit}$ is the parameter choice that we get from fitting the data $D_k$, $\bar a$ is what we would get by averaging $a_\textrm{fit}$ over an imagined ensemble of experiments, and $a$ represents the parameters that we are free to vary. Then if we substitute $a(t)$ given in Eq.~(\ref{eq:at}) for $a$, we get
\begin{align}
\label{eq:chisq5A}
\chi^2(D,a(t)) ={}& 
\sum_{ij}
\left[
D_i - \langle D_i \rangle  - R_i
\right]
\left[
D_j - \langle D_j \rangle  - R_j
\right]
C_{ij}
\notag
\\&
-   H_{\alpha\beta}
(a^\textrm{fit}_\alpha - \bar a_\alpha)  
(a^\textrm{fit}_\beta - \bar a_\beta)
\\&
+ \sum_n t_n^2
\notag
\;.
\end{align}
That is, varying $a$ from $a_\textrm{fit}$ in any direction $e^{(n)}_\alpha$ by $t_n = 1$ increases $\chi^2$ by 1.

The distribution (\ref{eq:adistribution}) of differences of $a_\textrm{fit}$ from $\bar a$ can be rewritten in the $e^{(n)}$ basis. We define
\begin{equation}
\label{eq:afitabar}
a^\textrm{fit}_\alpha - \bar a_\alpha
=
-
\sum_n t_n\,e^{(n)}_\alpha
\;.
\end{equation}
Then in Eq.~(\ref{eq:adistribution}), we can regard $f$ as a function of the eigenvector coordinates $t_n$ instead of the original parameters $a^\textrm{fit}_\alpha - \bar a_\alpha$, and we can change integration variables to the $t_n$, giving us
\begin{equation}
\begin{split}
\label{eq:vectdistribution}
\langle f \rangle
={}& 
\int_{-\infty}^\infty\!\frac{dt_1}{\sqrt{2\pi}}
\cdots \int_{-\infty}^\infty\!\frac{dt_{N_P}}{\sqrt{2\pi}}\
f(t_1,\dots,t_{N_P})
\\&\times
\exp\!\left(- \frac{1}{2}\sum_{n=1}^{N_P} t_n^2\right)
\;.
\end{split}
\end{equation}
Each coordinate $t_n$ follows an $\cN(0,1)$ distribution.

In particular, if we are interested only in the component of $\bar a - a_\textrm{fit}$ in the direction $e^{(1)}$, we can let the function $f$ in Eq.~(\ref{eq:vectdistribution}) depend only on $t_1$. Then
\begin{equation}
\label{eq:vectdistribution1}
\begin{split}
\langle f \rangle
={}& 
\int_{-\infty}^\infty\!\frac{dt_1}{\sqrt{2\pi}}\
f(t_1)\
\exp\!\left(- \frac{1}{2}\, t_1^2\right)
\;.
\end{split}
\end{equation}
Comparing to Eq.~(\ref{eq:chisq5A}), we see that a ``$2 \sigma$'' value $t_1$, that is $t_1 = 2$, increases $\chi^2$ by 4 from its best fit value.

It is of interest to understand the distribution $\rho(R^2,N_P)$ of $R^2 = \sum_n t_n^2$,
\begin{align}
\rho(R^2,N_P) ={}&
\int_{-\infty}^\infty\!\frac{dt_1}{\sqrt{2\pi}}
\cdots \int_{-\infty}^\infty\!\frac{dt_{N_P}}{\sqrt{2\pi}}\
\delta\left(\sum_{n = 1}^{N_P} t_n^2 - R^2\right)
\notag
\\&\times
\exp\!\left(- \frac{1}{2}\sum_{n = 1}^{N_P} t_n^2\right)
\;.
\end{align}
This is the $\chi^2$ distribution with $N_P$ degrees of freedom. In
fitting parton distributions, $N_P$ is quite large, say $N_P =
25$. The mean value of any of the $t_n^2$ is $t_n^2 = 1$, but the mean
value of $R^2$ is much larger, as we will see in
Sec.~\ref{sec:chisqstatistics}: $\langle R^2 \rangle = N_P$. 
Using the $t_n$ as coordinates, the hypersurface $\sum_{n = 1}^{N_P} t_n^2 = R^2$ is a sphere. In terms of the original parton parameters $\bar a_\alpha -
a^\textrm{fit}_\alpha$, it is an ellipsoid. With the values of $\bar a_\alpha -
a^\textrm{fit}_\alpha$ distributed according to
Eq.~(\ref{eq:adistribution}), in order for the ellipsoid to include
95\% of the points we should choose $R \approx 6.1$ for $N_P = 25$. In
contrast, if we look at just one $t_n$, then in order for the interval
$t_n^2 < R_n^2$ to include 95\% of the points $t_n$, we should choose
$R_n \approx 2$. But with $N_P = 25$, the fraction $P$ of points
inside an ellipsoid with $R = 2$ is $P \approx 5 \times 10^{-7}$. This
discussion illustrates that, when we discuss the uncertainties in the
determination of the parton parameters, we need to carefully
distinguish whether we are discussing an uncertainty interval in one
dimension or in 25 dimensions. 
As another consequence of the large-$N_P$ geometry, when the PDF
probability distribution is sampled by randomly varying the PDF
parameters, the overwhelming majority of such Monte-Carlo
parameter replicas are likely to be bad fits
with $P \sim 0$  \cite{Hou:2016sho}. Thus, 
though the estimates of
the first and second moments of the $N_P$-dimensional probability
distribution from the Monte-Carlo sample converge to their true values
with about 100-1000 replicas,
the Monte-Carlo method tends to be highly inefficient for exploring the
neighborhood $\sum_n t_n^2 \leq R^2$ with $R$ of order unity. In
contrast, the analytic minimization of $\chi^2$ by the gradient descent,
as implemented in CTEQ and MMHT fits, directly finds
the neighborhood $\sum_n t_n^2 \leq R^2$ around the global minimum and
renders the probability distribution within this neighborhood. The
analytic minimization and Monte-Carlo sampling approaches thus offer
complementary strengths when examining the multidimensional probability
distribution of PDF parameters and associated PDF uncertainties. 

\subsection{Calculation of a cross section}
\label{sec:crosssection}

We can now ask another question. Suppose that $\sigma$ is a cross section that is determined by the PDFs. Then $\sigma$ is a function $\sigma(a)$ of the parton parameters $a$. We would need data with no errors to determine the ideal parameters $\bar a$, so we never know $\sigma(\bar a)$ exactly. However, we can fit the parameters and estimate $\sigma(\bar a)$ by $\sigma(a_\textrm{fit})$. 

What is the expected error resulting from using the fit parameters? To analyze this, we begin by defining the shift in the cross section,
\begin{equation}
\label{eq:Deltasigmadef}
\Delta\sigma
= \sigma(\bar a) - \sigma(a_\textrm{fit})
\;.
\end{equation}
Then with a linear approximation we write
\begin{equation}
\label{eq:Deltasigma1lin}
\Delta\sigma = (\bar a_\alpha - a^\textrm{fit}_\alpha)\,\sigma^\prime_\alpha
\;,
\end{equation}
with $\sigma^\prime_\alpha \equiv \partial \sigma(a)/\partial a_\alpha$.

Now we define a special vector in the space of parameters according to
\begin{equation}
\label{eq:e1forsigma}
e(\sigma)_\alpha = 
\frac{H^{-1}_{\alpha \beta}\,\sigma^\prime_\beta}
{\sqrt{\sigma^\prime_\gamma H^{-1}_{\gamma \delta}\sigma^\prime_\delta}}
\;.
\end{equation}
This vector is normalized to
\begin{equation}
\label{eq:enormalization}
e(\sigma)_\alpha H_{\alpha\beta} e(\sigma)_\beta = 1
\;.
\end{equation}

It is useful to define more vectors, $\{e(\sigma)^{(2)}, \dots, e(\sigma)^{(N_P)}\}$ such that these vectors, together with $e(\sigma) \equiv e(\sigma)^{(1)}$, form a basis for the parameter space and such that the basis vectors are orthogonal and normalized using the metric tensor $H_{\alpha\beta}$ as in Eq.~(\ref{eq:enproducts}).

With the aid of these basis vectors, we can write a general parameter vector as
\begin{equation}
\label{eq:afromtn}
a_\alpha = a^\textrm{fit}_\alpha + t_1\, e(\sigma)_\alpha 
+ \sum_{n\ge 2} t_n e(\sigma)^{(n)}_\alpha
\;.
\end{equation}
The corresponding change in $\chi^2$ is, using Eq.~(\ref{eq:chisq5A}),
\begin{equation}
\begin{split}
\label{eq:chisqshifteda}
\chi^2 \Big(
D,a_\textrm{fit} +  t_1 e(\sigma)
&+\sum_{n \ge 2} t_n e(\sigma)^{(n)} \Big) 
\\
={}& \chi^2(D,a_\textrm{fit})
+ t_1^2
+ \sum_{n\ge 2} t_n^2
\;.
\end{split}
\end{equation}
Let us set $a \to \bar a$ in the definition Eq.~(\ref{eq:afromtn}), so that
\begin{equation}
\label{eq:afromtnbis}
\bar a_\alpha - a^\textrm{fit}_\alpha 
= t_1\, e(\sigma)_\alpha + \sum_{n\ge 2} t_n e(\sigma)^{(n)}_\alpha
\;.
\end{equation}
Then, according to Eq.~(\ref{eq:vectdistribution}), as $a_\textrm{fit}$ varies in an ensemble experiment sets, the expansion parameters $\{t_1,\dots,t_{N_P}\}$ fluctuate following $\cN(0,1)$ distributions.

We can use this result to analyze the fluctuations in the cross section from Eq.~(\ref{eq:Deltasigma1lin}):
\begin{equation}
\label{eq:DeltasigmaA}
\Delta\sigma = t_1 e(\sigma)_\alpha\,\sigma^\prime_\alpha
+ \sum_{n \ge 2} t_n e(\sigma)^{(n)}_\alpha\,\sigma^\prime_\alpha
\;,
\end{equation}
Using Eq.~(\ref{eq:e1forsigma}), this becomes
\begin{equation}
\begin{split}
\label{eq:DeltasigmaB}
\Delta\sigma = {}&
\sqrt{\sigma^\prime_\gamma H^{-1}_{\gamma \delta}\sigma^\prime_\delta}
\Big\{
t_1\ 
e(\sigma)_\alpha\,H_{\alpha\beta} e(\sigma)_\beta
\\&\qquad
+ 
\sum_{n\ge 2} t_n\ e(\sigma)^{(n)}_\alpha\,H_{\alpha\beta} e(\sigma)_\beta
\Big\}
\;.
\end{split}
\end{equation}
Because of the orthonormality condition (\ref{eq:enproducts}), only the first term survives and we obtain
\begin{equation}
\label{eq:DeltasigmaBB}
\Delta\sigma = 
\sqrt{\sigma^\prime_\gamma H^{-1}_{\gamma \delta}\sigma^\prime_\delta}\
t_1
\;.
\end{equation}
Thus the fluctuations in the cross section are given entirely by the fluctuations of the parameters along the special direction $e(\sigma)^{(1)}$. There is a coefficient, $[{\sigma^\prime_\gamma H^{-1}_{\gamma \delta}\sigma^\prime_\delta}]^{1/2}$ that is larger when the cross section is a fast varying function of the parton parameters. The remaining factor, $t_1$ fluctuates following an $\cN(0,1)$ distribution as the data fluctuate. That means that if we want $\Delta \sigma$ to represent, say, a two standard deviation error on $\sigma$, we set $t_1 = 2$. Furthermore, $t_1$ has the property that when the parameters vary from the best fit parameters according to $a = a_\textrm{fit} + t_1\, e(\sigma)^{(1)}$, the $\chi^2$ increases by $t_1^2$.

{\bf Formulas to compute PDF uncertainties.} There is a standard practical method for calculating $\Delta \sigma$. We choose basis vectors that are not specially adapted to the cross section $\sigma(a)$. We choose the basis vectors $e^{(n)}$, $n = 1,\dots,N_P$ to obey the orthonormality condition Eq.~(\ref{eq:enproducts}). Typically, the basis vectors $e^{(n)}$ are chosen to be eigenvectors of the Hessian matrix $H$.  Commonly, there are $2N_P$ error fits,  $a_\textrm{fit} \pm \tilde t e^{(n)}$, that come with a set of published PDFs. Here $\tilde t$ is defined by the published analysis. If we assume that linear approximations are adequate, then we need only $N_P$ error fits, $a_\textrm{fit} + \tilde t e^{(n)}$, with positive $\tilde t$. Error fits with two signs \cite{Nadolsky:2001yg} allow for a more complete treatment, beyond what we give here, that allows for nonlinear contributions. 

One can use the basis vectors $e^{(n)}$ to evaluate the uncertainty in $\sigma(a_\textrm{fit})$. For each direction $n$, define 
\begin{equation}
\label{eq:Deltasigman}
\Delta\sigma^{(n)} = \sigma(a_\textrm{fit} + t e^{(n)}) 
- \sigma(a_\textrm{fit})
\;.
\end{equation}
Recall from Eq.~(\ref{eq:chisq5A}) that if we set $a = a_\textrm{fit} + t e^{(n)}$, then $\chi^2(D,a)$ increases by $t^2$ compared to $\chi^2(D,a_\textrm{fit})$.   As long as we use a linear approximation, we have
\begin{equation}
\Delta\sigma^{(n)} = t\,\sigma^\prime_\alpha\, e^{(n)}_\alpha
\;.
\end{equation}
Now sum the squares of the $\Delta\sigma^{(n)}$:
\begin{equation}
\sum_n (\Delta\sigma^{(n)})^2
= t^2 \sum_n \sigma^\prime_\alpha\, e^{(n)}_\alpha e^{(n)}_\beta\, \sigma^\prime_\beta 
\;.
\end{equation}
Using the completeness relation (\ref{eq:encompleteness}) for the basis vectors $e^{(n)}$, this is
\begin{equation}
\sum_n (\Delta\sigma^{(n)})^2
= t^2 \sigma^\prime_\alpha\, H^{-1}_{\alpha\beta}\, \sigma^\prime_\beta
\;.
\end{equation}
According to Eq.~(\ref{eq:DeltasigmaBB}), we can estimate the error in $\sigma$ by
\begin{equation}
(\Delta \sigma)^2 = t^2 \sigma^\prime_\alpha\, H^{-1}_{\alpha\beta}\, \sigma^\prime_\beta
\;,
\end{equation}
where, for example, we would choose $t = 2$ if we want a ``$2\sigma$'' error estimate. Thus we can obtain $\Delta \sigma$ by using variations in the eigenvector directions $e^{(n)}$:
\begin{equation}
(\Delta \sigma)^2 = \sum_n (\Delta\sigma^{(n)})^2
\;.
\end{equation}
That is, we need to calculate $N_P$ error contributions $\Delta\sigma^{(n)}$ by using the parton error sets according to Eq.~(\ref{eq:Deltasigman}). Then adding the errors $\Delta\sigma^{(n)}$ in quadrature gives the total error $\Delta \sigma$.

There is a second standard practical method for calculating $\Delta \sigma$. This method derives from the publications of the NNPDF group \cite{Ball:2010de, Ball:2012cx, Ball:2014uwa, Ball:2017nwa}. With the NNPDF approach, there is effectively a very large number of parameters, and the distribution of the parameters is not strictly Gaussian. The distribution of results is represented by giving a large sample of parton distribution sets. Within the linear approximations that we use in this review, one would generate a large sample of parton distributions based on parameters $a_\alpha = a^\textrm{fit}_\alpha + \sum_n t_n\,e^{(n)}_\alpha$ as in Eq.~(\ref{eq:at}), with Gaussian random variables $t_n$. Given this sample, one calculates $\sigma(a)$ for each example and, thus, obtains a corresponding sample of $\sigma(a)$ values, from which one obtains the statistical properties of the sample such as $\langle \sigma \rangle$ and $(\Delta \sigma)^2$.

We have spoken of $\sigma(a)$ as being a cross section. More broadly, $\sigma(a)$ in this section could be any physical quantity that depends on the parton parameters $a$. In particular, $\sigma(a)$ could be a parton distribution function $f_{a/A}(x,\mu^2)$ for a particular parton flavor $a$, evaluated at a particular momentum fraction $x$ and a particular scale $\mu$. Then the calculation presented above gives us an error estimate $\Delta f_{a/A}(x,\mu^2)$.

\subsection{Expectation value and variance of $\chi^2$}
\label{sec:chisqstatistics}

In this section, we investigate the value of $\chi^2$ obtained in the fit. Start with Eq.~(\ref{eq:chisq3}) for $\chi^2(D,a)$. We fit the parameters $a$ to minimize $\chi^2(D,a)$ for given data $D$. Using the fit parameters gives
\begin{align}
\label{eq:chisq5}
\chi^2(D,a^\textrm{fit}) ={}& 
\sum_{ij}
\left[
D_i - \langle D_i \rangle  - R_i
\right]
\left[
D_j - \langle D_j \rangle  - R_j
\right]
C_{ij}
\notag
\\&
-   H_{\alpha\beta}
(a^\textrm{fit}_\alpha - \bar a_\alpha)  
(a^\textrm{fit}_\beta - \bar a_\beta)
\;.
\end{align}
Using Eq.~(\ref{eq:theoryT44}) for $a^\textrm{fit}_\alpha - \bar a_\alpha$, this is
\begin{equation}
\label{eq:chisq6}
\begin{split}
\chi^2 ={}&
\sum_{ij}
\left[
D_i - \langle D_i \rangle  - R_i
\right]
\left[
D_j - \langle D_j \rangle  - R_j
\right]
C_{ij}
\\&
- \sum_{ij}
\left[
D_i - \langle D_i \rangle
\right]
\left[
D_j - \langle D_j \rangle
\right]
M_{ij}
\;,
\end{split}
\end{equation}
where the matrix $M$ is 
\begin{equation}
\label{eq:Mdef}
M = C T H^{-1} T^\LT C
\;.
\end{equation}
Note that here we have eliminated the parameters $a_\alpha$ entirely.

We can use this to evaluate the expectation value $\langle \chi^2 \rangle$ of $\chi^2$ and its variance
\begin{equation}
\langle (\chi^2 - \langle \chi^2\rangle)^2 \rangle
= 
\langle (\chi^2)^2 \rangle -\langle \chi^2\rangle^2
\;. 
\end{equation}
We use Eq.~(\ref{eq:expectation2}) for the probability distribution of the data $D$. This gives
\begin{equation}
\begin{split}
\label{eq:DiDj}
\Big\langle\left[
D_i - \langle D_i \rangle
\right]
\left[
D_j - \langle D_j \rangle
\right]
\Big\rangle ={}& C^{-1}_{ij}
\end{split}
\end{equation}
and
\begin{equation}
\begin{split}
\label{eq:DiDjDkDl}
\Big\langle \!
\left[
D_i - \langle D_i \rangle
\right]&
\left[
D_j - \langle D_j \rangle
\right]
\left[
D_k - \langle D_k \rangle
\right]
\left[
D_l - \langle D_l \rangle
\right]\!
\Big\rangle 
\\={}& 
C^{-1}_{ij} C^{-1}_{kl}
+ C^{-1}_{ik}C^{-1}_{jl}
+ C^{-1}_{il}C^{-1}_{jk}
\;.
\end{split}
\end{equation}
To derive this, one can change variables in Eq.~(\ref{eq:expectation2}) to $x_i = \sum_j (\sqrt C)_{ij} (D_j - \langle D_j \rangle )$. Then the symmetries of the integrand imply that $\langle x_i x_j\rangle \propto \delta_{ij}$ and $\langle x_i x_j x_k x_l\rangle \propto \delta_{ij}\delta_{kl} + \delta_{ik}\delta_{jl} + \delta_{il}\delta_{jk}$. The coefficients of proportionality are simple to evaluate, giving Eqs.~(\ref{eq:DiDj}) and (\ref{eq:DiDjDkDl}).

Now a certain amount of algebra with the matrices leads to results containing factors
\begin{equation}
\sum_{i = 1}^{N_D} \delta_{ii} ={} N_D,\qquad \sum_{\alpha = 1}^{N_P} \delta_{\alpha \alpha} = N_P.
\end{equation}
The results are 
\begin{equation}
\label{eq:chisqexpectation}
\langle \chi^2 \rangle = N_D - N_P 
+ \sum_{ij}
R_i C_{ij} R_j
\end{equation}
and
\begin{equation}
\label{eq:chisqvariance}
\langle (\chi^2 - \langle \chi^2\rangle)^2 \rangle = 
2 (N_D - N_P)
+ 4 \sum_{ij} R_i C_{ij} R_j
\;.
\end{equation}

These results are often used to provide an indication of whether a good fit has been found. The parameters $R_i$ that we have introduced represent an imperfection in the theory: if the parton distributions do not have enough available parameters, we expect $R_i \ne 0$. If there are enough parameters and if the rest of the theoretical model is correct, then the $R_i$ should vanish. In that case, $\langle \chi^2\rangle$ should be close to $N_D - N_P$. For example, if $N_D = 3000$ and $N_P = 25$, then $\chi^2$ should be around 2975. The square root of the variance of $\chi^2$ in this case is $\sqrt{5950} \approx 77$. Thus we expect to find $\chi^2 \approx 2975 \pm 77$. The fit is bad if $\chi^2$ is too high or too low.

%-------------------------------------------------------------------

% !TEX root = main.tex
\section{Tests of performance of the fit}\label{sec:goodness}

In Sec.~\ref{sec:theorypredictions}, we examined the function $\chi^2(D,a)$ defined in Eq.~(\ref{eq:chisqDa}) that is minimized to determine PDF parameters $a$ from experimental data $D$, giving values $a_\textrm{fit}$. However, this fitting procedure produces correct results only if the data $D$ are reliable within their errors as given by the experiments, and if the adopted theory is actually a good description of nature for some parameter combination $a_\mathrm{fit}$. If $\chi^2(D,a_\textrm{fit})$ does not lie within certain limits, one can conclude that something is wrong with the fit. However, it has been realized since the inception of the global QCD analysis in late 1980's that the value of the global $\chi^2(D,a_\textrm{fit})$ is an essential, but far from sufficient, measure of the goodness of fit. [See, e.g., \cite{Morfin:1990ck}]. 

In this section, we argue that the PDF fit should pass a number of tests in order to fulfill what one might call a {\it strong set of goodness-of-fit criteria}. Several of these tests involve looking at quantities derived from the fitting procedure that should follow a predicted distribution if the statistical assumptions on which the fit is based are valid. One can then test whether the quantities are in fact distributed as predicted. We include the distribution of the nuisance parameters, the distribution of the residuals for the fitted data, and the distribution of $\chi^2$ values for many subsets of the data. Another test looks at whether individual subsets of the data are statistically consistent with the global $a_\mathrm{fit}$ in individual directions in the space of parameters. In Sec.~\ref{sec:ClosureTest}, we discuss the \textit{closure test}, which is a powerful test of the fitting methodology and is used, in particular, by NNPDF  \cite{Ball:2014uwa}.

These tests, taken together, are more constraining and difficult to satisfy than the standard {\it weak} goodness-of-fit criterion based on the value of the global $\chi^2(D,a)$. The point of the strong set of goodness-of-fit criteria is to find places where the statistical assumptions used in the fitting procedure break down. We will find, in examples, that some of the assumptions used in the fit do break down. When this happens, the PDF error estimates that emerge from the fit cannot be realistic. One possibility is that any observed problems are the result of understated estimates of the experimental systematic errors. We propose ways to deal with this in Sections~\ref{sec:consistency} and \ref{sec:conservativeErrorAdjustment}. The method usually used to adjust the final PDF error estimates is to apply a {\em tolerance} criterion. This is a complicated subject, which we address briefly in Sec.~\ref{sec:tolerance}.

One of the tests, which we investigate in Sec.~\ref{sec:experimentchisqall}, looks at $\chi^2$ for subsets of the data. Data subsets are often examined visually to rule out systematic discrepancies by comparing data and theory predictions in the figures. This is a reasonable, but slow and imprecise test. It can be  made more precise using the procedure in Secs.~\ref{sec:experimentchisq} and \ref{sec:experimentchisqall}. 

A standard goodness-of-fit criterion based on the value of the overall $\chi^2(D,a_\textrm{fit})$ is called the {\it hypothesis-testing criterion} \cite{Collins:2001es}. We have seen in Eq.~(\ref{eq:chisqexpectation}) that, if the theory is perfect so that the $R_j = 0$, the expectation value of this quantity is  $\langle \chi^2 (D,a_\textrm{fit}) \rangle = N_D - N_P$. However,  Eq.~(\ref{eq:chisqvariance}) shows that  $\chi^2(D,a_\textrm{fit})$ is expected to fluctuate by about  $\sqrt{2 (N_D - N_P)}$. Thus we surely have a bad fit at the  $2\sigma$ level if  $\chi^2(D,a_\textrm{fit}) - (N_D - N_P)$ is bigger than $2\times\sqrt{2 (N_D - N_P)}$, that is, about 150 for $N_D - N_P \approx 3000$.

However, if we fit the parameters $a$ without changing the  functional form, a small difference in $\chi^2$ values  of order $2^2 = 4$ is already significant, while $2\times\sqrt{2 (N_D - N_P)}\approx  150$ is far too large. In this restricted situation, the {\it parameter-fitting criterion}, which assigns the 95\% (or 68\%) probability level to the increase $\Delta \chi^2 = 4$ (or 1), adequately estimates the uncertainty on parameters of this fixed model, as long as the statistical assumptions on which the fit is based are all valid.  Multiple functional forms can give good fits to the data. To the parameter-fitting uncertainty, one must add the uncertainty due to the choice of the functional form of PDFs, as is discussed in Sec.~\ref{sec:functionalform}.
 
As outlined above, we will look at quantities derived from the fitting procedure that should follow a predicted distribution. Call the quantities $q_j$. When testing for a possible systematic deviation from the predicted distribution for the $q_j$, we find it useful to transform the observed quantity to a form $x_j = x(q_j)$ such that expected distribution of the $x_j$ is  $\cN(0,1)$,  the standard normal (Gaussian) distribution with the mean of zero and variance of one. For a quantitative estimate of the probability that the observed distribution of the $x_j$ was sampled from ${\cal N} (0,1)$, we can apply the standard Anderson-Darling test \cite{anderson1952}. The test yields a ``distance''$A_\textrm{obs}$ of the observed cumulative probability distribution of $x_j$ values from that for ${\cal  N}(0,1)$. Then it calculates the probability $P_\textrm{A-D}$ that the the same number of randomly drawn $x_j$ values from ${\cal N}(0,1)$ will have a distance $A$ with $A > A_\textrm{obs}$. With this test, a) $P_\textrm{A-D}$ always lies between 0 and 1, b) $P_\textrm{A-D}$ is close to 1 (or 0) if the histogram matches the ${\cal N} (0,1)$ distribution closely (or poorly), and c) if we repeat the sampling procedure many times with data actually drawn from the ${\cal N} (0,1)$ distribution, the values of $P_\textrm{A-D}$ will be uniformly distributed between 0 and 1.     

We begin with a preliminary question: do we have enough fitting parameters to obtain a good fit to the data? 

\subsection{Testing with resampled data}
\label{sec:differentdata}

The PDFs $f_{a/p}(\xi,\mu^2)$
must use a sufficiently flexible functional form to reproduce
only regular, but no random, features of the hadronic data. However,
the functional form for PDFs is known only semi-qualitatively based on
considerations like the positivity of cross sections, asymptotic
limits at small and large $x$, and nucleon sum rules. One resorts to
a phenomenological form $f_{a/p}(\xi,\mu^2_0)$ for the PDFs at the
initial scale $\mu_0^2$ and must decide how many parameters $a_\alpha$
to use. If the number $N_P$ of parameters is too small, the theory may not be
perfect. If too many, no global minimum of $\chi^2$ may exist, or we may overfit the
data. In this section, we will explore the dependence on the number of free parameters for a given parameterization form. One can also vary the PDF functional forms, as discussed in Sec.~\ref{sec:functionalform}.

\begin{figure}
\centerline{\includegraphics[width=0.48\textwidth]{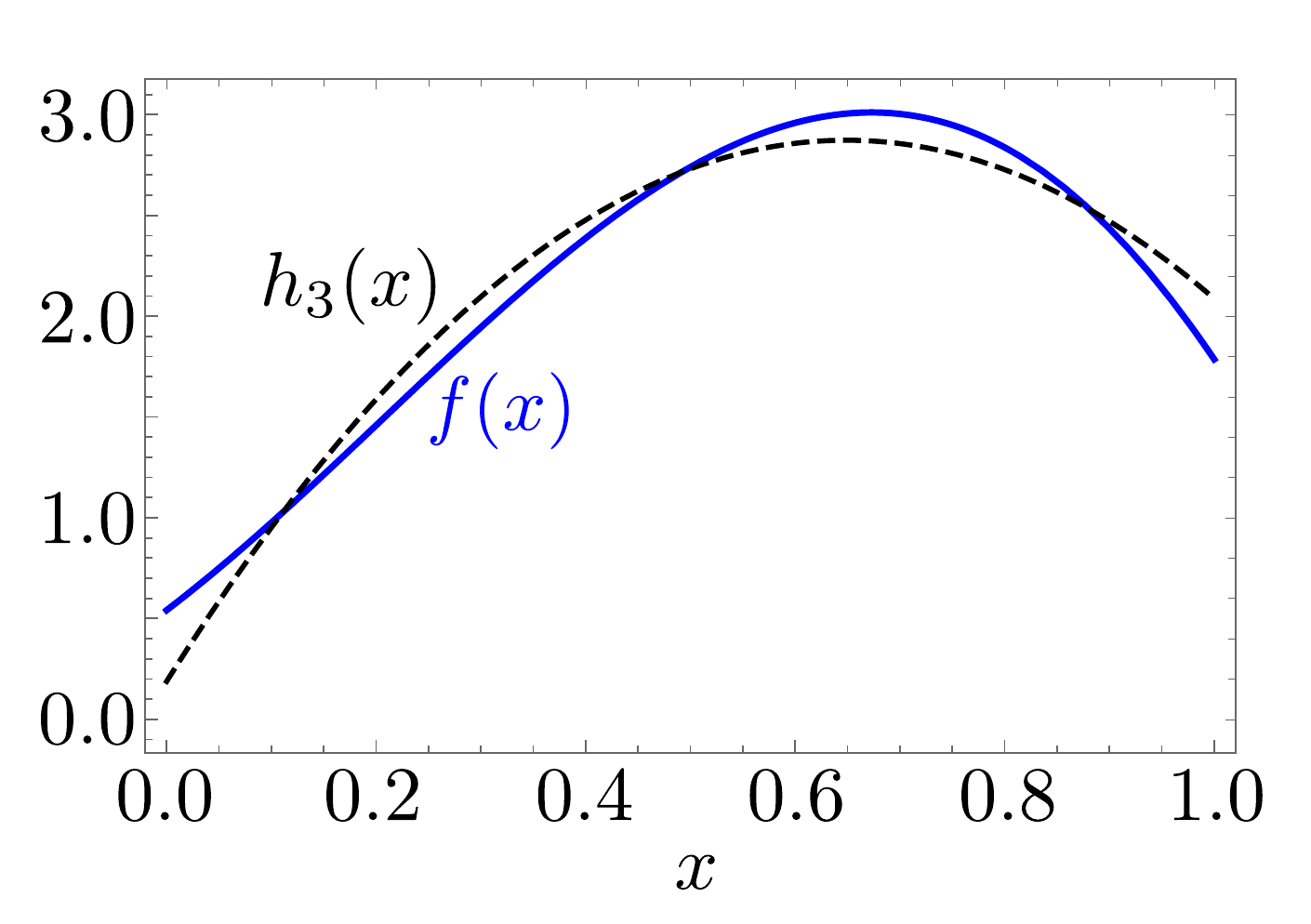}}
\caption{The function $f(x)$ in Eq.~(\ref{eq:fexample}) and a three parameter polynomial fit $h_3(x)$, Eq.~(\ref{eq:polynomialhn}), to this function. Here there are not enough parameters to get a good fit.}
\label{fig:fits1}
\end{figure}

\begin{figure*}[tb]
\center{\includegraphics[width=0.48\textwidth]{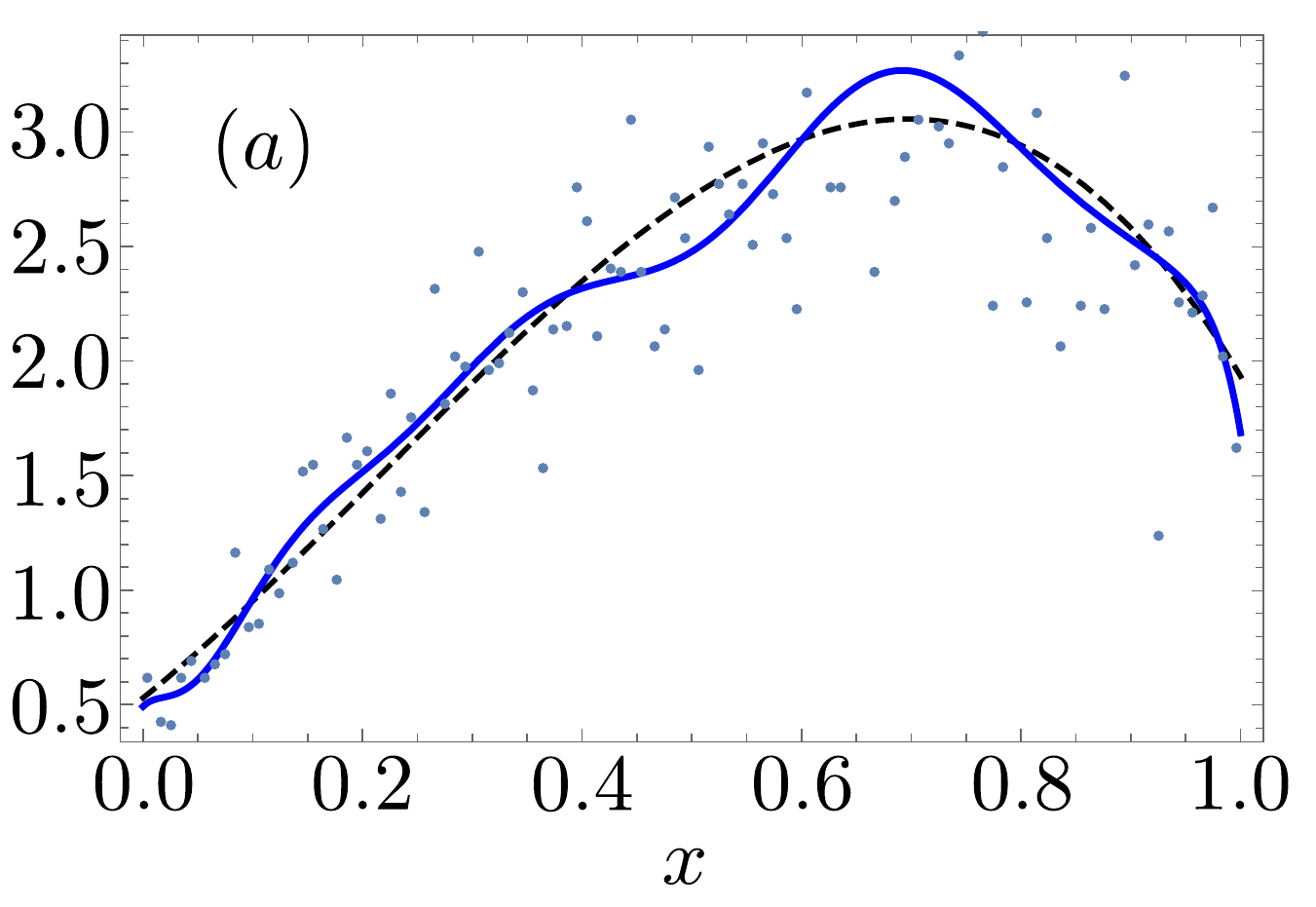}\quad \includegraphics[width=0.48\textwidth]{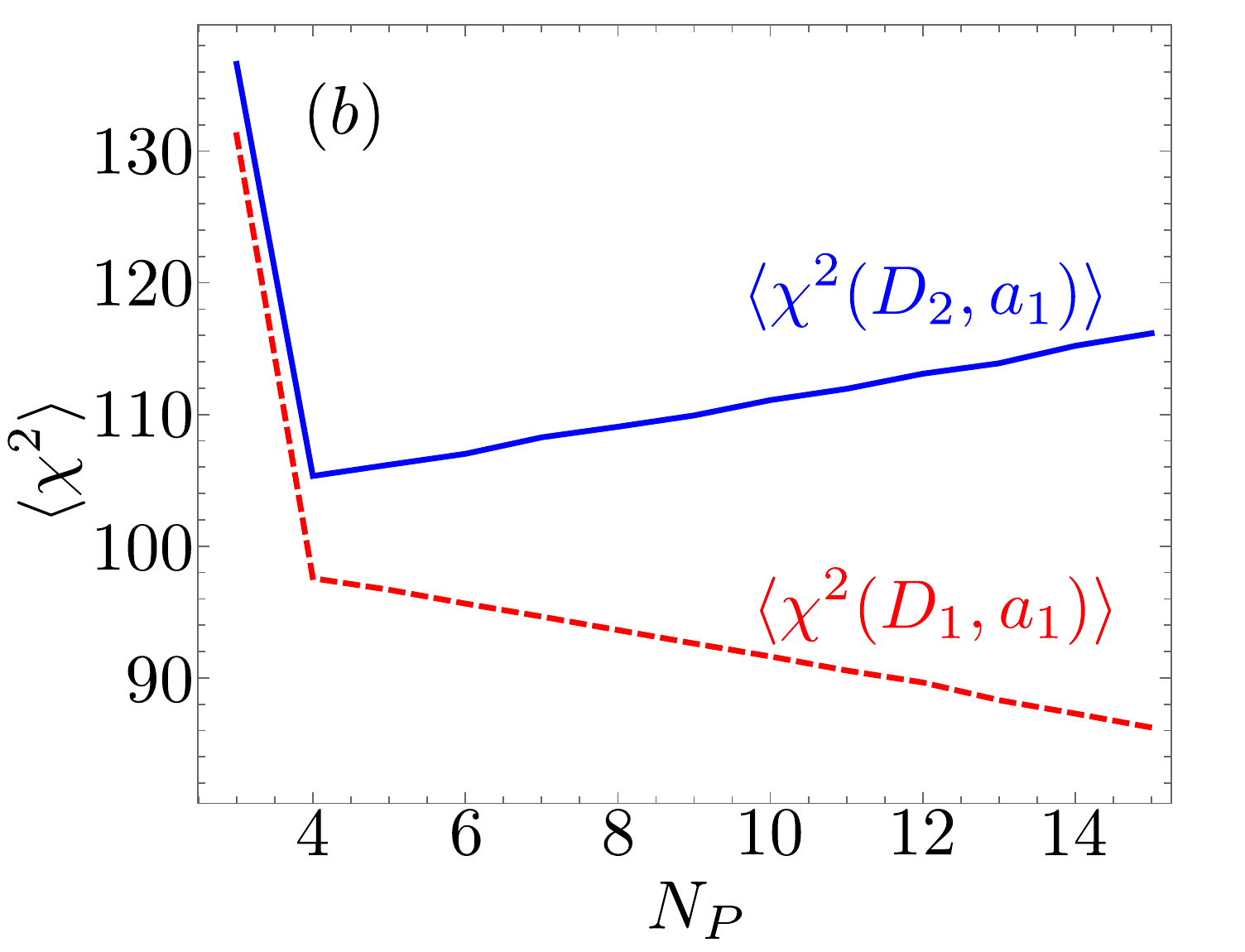}}
\caption{
(a) Data $\{y_i,x_i\}$ generated from $f(x)$, Eq.~(\ref{eq:fexample}), shown with a 4-parameter (dashed curve) and 13-parameter (solid curve) polynomial fits to the data. 
(b) Average over a large number of trials of  $\chi^2(D_1,a_1)$ and $\chi^2(D_2,a_1)$ as a function of the number of parameters $N_P$. In each trial, a polynomial with $N_P$ parameters is fit to data $D_1$ generated from $f(x)$, then $\chi^2(D_2,a_1)$ is measured for that polynomial compared to an independent data sample $D_2$ generated from $f(x)$.}
\label{fig:fits2}
\end{figure*}

To estimate the optimal number of parameters, let us return to $\chi^2$, using Eq.~(\ref{eq:theoryT44}) in Eq.~(\ref{eq:chisq3}):
\begin{align}
\label{eq:chisqAbis}
\chi^2(D,a) ={}&
\sum_{ij}
\left[
D_i - \langle D_i \rangle  - R_i
\right]
\left[
D_j - \langle D_j \rangle  - R_j
\right]
C_{ij}
\notag
\\&
-2 \sum_{ij}
\left[
D_i - \langle D_i \rangle
\right]
C_{ij} T_{j\beta}
(a_\beta-\bar a_\beta)
\\&
+ H_{\alpha\beta} (a_\alpha-\bar a_\alpha)
(a_\beta-\bar a_\beta)
\;.
\notag
\end{align}
Suppose that we obtain a set of parameters $(a_\alpha)_1$ by minimizing $\chi^2(D_1,a_1)$ for the {\em fitted data sample} $D_1$ with $N_D$ data points. Then we use the same parameters to calculate $\chi^2(D_2,a_1)$ for a {\em control data sample} $D_2$ that is obtained by repeating the experiment with different random fluctuations. [$N_D$, $R_i$, $C_{ij}$ are the same for $D_1$ and $D_2$.]  
What do we get? From Eqs.~(\ref{eq:Dkexpectations}), (\ref{eq:expectationa}),  and (\ref{eq:expectationaa}), the $\chi^2$ expectation for sample $D_2$, but using the parameters $a_1$ fitted to $D_1$, is given by
\begin{equation}
\begin{split}
\label{eq:chisq12}
\langle\chi^2(D_2,a_1)\rangle 
 ={}& \sum_{ij}
\sum_{ij}
C^{-1}_{ij}\,
C_{ij}
+ \sum_{ij} R_i R_J C_{ij}
\\&
+ H_{\alpha\beta} H^{-1}_{\alpha\beta}
\\={}& 
N_D + N_P +  \sum_{ij} R_i R_J C_{ij}\; .
\end{split}
\end{equation}
This is bigger by $+2N_P$ than 
\begin{equation}
\label{eq:chi2D2a2}
\langle \chi^2(D_1,a_1) \rangle = N_D - N_P 
+ \sum_{ij}
R_i R_j
C_{ij}
\end{equation}
obtained by using the data sample $D_1$ to evaluate the fit $a_1$ obtained from this data. If there are more than just a few parameters, this is a big change.

Although having ``too many'' parameters makes $\langle \chi^2(D_2,a_1) \rangle$ larger, we could imagine that, with ``too few'' parameters, we get a bad fit because the fit cannot get close to the true PDFs. In our treatment here, with ``too few'' parameters, $\sum_{ij} R_i R_j C_{ij}$ is large. 

We can illustrate this with a toy example, showing what may happen in parameterization studies with real parton distributions, such as the one in \cite{Martin:2012da}.
Suppose we fit a test function $h_{N_P}(x)$ with $N_P$ parameters to pseudodata generated by random fluctuations around the function
\begin{equation}
\label{eq:fexample}
f(x) = 3 (x + 0.2)^{1.2} (1.2 - x)^{1.2}(1 + 2.3 x)
\;.
\end{equation}
For the fit, we use a polynomial,
\begin{equation}
\label{eq:polynomialhn}
h_{N_P}(x) = \sum_{i=1}^{N_P} a_i x^{i-1}
\;.
\end{equation}
If we use $h_3(x)$, with just three parameters, we do not get a good fit, as we see in Fig.~\ref{fig:fits1}. Here for typical choices $x_j$ of $x$, the measures $R_j = f(x_j) - h_3(x_j)$ of how well the ``theory'' matches the exact function are typically of order 0.1. However, if we use 4 parameters to form $h_4(x)$, we can get a good fit, with $R_j \sim 0.01$. If we increase $N_P$ beyond 4, the measures $R_j$ are even smaller. Then increasing $N_P$ when fitting data, with its fluctuations, does not help produce a better fit to the true function $f(x)$. However, increasing $N_P$ does make $\chi^2$ smaller --- because we start better fitting the random fluctuations! 

To illustrate what happens, let us generate ``toy data'' $y_i = f(x_i)
+ 0.2 f(x_i)\,r_i$ at coordinate values $x_i = 0.01 i - 0.05$ for $i = 1,\cdots,N_D$, where $N_D = 100$. These are shown as scattered points in Fig.~\ref{fig:fits2}(a). The $r_i$'s are random numbers sampled from ${\cal N}(0,1)$. Then fitting the data using $h_4(x)$ produces the black dashed curve in Fig.~\ref{fig:fits2}(a). This is already a fairly good fit to $f(x)$, with $R_j \sim 0.1$; but we can allow ourselves even more parameters, for example by fitting $h_{13}(x)$ with 13 parameters. The 13-parameter fit is shown as the solid blue curve in Fig.~\ref{fig:fits2}(a). This produces a smaller value of $\chi^2$, but not a better fit to $f(x)$. 

It appears from Fig.~\ref{fig:fits2}(a) that what we are doing is fitting the fluctuations in the data. To test this, we can generate a second set of data $D_2$ using the same $f(x)$ and a different set of $r_i$. We measure $\chi^2(D_2,a_1)$ of the original fit $h_{13}(x)$  to the new data sample $D_2$. In Fig.~\ref{fig:fits2}(b), for each $N_P$, we repeat this procedure many times and show $\chi^2(D_2,a_1)$ averaged over many such trials as a function of $N_P$. We see that $N_P = 3$ is not enough: there is a substantial decrease in $\langle \chi^2(D_2,a_1)\rangle$ if we increase $N_P$ to $4$. However, beyond $N_P = 4$, $\langle\chi^2(D_2,a_1)\rangle$ increases with $N_P$, in agreement with Eq.~(\ref{eq:chisq12}). The rise of $\chi^2(D_2,a_1)$ for $N_P>4$ is suggestive of overfitting the randomly fluctuating data: while increasing $N_P$ improves $\chi^2(D_1,a_1)$ for the fitted sample, when $N_P$ is too large, it increases $\chi^2(D_2,a_1)$ for the control sample, indicating that the fit adapts to random fluctuations in $D_1$. 

Of course, the optimal number of fitted parameters will depend on the size of the fluctuations in the data. Having 20\% fluctuations in the data is not representative of what one finds in data used in PDF global fits. We can repeat this same exercise using 2\% fluctuations. Then the minimum of $\chi^2(D_2,a_1)$ (as in Fig.~\ref{fig:fits2}(b)) occurs at $N_P = 5$ instead of $N_P = 4$.

\subsection{Dependence on the number of PDF parameters}
\label{sec:numberofparameters}

\begin{figure}[tb]%--------------------------
\begin{center}\includegraphics[width=0.48\textwidth]{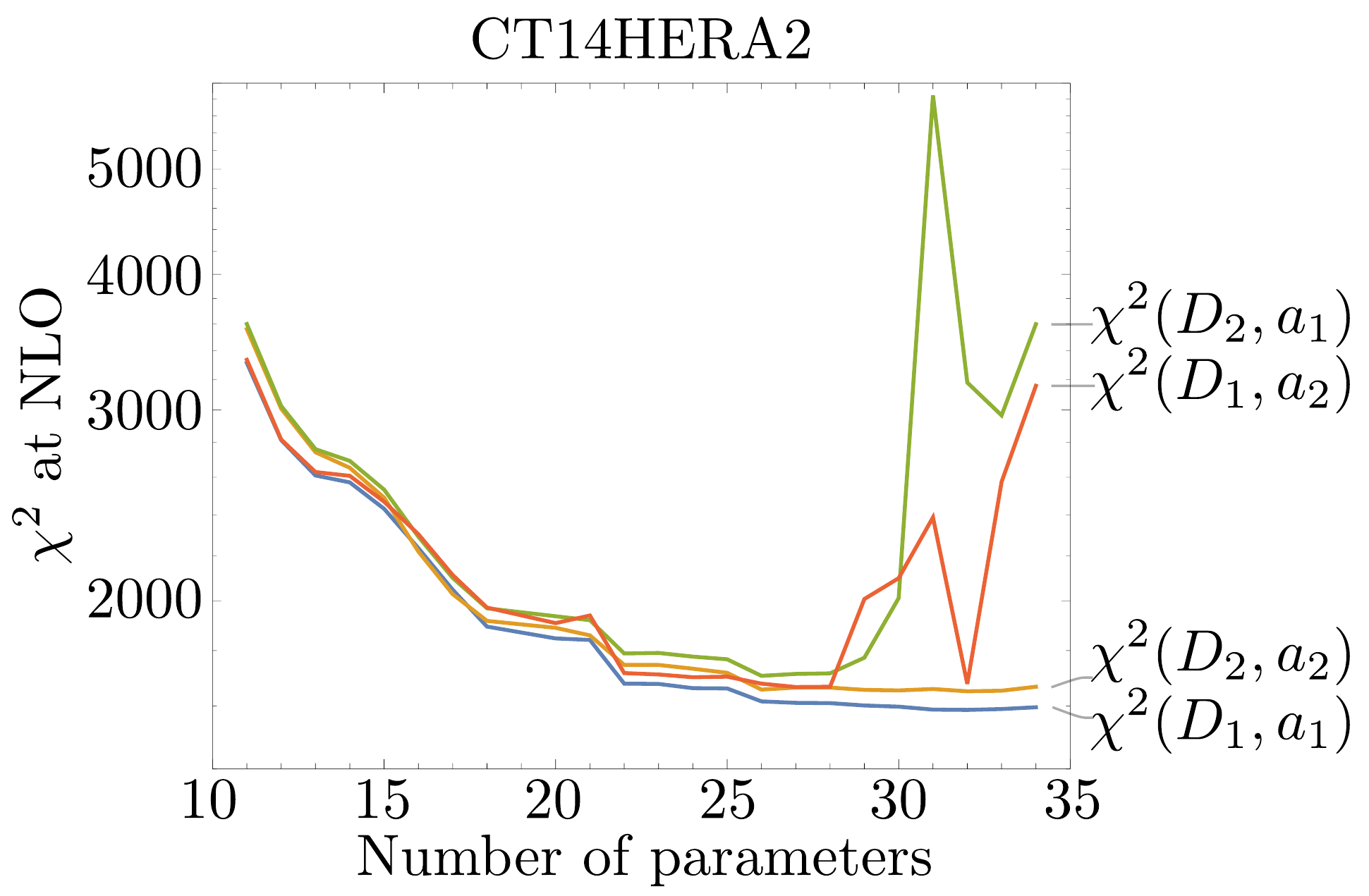}
\end{center}
\caption{$\chi^2$ from the fitted and control samples of data from the CT14HERA2 NLO resampling exercise. \label{fig:resampling}}
\end{figure}%--------------------------

A strategy comparing $\chi^2$ values of the fitted and control samples is routinely employed to prevent overfitting of the data in the approach utilizing neural-network (NN) parton distribution functions \cite{Ball:2010de, Ball:2012cx, Ball:2014uwa, Ball:2017nwa}. The PDF of each flavor is given by a NN of a certain configuration, which behaves essentially as a very flexible function, with its optimal number of parameters selected so as to satisfy a two-fold condition that the resulting PDFs render acceptable fits to the fitted and control samples in each run, or replica, of the global analysis. Both the fitted sample $D_1$ and control sample $D_2$  are obtained by randomly fluctuating the central values of the data according to the Gaussian distributions provided by the standard deviations of the data. The fit consists in training the NN to maximize agreement with the fitted sample. When training the NN on sample $D_1$, $\chi^2(D_1, a_1)$ is improved to an arbitrary accuracy by training the neural network long enough. For the control sample $D_2$, the $\chi^2(D_2, a_1)$ initially decreases and then grows after some number of training cycles. The training is stopped when $\chi^2(D_2, a_1)$ starts growing. The NN obtained at this point most optimally approximates both $D_1$ and $D_2$ samples without overfitting $D_1$.  

In the traditional approach used by the groups other than NNPDF, the PDFs are parameterized by a set of fixed functional forms; if the number of free parameters $N_P$ is too small or too large, the fit is too poor or unstable. For example, the recent MMHT fits approximate each PDF by a 6-degree Chebyshev polynomial \cite{Martin:2012da, Thorne:2019mpt}. CT18 error PDFs \cite{Hou:2019efy} are parameterized with functional forms containing Bernstein polynomials and up to five free parameters. 

We can illustrate this behavior with an example using real data. The CT14HERA2 parton distributions \cite{Hou:2016nqm} are parameterized using a generic form 
\begin{equation}
    f_a(x,Q_0) = A_0 x^{A_1} (1-x)^{A_2} P(x; A_3, A_4, ...).
    \label{eq:pdfCT14HERA2}
\end{equation}
The $x^{A_1}$ and $(1-x)^{A_2}$ prefactors capture the typical behavior in the $x\rightarrow 0$ and $x\rightarrow 1$ limits, respectively. The function $P(x; A_3, A_4, ...)$ is constructed as a linear combination of Bernstein basis polynomials, 
\begin{equation}
 P(x;A_3, A_4, ...)=\sum_{k=1,2,...}A_{k+2} b_{n,k}(x),
\end{equation}
with $b_{n,m}(x)=\left({\begin{smallmatrix}
    n &  \\
    m & 
\end{smallmatrix}}\right) x^m (1-x)^{n-m}$. Up to four Bernstein polynomials per flavor are introduced in the CT14HERA2 parameterization, with a total of $N_P=26$ parameters. However, we can try to choose  $N_P \le 26$ or $N_P > 26$. The number $N_P$ can be easily varied by adding or removing Bernstein polynomials with non-zero coefficients in the functions $P(x;A_3,A_4,...)$ in $f_a(x,Q_0)$.

We divide the CT14HERA2 data set into two equal parts, assigning each
datum to the half set $D_1$ or half set $D_2$ at random. Then we fit
the $N_P$ parameters to data set $D_1$, giving parameters $a_1$. We
measure $\chi^2(D_1,a_1)$ for this fitted data using the data $D_1$ to
which it was fitted. We also measure the $\chi^2(D_2,a_1)$ for the
fitted parameters $a_1$ using the second half set $D_2$. Alternatively, we fit the $N_P$ parameters to data set $D_2$, giving parameters $a_2$.  We measure $\chi^2(D_2,a_2)$ for this fitted data using the data $D_2$ to which it was fitted.   Then we also measure the $\chi^2(D_1,a_2)$ for this fit using the other data set, $D_1$.

We show the results in Fig.~\ref{fig:resampling}. As we increase the
number of parameters, $\chi^2(D_1,a_1)$ and $\chi^2(D_2,a_2)$
decrease. For $N_P < 28$, $\chi^2(D_2,a_1)$ and $\chi^2(D_1,a_2)$ also
decrease, although there is not much decrease beyond $N_P = 26$. For
$N_P > 28$, the behavior is different. For the fitted samples, the $\chi^2$ values continue to decrease, although the numerical minimization by the MINUIT program \cite{James:1975dr} becomes less stable. On the other hand, $\chi^2(D_2,a_1)$ and $\chi^2(D_1,a_2)$ exhibit large fluctuations for $N_P > 28$ and, additionally, their values increase dramatically for the largest $N_P$ in the figure.

We could also examine $\chi^2(D,a)$ for the whole data set, $D=D_1+D_2$. In this case, $\chi^2(D,a)$ also initially decreases as $N_P$ increases, but for large enough $N_P$, the fits become unstable: first, the $\chi^2$ is no longer a well-behaved function of PDF parameters; for even higher $N_P$, a unique best-fit PDF set itself may not be found. This behavior is an indication of overfitting. The number $N_P$ beyond which overfitting happens depends on the functional forms of PDFs and other factors. With the resampling exercise, we detect traces of overfitting in a half of the full data set. The presented exercise and other parameterization studies indicate that, with the parameterization forms adopted in the latest CT PDFs, about 25-30 free parameters is optimal for describing the fitted data set. 

A way to understand the results in Fig.~\ref{fig:resampling} is to note that the analysis in Sec.~\ref{sec:differentdata} of $\langle \chi^2 \rangle$ was based on the assumptions (from  Sec.~\ref{sec:deptheopar}) that one needed to consider only a small region of the parameter space, and that, within this region, the theory predictions $T_k(a)$ are approximately linear functions of the parameters $a$. The analysis then predicts a slow change of $\chi^2(D_2,a_1)$ (and $\chi^2(D_1,a_2)$) according to Eq.~(\ref{eq:chisq12}), as we increase the number of parameters $N_P$. This is not what we see in Fig.~\ref{fig:resampling}. 

To help interpret this, suppose that we use the data $D_1$ and determine fit parameters $a_1$ from $D_1$. We see in Fig.~\ref{fig:resampling} a large increase of $\chi^2(D_2,a_1)$ and large fluctuations for $N_P > 28$. What apparently happens is that, when there are too many parameters, the data do not strongly constrain the components of $a$ in one or more ``flat'' directions in the parameter space.  The value of $\chi^2(D_1,a)$ changes little as $a$ varies in these directions. One needs to move through a wide range of $a$ in a flat direction before  $\chi^2(D_1,a)$ increases significantly. In this wide range, the $T_k(a)$ are not approximately linear functions of $a$, and $\chi^2(D_1,a)$ is not approximately quadratic. Then $\chi^2(D_1,a)$ may have multiple minima at locations that correspond to quite unreasonable behavior of the PDFs. The best fit parameters $a_1$ can be at an unphysical location that strongly depends on the fluctuations in the data $D_1$. We can test how reasonable the fit $a_1$ is by testing it against the second data set, $D_2$. With too many parameters, we can get the large values of $\chi^2(D_2,a_1)$ that we see in Fig.~\ref{fig:resampling}. We conclude that, if one uses a family of functions for the PDFs that have a variable number of parameters, then a simple test like that shown in Fig.~\ref{fig:resampling}, ideally repeated for several random partitions of the data in two halves, can estimate how many parameters one should allow.

%%--------------------------------------------------
\subsection{Dependence on the PDF functional form}
\label{sec:functionalform}

There is another sort of test available. The PDFs are unknown functions, but one represents them using fixed functional forms with a finite number of unknown parameters.\footnote{The neural net approach avoids this limitation.} 
PDF parameterization studies thus constitute an essential step of the global analyses. 
In the previous section, we have examined the dependence of the fit results on the number of parameters within one general functional form. Evidently, one can also change the general functional form. Thus one should examine the tests listed above and later in this section for a large class of PDF functional forms. 

Universal approximation theorems \cite{cybenko_approximation_1989, hornik_universal_1990, hornik_approximation_1991} demonstrate that feed-forward neural networks with a single or multiple layers can approximate any continuous function and its derivatives to arbitrary accuracy. This fundamental result motivates the Neural Network PDF approach that parameterizes parton distribution functions by neural networks trained on the fitted data.

\begin{figure}%--------------------------
\begin{center}\includegraphics[width=0.48\textwidth]{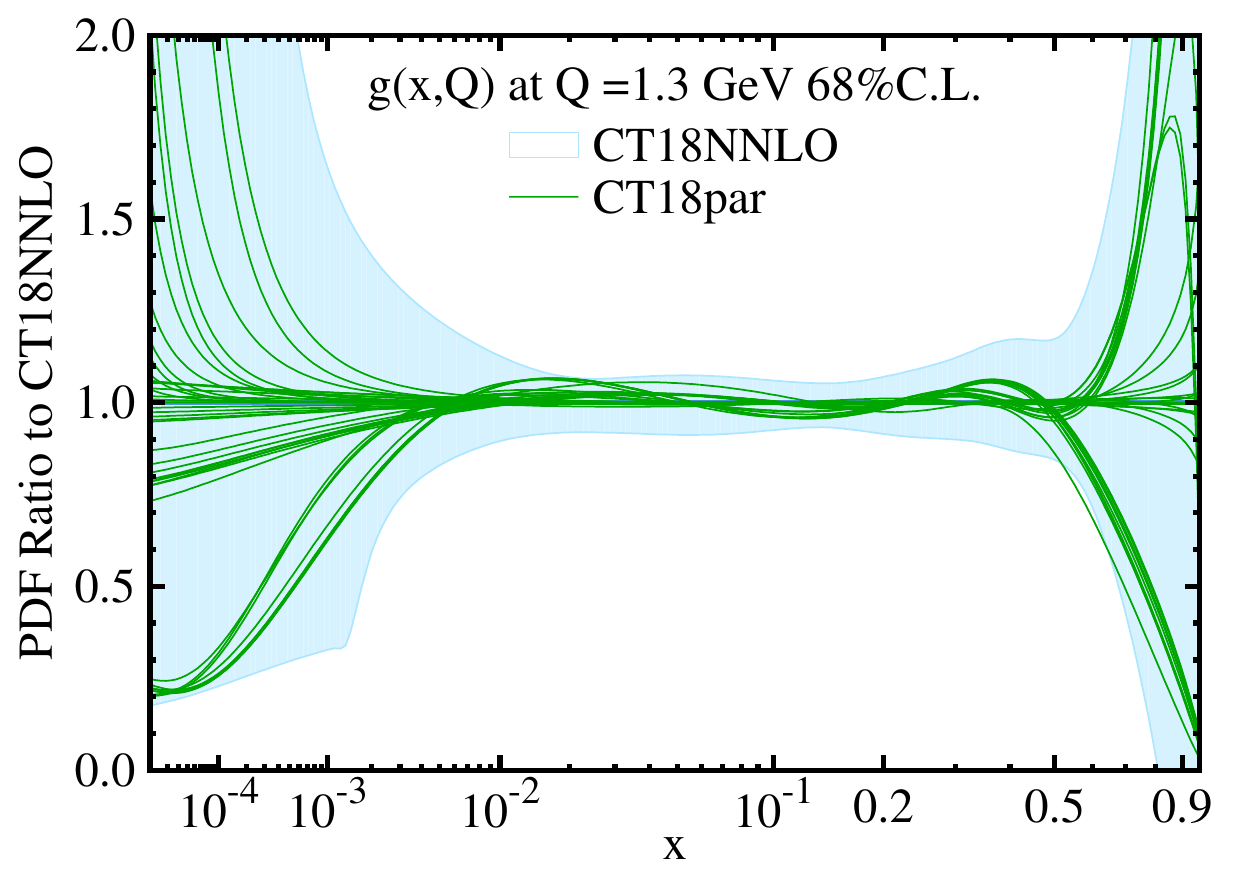}
\end{center}
\caption{Green solid lines: the gluon PDFs from candidate fits of the CT18 NNLO analysis, obtained using alternative parameterization forms and plotted as ratios to the default CT18 NNLO $g(x,Q)$. Light blue band: the 68\% C.L. uncertainty band of the published CT18 NNLO PDFs. From \cite{Hou:2019efy}. \label{fig:params}}
\end{figure}%------------------------
Alternatively, modern parameterizations based on Chebyshev or Bernstein polynomials also allow one to parameterize a variety of functional behaviors \cite{Pumplin:2009bb,Glazov:2010bw, Martin:2012da, Gao:2013bia, Hou:2019efy}.
Suppose that two choices for functional forms give good fits in the sense of reasonably meeting the array of goodness-of-fit criteria discussed in this section, but they also give two sets of PDFs that differ outside of their parameter-fitting error estimates. 
This would suggest that there is a theoretical systematic error associated with the choice of PDF functional form that needs to be added to the analysis. In a global analysis such as CT18, multiple parameterization forms are tried to estimate this source of PDF uncertainty. For example, the green solid lines in Fig.~\ref{fig:params} illustrate predictions for the best-fit gluon PDFs obtained with about 100 candidate CT18 fits based on different parameterization forms. Notice that the solid lines show large deviations from each other for large $x$ and small $x$, regions in which the gluon distribution is not well constrained by data.  The PDF uncertainty of the {\it published} CT18 NNLO set, shown as a blue band, is computed using a prescription that is broad enough to cover variations due to the choice of the parameterization form. For details, see \cite{Hou:2019efy}.

\subsection{Closure test \label{sec:ClosureTest}}

Given a PDF functional form depending on parameters $a$, PDF fitting of data determines best fit parameters $a_\textrm{fit}$. The fitting procedure is based on Eq.~(\ref{eq:PDTa}), which asserts that if the theory represented by parameters $\bar a$ is correct, then, in an ensemble of trials, the data $D_i$ will be distributed with a probability $P(D|T(\bar a))$ that involves the theory predictions $T_i(\bar a)$ and the covariance matrix and the covariance matrix $C_{ij}$. Then the parameters $a^\mathrm{fit}$ obtained from this data will be distributed according to Eq.~(\ref{eq:adistribution}). We can test this.  We can generate a set of pseudodata $D^\prime_k$ using Eq.~(\ref{eq:Dkdef}) with the truth values $\langle D_k^\prime\rangle$ set equal to theory predictions $T_i(a)$ for some realistic parameters $a$, and with $\sigma_k$ and $\beta_{kJ}$ obtained from the experimental groups. Then we can run the fitting procedure with the generated data $D^\prime_k$, giving a new fit $a^\prime_\textrm{fit}$. We should find that the new fit $a^\prime_\textrm{fit}$ agrees with the original parameters $a$ within the errors generated by the fit. If they do not agree, we should understand why. This ``closure test'' is not commonly carried out by fitting groups who use the Hessian method described in this review, but it is a feature of fits by the NNPDF group \cite{Ball:2014uwa}.

%%--------------------------------------------------
\subsection{Test of the nuisance parameters}
\label{sec:nuisance}

We now describe a goodness-of-fit test based on the distribution of the nuisance parameters. For this test, it is useful to use the form of $\chi^2$ in which
nuisance parameters $\lambda_J$ appear explicitly, $\chi^2(D,a,\lambda)$ as given in Sec.~\ref{sec:altchisq}, Eq.~(\ref{eq:chisqDalambda}). Then we can fit values
$\lambda_\textrm{fit}$ of $\lambda$ by minimizing $\chi^2(D,a,\lambda)$. The minimum value is $\chi^2(D,a)$, expressed in terms of the covariance matrix $C_{ij}$ in Eq.~(\ref{eq:chisqDa}). We have $ \chi^2(D,a_\textrm{fit},\lambda_\textrm{fit}) = 
\chi^2(D,a_\textrm{fit})$. We also argued after Eq.~(\ref{eq:lambdafit}) that
$\lambda^\textrm{fit}_J \approx \bar \lambda_J$ in an accurate fit with enough data. Since the $\bar\lambda_J$ are independent random variables distributed according to ${\cal N}(0,1)$, the $\lambda^\textrm{fit}_J$ are expected to follow the ${\cal  N}(0,1)$ distribution, too. We can test these assumptions by making a histogram of $\lambda^\textrm{fit}_J$.

\begin{figure}%--------------------------
\centerline{\includegraphics[width=0.48\textwidth]{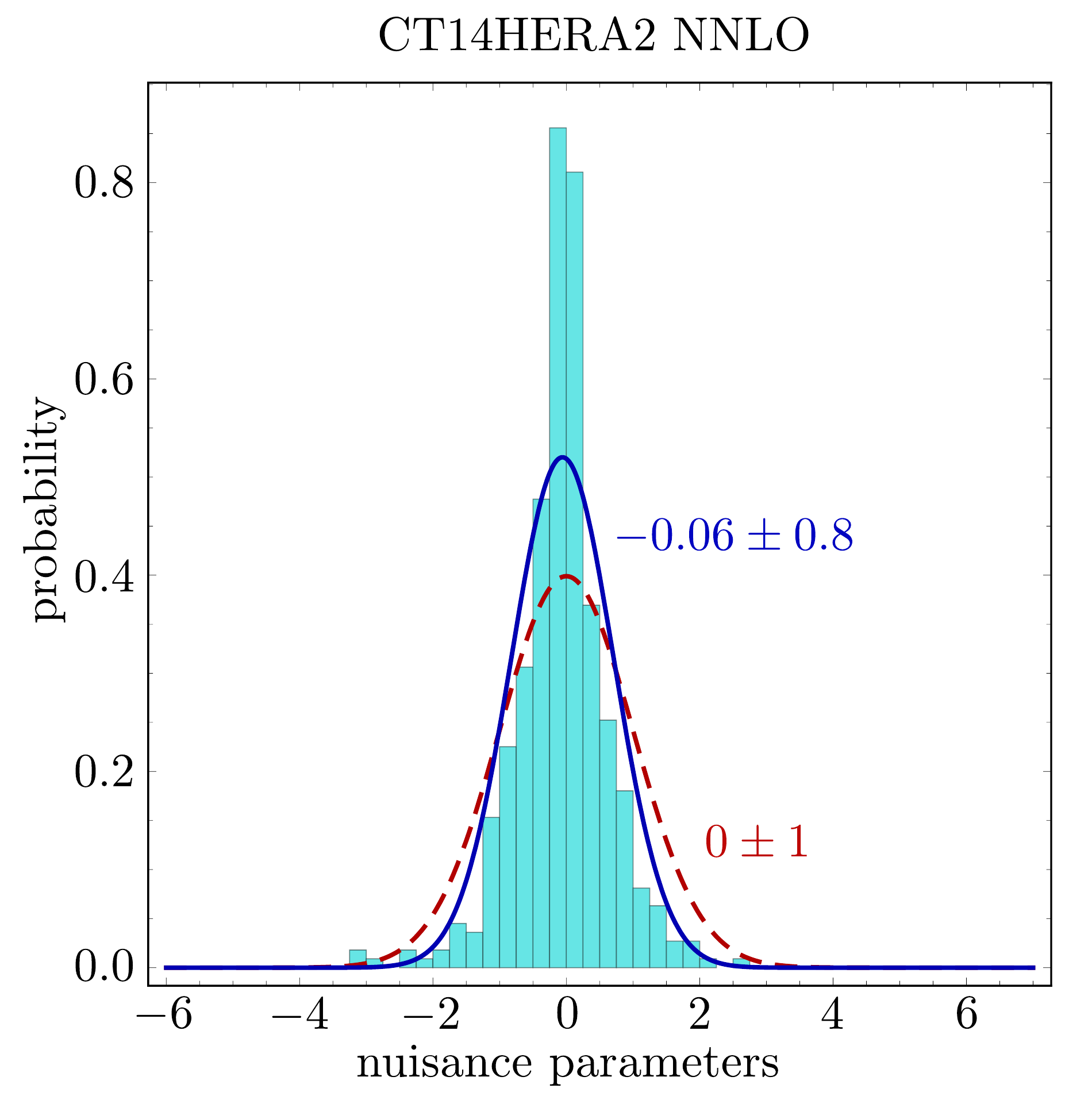}}

\caption{Distribution of nuisance parameters $\lambda_J^\textrm{fit}$
  for the CT14 HERA2 NNLO parton distributions fit. \label{fig:nuisance}}
\end{figure}%--------------------------

We show in Fig.~\ref{fig:nuisance} a histogram of the best fit nuisance parameters $\lambda^\textrm{fit}_J$ for the CT14 HERA2 NNLO fit along with a dashed, red curve showing the expected Gaussian distribution. The correlated systematic errors are implemented in the CT14HERA2 fit using the matrices $\beta_{kJ}$ provided with the experimental data sets. Evidently, the observed distribution is substantially narrower than the expected distribution. The mean for the observed distribution is $-0.06$, which is very close to 0, but its standard deviation is 0.8, noticeably smaller than 1. We also show a blue, solid curve giving a Gaussian distribution with this mean and standard deviation. This curve also does not match the observed distribution well.

For a quantitative estimate of the probability that the observed distribution was sampled from ${\cal N} (0,1)$, we can apply the standard Anderson-Darling test\cite{anderson1952} described at the beginning of Sec.~\ref{sec:goodness}. For the histogram in Fig.~\ref{fig:nuisance}, we find that $P_\textrm{A-D} \sim 10^{-6}$. This indicates that it is very unlikely that the $\lambda_J^\textrm{fit}$ values were generated from the expected ${\cal N} (0,1)$ distribution, as was self-evident from the figure. 
 
There are a few more $|\lambda^\textrm{fit}_J|$ that are larger than 2
than would be expected. However, the main feature that we see in
Fig.~\ref{fig:nuisance} is that too many $|\lambda^\textrm{fit}_J|$
are very small, indicating possibly that the corresponding $\beta_{kJ}$ values
are overestimated. This could indicate that the estimates are
conservative in the sense that, if one suspects that the $\beta_{kJ}$
values for a source $J$ of systematic error should be smaller but one
cannot prove it with solid evidence, then the conservative approach is
to leave these $\beta_{kJ}$ values unchanged. 
 
There is another reason that the values of some of the $\lambda_J^\mathrm{fit}$ might be smaller than expected. The expectation from Eq.~(\ref{eq:lambdafit}) that $\lambda_J^\mathrm{fit} \approx \bar \lambda_J$ was based on the assumption that, for each data set $E$, there are many data and few nuisance parameters. What happens if this is not the case, as suggested near the end of Sec.~\ref{sec:altchisq}? Suppose that we add a spurious nuisance parameter $\lambda_L$ with $\beta_{kL} = 0$ for all data indices $k$. Then, in Eq.~(\ref{eq:chisqDalambda}) for $\chi^2(D,a,\lambda)$, the spurious $\lambda_L$ does not contribute to the first term, which involves data, and contributes only to the ``penalty'' term $\sum_J \lambda_J^2$. Then minimizing $\chi^2(D,a,\lambda)$ with respect to the $\lambda_J$ gives $\lambda_L^\mathrm{fit} = 0$. If a certain data set $E$ has many systematic errors and thus many nuisance parameters, it may be that some of linear combinations of them are spurious in this sense. Then the same linear combinations of the fitted values $\lambda_J^\mathrm{fit}$ will vanish, so that the fitted values do not spread out over the whole $\lambda_J$ space but only a subspace. In this case, the absolute values of these $\lambda_J^\mathrm{fit}$ will be smaller than would be given by an $\cN(0,1)$ distribution. In the case that there are many nuisance parameters, some of which have little effect, one can approximate the matrix $C^{-1}$ in Eq.~(\ref{eq:Cinversedef}) so as to leave effectively fewer, but better behaving, nuisance parameters, as described in Appendix B of \cite{Hou:2019efy}.
\\ \\
We suggest that, when performing a fit, it is useful to check the $\lambda^\textrm{fit}_J$ distribution. Values of $\lambda^\textrm{fit}_J$ that are smaller than expected from an $\cN(0,1)$ distribution suggest problems with the systematic errors used in the fit but are perhaps not of great concern. An excessive number of large $\lambda^\textrm{fit}_J$ values could indicate a more serious difficulty that needs to be resolved.

\subsection{Test of data residuals}
\label{sec:residuals}

Eq.~(\ref{eq:chisqDalambda}) gives $\chi^2(D,a_\textrm{fit},\lambda_\textrm{fit})$ as
\begin{equation}
\chi^2(D,a_\textrm{fit},\lambda_\textrm{fit})
= \sum_k \big[r_k(a_\textrm{fit},\lambda_\textrm{fit})\big]^2 
+ \sum_J \big[\lambda_J^\textrm{fit}\big]^2
\;,
\end{equation}
where
\begin{equation}
r_k(a_\textrm{fit},\lambda_\textrm{fit})
= 
\frac{D_k - T_k(a_\textrm{fit})}{\sigma_k} - \sum_I \beta_{kI}\lambda^\textrm{fit}_I
\end{equation}
is called the {\em residual} for datum $k$ obtained in the fit. Using Eqs.~(\ref{eq:Dkdef}), (\ref{eq:theoryT1}), and (\ref{eq:theoryT2}) with $R_k = 0$, this is
\begin{equation}
\begin{split}
r_k(a_\textrm{fit},\lambda_\textrm{fit})
& = \Delta_k 
-\frac{T_{k\alpha}}{\sigma_k}\, 
(a^\textrm{fit}_\alpha - \bar a_\alpha) \\ 
& 
- \sum_I \beta_{kI}(\lambda^\textrm{fit}_I - \bar\lambda_I)
\;.
\end{split}
\end{equation}
The $\Delta_k$, introduced in Eq.~(\ref{eq:Dkdef}),  follow independent $\cN(0,1)$ distributions. The values
$(a^\textrm{fit}_\alpha - \bar a_\alpha)$ and $(\lambda^\textrm{fit}_I
- \bar\lambda_I)$ have expectation values zero, but they have
fluctuations that arise from the fluctuations in the data. If there
are enough data, we expect the fluctuations of $(a^\textrm{fit}_\alpha
- \bar a_\alpha)$ and $(\bar\lambda_I - \lambda^\textrm{fit}_I)$ to be
small. Then the residuals $r_k$ should also be approximately
distributed as ${\cal N}(0,1)$. 

\begin{figure}%--------------------------
\centerline{\includegraphics[width=0.48\textwidth]{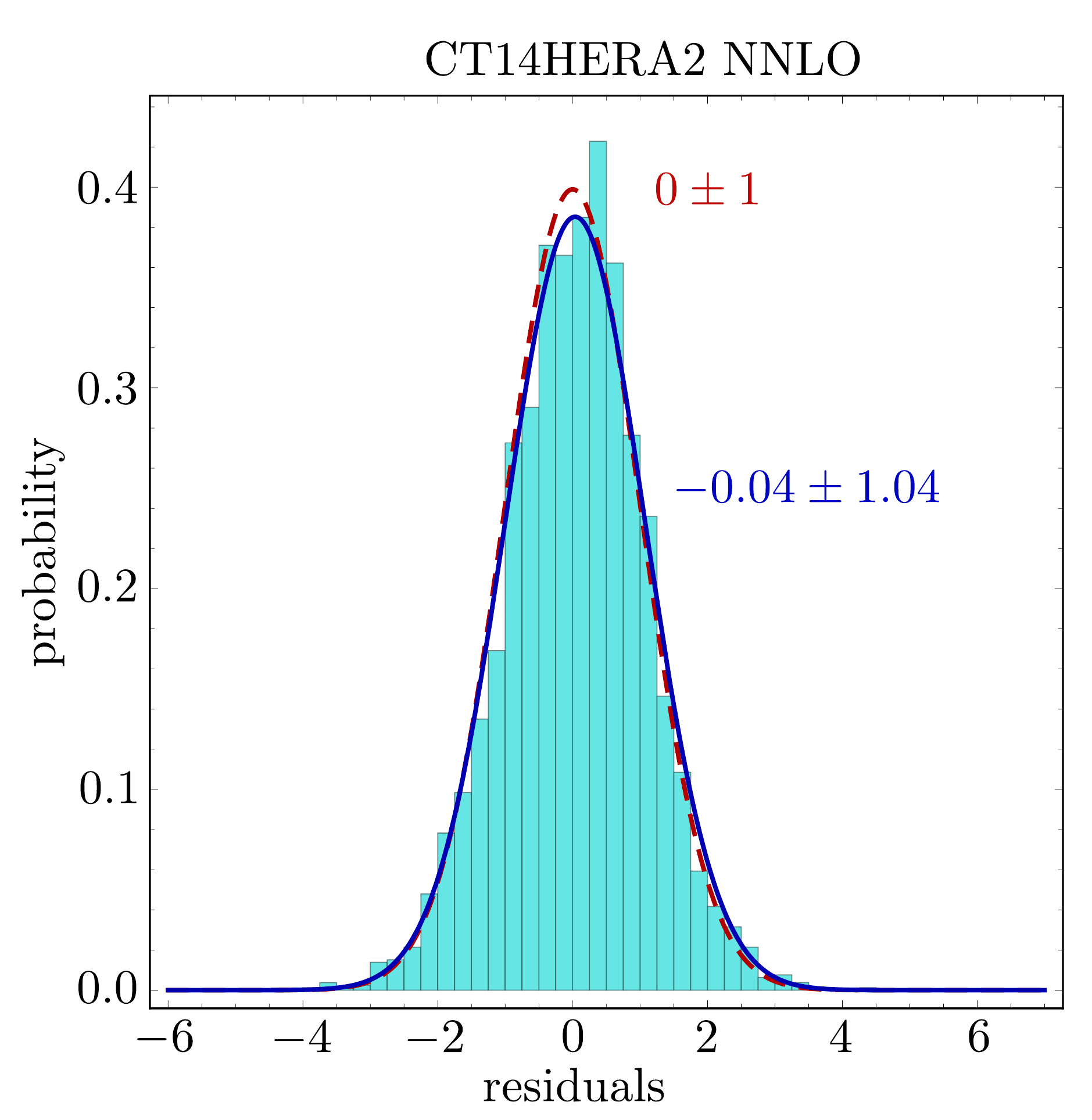}}

\caption{Distribution of residuals for the CT14 HERA2 NNLO parton distributions fit. \label{fig:residuals}}
\end{figure}%--------------------------

We can test these assumptions by making a histogram of the values
$r_k$ obtained. We show in Fig.~\ref{fig:residuals} a histogram of the
residuals $r_k$ for the CT14 HERA2 NNLO fit along with a curve showing
the expected Gaussian distribution. Comparing these using the
Anderson-Darling test gives $P_\textrm{A-D} = 5.7\times 10^{-3}$. Thus
we can conclude with some confidence that the observed distribution of
residuals was not drawn from exactly ${\cal N}(0,1)$. However, we
judge the difference between the two distributions to be not
physically significant. After all, we expect the observed distribution
to be  only approximately an ${\cal N} (0,1)$ distribution. The mean
for the observed distribution is $0.04$, which is very close to 0, and
its standard deviation is 1.04, which is quite close to 1. We also
show a blue, solid curve giving a Gaussian distribution with this mean
and standard deviation. 

The distribution of residuals is another indicator that should be
checked when performing a fit.
In this case, no large discrepancies are observed.

\subsection{Value of $\chi^2$ from an individual experiment}
\label{sec:experimentchisq}

In this subsection, we describe a goodness-of-fit test based on a decomposition of the data into subsets. We will sometimes refer to a subset of the data as an ``experiment'', although we could divide the data into subsets in different ways.

Label the subset of the data that we wish to consider by an index
$E$. Let $D(E)$ refer to the data in subset $E$, that is, all data points
$D_i$ for $i \in E$. Recall Eq.~(\ref{eq:PDTa}), giving the probability to find data $D$ if the theory $T(a)$ is correct. The analogue of this that gives the probability to find data $D(E)$ if the theory $T(a)$ is correct is
\begin{equation}
\label{eq:PDETa}
P(D(E)|T(a))  = d\mu(D) 
\exp\!\left(- \frac{1}{2}\,\chi^2(D(E),a)
\right)
,
\end{equation}
where
\begin{equation}
\begin{split}
\label{eq:chisqE}
\chi^2&(D(E),a) 
\\&= \sum_{i,j \in E}
(D_i - T_i(a))(D_j - T_j(a))\, C_{ij}
\;,
\end{split}
\end{equation}
$d\mu(D(E))$ is the measure
\begin{equation}
\label{eq:muD2}
d\mu(D(E)) = (2\pi)^{- N_E/2}\sqrt{\det C_E}\
d^{N_E} D
\;,
\end{equation}
$C_E$ is the matrix $C_{ij}$ for $i,j \in E$ and $N_E$ is the number of data in the subset $E$.

If we have a good fit, then the probability $P(D(E)|T(a))$ should be not too small for each subset $E$ of the data. That is, for each subset $E$, $\chi^2(D(E),a)$ should not be too large.

As we already mentioned, this is a much stronger criterion than the hypothesis-testing criterion. An individual experiment $E$ may be very badly fit in a large global
fit (have an unacceptably high $\chi^2(D(E),a)$), even while the total $\chi^2(D,a)$ may look reasonable. In this section we ask, how large is too large for $\chi^2(D(E),a)$? What should the distribution of this quantity be?

Consider structure of the covariance matrix $C_{ij}$, defined by Eq.~(\ref{eq:Cinversedef}). Suppose first that each experiment has independent systematic errors that are not shared among the experiments, so that each source $J$ of systematic error is associated with just one experiment $E_J$. Then $\beta_{i J} = 0$ unless $i \in E_J$. The covariance matrix is then block-diagonal: $C_{ij} = C^{-1}_{ij} = 0$ unless $i \in E$ and $j \in E$ for the same experiment label $E$. Then the total $\chi^2(D,a)$ is a sum of contributions $\chi^2(D(E),a)$ from the separate experiments:
\begin{equation}
\label{eq:chisqsum}
\chi^2(D,a) = \sum_E \chi^2(D(E),a)
\;.
\end{equation}

It is, however, not necessary that errors for experiment $E$ are uncorrelated with the errors from other experiments. If, for some of the data sets $E$, the covariance matrix has elements $C_{ij}$ that are non-zero for $i \in E$ and $j \notin E$, then Eq.~(\ref{eq:chisqsum}) will fail. However the probability to find data $D(E)$ if the theory $T(a)$ is correct is still given by Eq.~(\ref{eq:PDETa}).

We return to Eq.~(\ref{eq:chisqE}). There are $N_E$ data in the set $E$. Then, with the parameters $a$ fixed to ideal values, ($a=\bar a$, as in Sec.~\ref{sec:deptheopar}), and flawless theory ($R_k =0$), we find that $T_k(\bar a) = \langle D_k \rangle$, and the distribution of
$\chi^2(D(E),\bar a)$ is the standard $\chi^2$ distribution with $N_E$
degrees of freedom.  

If, however, we use the best-fit parameters $a_\textrm{fit}$ that are constrained by experiment $E$ as well as the rest of experiments, we find that $\chi^2(D(E),a_\textrm{fit})$ is approximately equal to the standard $\chi^2$ distribution with $N_E$ degrees of freedom up to a subleading term that can be determined as follows. 

From Eqs.~(\ref{eq:theoryT1}), (\ref{eq:theoryT2}), and (\ref{eq:chisqDa}), we find that 
\begin{equation}
\label{eq:theorybis}
T_k(a) = \langle D_k \rangle + R_k +  T_{k\alpha} (a_\alpha - \bar a_\alpha)
\end{equation}
and
\begin{equation}
\begin {split}
\label{eq:chisqE2}
\chi^2(&D(E),a) 
\\& = \sum_{i,j \in E}
(D_i - \langle D_i\rangle - T_{i\alpha} (a_\alpha - \bar a_\alpha) - R_i)
\\&\qquad\ \times
(D_j - \langle D_j\rangle - T_{j\beta} (a_\beta - \bar a_\beta) - R_j)\, C_{ij}
\;.
\end{split}
\end{equation}
What is the expectation value of this at $a = a_\textrm{fit}$? We can refer to Eqs.~(\ref{eq:expectationa}), (\ref{eq:expectationaa}), (\ref{eq:Dkexpectations}) and (\ref{eq:daexpectation}) for $a = a_\textrm{fit}$ to arrive at
\begin{equation}
\begin{split}
\label{eq:chiEexpectation}
\langle\chi^2(D(E),a_\textrm{fit})\rangle ={}& 
N_E
- N_\LP(E)
+\sum_{i,j \in E} R_i R_j C_{ij}
\;.
\end{split}
\end{equation}
where
\begin{equation}
\label{eq:NPE}
N_\LP(E) = \sum_{i,j \in E}
T_{i\alpha} 
T_{j\beta}\, C_{ij}
H^{-1}_{\alpha\beta}
\end{equation}
The first term in Eq.~(\ref{eq:chiEexpectation}) is just the number of data, $N_E$, in the data set. The third term is small if the theory is good ($R_i \approx 0$). The second term $- N_\LP(E)$, requires some analysis. We interpret $N_\LP(E)$ as the effective number of parameters constrained by data set $E$. To support this interpretation, we note from  Eqs.~(\ref{eq:Halphabeta}) and (\ref{eq:chisqexpectation}) that if $N_\LP(E)$ is summed over all data sets $E$, it gives the number $N_\LP$ of parameters. Now, the number of parameters, $N_\LP$ is much smaller than the total number of data, $N_\LD$.  This makes it plausible that $N_\LP(E)$ is small compared to the number $N_E$ of data in data set $E$: $N_E \gg N_\LP(E)$. Then the first term in Eq.~(\ref{eq:chiEexpectation}) dominates.

\subsection{Test of $\chi^2$ from individual experiments}
\label{sec:experimentchisqall}

What can we do with $\chi^2(D(E),a_\textrm{fit})$? Its
value is given by Eq.~(\ref{eq:chisqE2}) with $a = a_\textrm{fit}$.
When the best-fit parameters are close to the true ones
($a_\textrm{fit}\approx \bar a$), and theory is nearly perfect
($R_k\approx 0$), the $\chi^2(D(E),a_\textrm{fit})$ distribution
reduces to the form
\begin{equation}
\begin {split}
\label{eq:chisqDapprox}
\chi^2(& D(E),a_\textrm{fit}) \\
 & = \sum_{i,j \in E}
(D_i - \langle D_i\rangle)(D_j - \langle D_j\rangle)\, C_{ij} +...
\notag
\end{split}
\end{equation}
that, as we already know, obeys the $\chi^2$ distribution with $N_E$
degrees of freedom. Now, $a_\textrm{fit}$ is not exactly $\bar a$, but is influenced by the data in data set $E$. Thus one might think that the number of degrees of freedom is $N_E - N_\LP(E)$ where $N_\LP(E)$, Eq.~(\ref{eq:NPE}), is the effective number of fit parameters associated with data set $E$. However, the reasoning in the previous subsection suggests that $N_\LP(E) \ll N_E$, so that the number of degrees of freedom is approximately just the number $N_E$ of data in data set $E$. This expectation can be tested by calculating $N_\LP(E)$ in the case that $\chi^2(D(E),a_\textrm{fit})$ appears anomalously small. 

We will now check if the distributions of the observed $\chi^2(D(E),a_\textrm{fit})$ values from the experiments $E$ in actual PDF fits are close to the ideal distributions. 

When $N_E$ is large, say, $N_E \gtrsim 30$, the $\chi^2$ distribution
with $N_E$ degrees of freedom approaches the Gaussian distribution
with with mean $\langle\chi^2(D(E),a_\textrm{fit})\rangle \approx N_E$
and standard deviation $\langle(\chi^2(D(E),a_\textrm{fit}) -
N_E)^2\rangle \approx 2 N_E$, as we have seen in
Sec.~\ref{sec:chisqstatistics}. For $N_E \lesssim 30$, the non-Gaussian
features are pronounced and $\chi^2$ distributions with different $N_E$
are not easily compared.
Conveniently for our purpose, the variable
\begin{equation}
\label{eq:Sdef}
S_E \equiv \sqrt{2\chi^2(D(E),a_\textrm{fit})}-\sqrt{2N_{E}-1}
\end{equation}
fluctuates with a distribution that is quite accurately\footnote{Other
  definitions \cite{Lewis:1988} of $S_E$ are more accurate but with the very simple form
  (\ref{eq:Sdef}), the distribution function $\rho(S_E)$ matches the
  Gaussian distribution $(2\pi)^{-1/2} \exp(- S_E^2/2)$ to within 0.04
  for $N_E = 5$ and to within 0.01 for $N_E = 50$.} an ${\cal N}(0,1)$ distribution \cite{Lai:2010vv, Fisher:1925}, namely
\begin{equation}
\label{eq:SEdistribution}
\rho(S_E) \approx (2\pi)^{-1/2}\exp(-S_E^2/2)
\;.
\end{equation}
Note that the $S_E$ distribution is independent of $N_E$. The original $N_E$ dependence for the distribution of $\chi^2$ was absorbed into the definition of $S_E$.

To test the quality of the fit, we can plot a histogram of the $S_E$ values for all of the experiments (or data sets) $E$ contributing to the fit. The histogram should match the Gaussian distribution (\ref{eq:SEdistribution}).

\begin{figure*}%--------------------------
\includegraphics[width=0.48\textwidth]{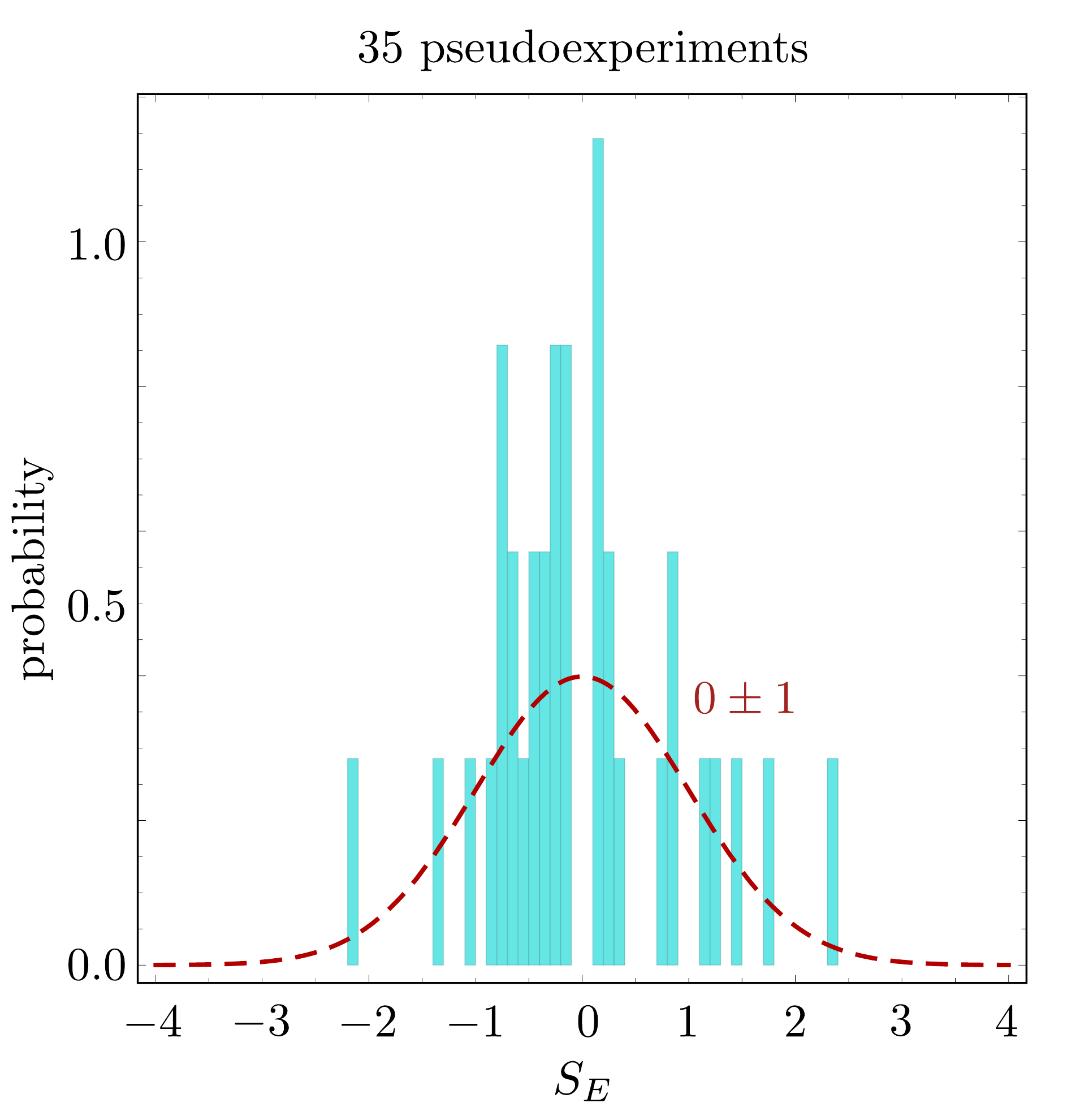}\includegraphics[width=0.48\textwidth]{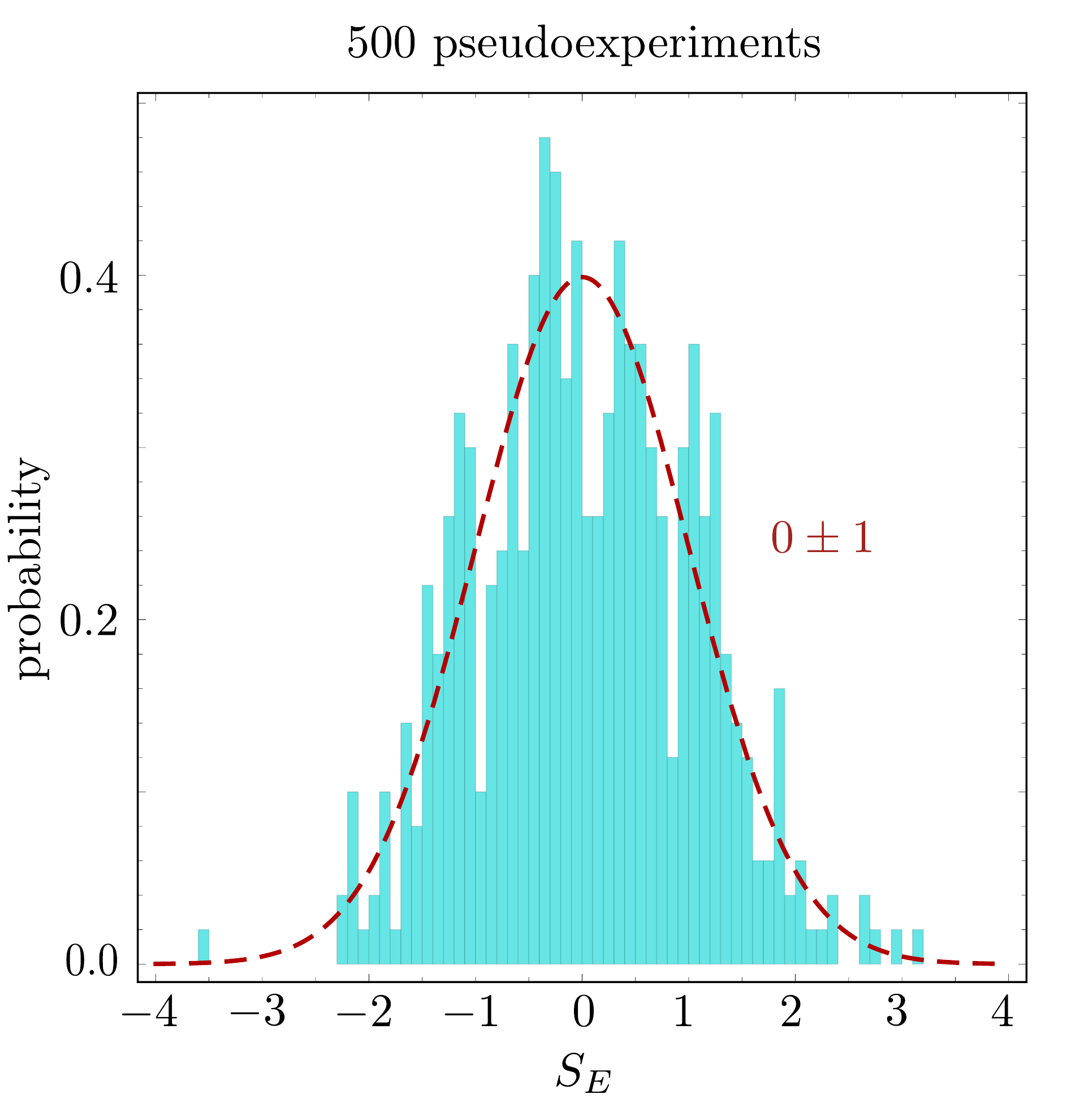}

\caption{
The probability distributions of
$S_E=\sqrt{2\chi^2(D(E),a_\textrm{fit})}-\sqrt{2N_{E}-1}$ for 35 and
500 random pseudoexperiments, each of which has the number $N_E$ of
data chosen at random in the range $0\protect\leq N_{E}\protect\leq
3000$. The red dashed line shows the ${\cal N}(0,1)$ Gaussian distribution, which describes well the observed probabilities. \label{fig:sqrt2chi2}}
\end{figure*}

To see how this should work, we can generate $S_E$ values for a number of randomly generated pseudoexperiments. The number $N_E$ of data for each pseudoexperiment is chosen at random between 0 and 3000. For each pseudoexperiment, we generate a value of $\chi_E^2$ at random according to the standard $\chi^2$ distribution with $N_E$ degrees of freedom. Then we define $S_E$ for that pseudoexperiment by $S_E = \sqrt{2\chi^2_E} - \sqrt{2N_E - 1}$. In the left-hand plot in Fig.~\ref{fig:sqrt2chi2}, we show the resulting histogram of $S_E$ values obtained for 35 pseudoexperiments, along with the expected Gaussian distribution (\ref{eq:SEdistribution}). In the right-hand plot in Fig.~\ref{fig:sqrt2chi2}, we show the analogous histogram for 500 pseudoexperiments.

The histogram in Fig.~\ref{fig:sqrt2chi2} for 500 pseudoexperiments is
evidently pretty close to the expected distribution
(\ref{eq:SEdistribution}). For the histogram for 35 pseudoexperiments,
it is not so clear merely by eye. For a quantitative estimate, we can
apply the standard Anderson-Darling test, described in
Sec.~\ref{sec:nuisance}, of the probability that the observed distribution matches
${\cal N}(0,1)$. For the left-hand histogram in
Fig.~\ref{fig:sqrt2chi2}, we find that $P_\textrm{A-D} = 0.53$ and, for
the right-hand histogram, we find that $P_\textrm{A-D} = 0.44$. These
values indicate that it is quite plausible that the $S_E$ values were
generated from ${\cal N}(0,1)$, which, to a good approximation, they were.
 
%--------------------------

\begin{figure*}%--------------------------
\includegraphics[width=0.48\textwidth]{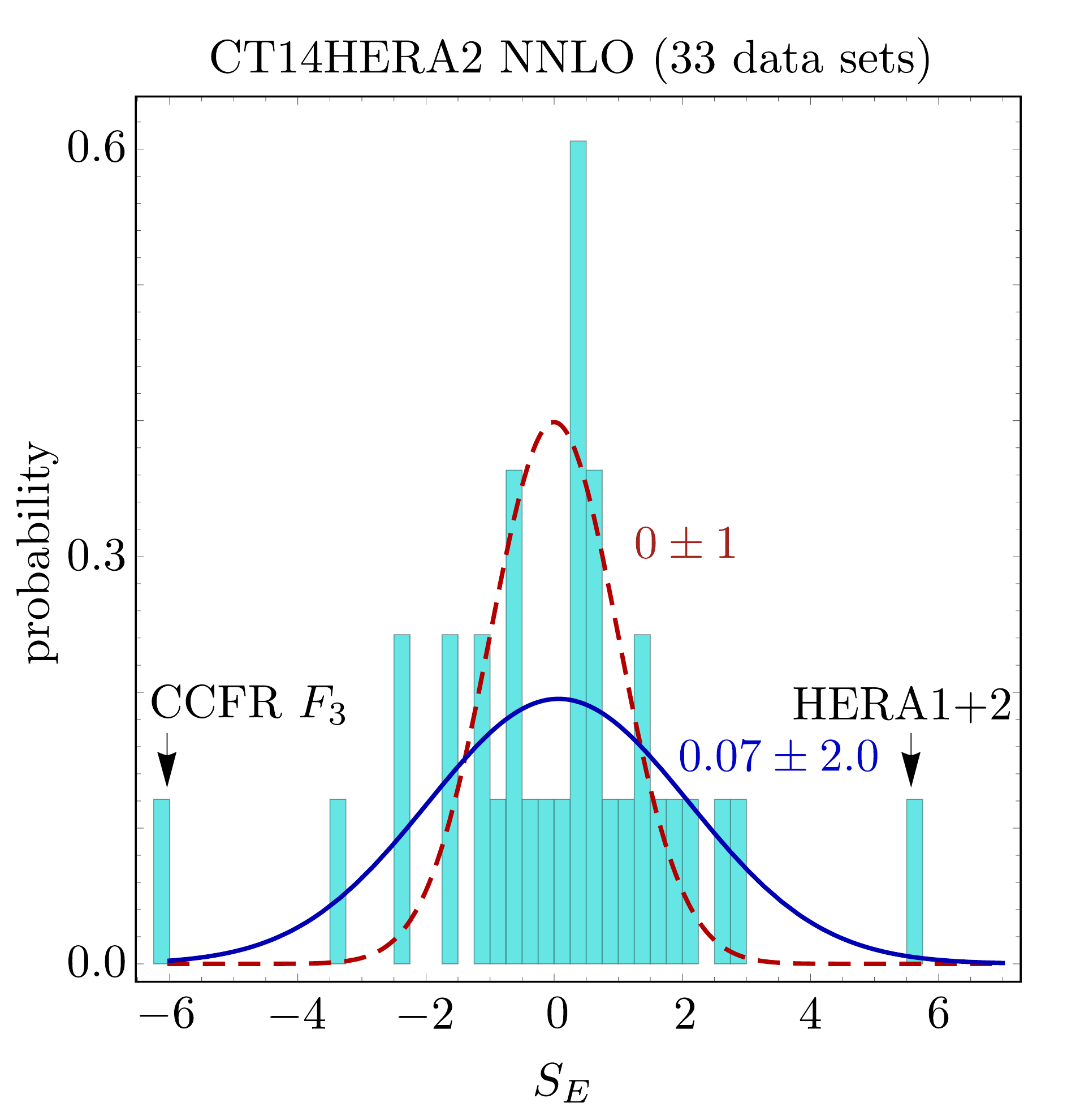}\quad 
\includegraphics[width=0.48\textwidth]{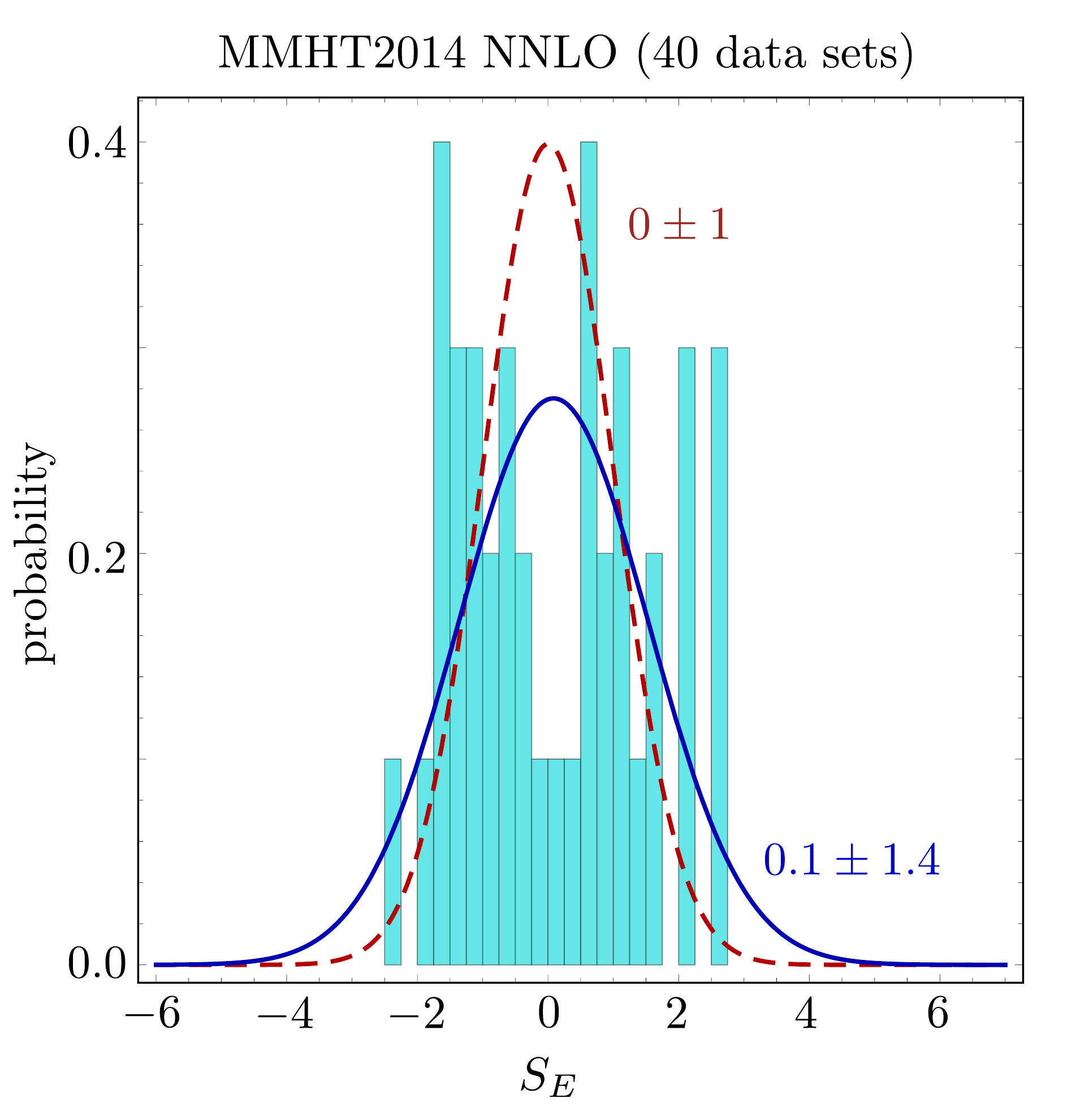}\\
\quad\\
\includegraphics[width=0.48\textwidth]{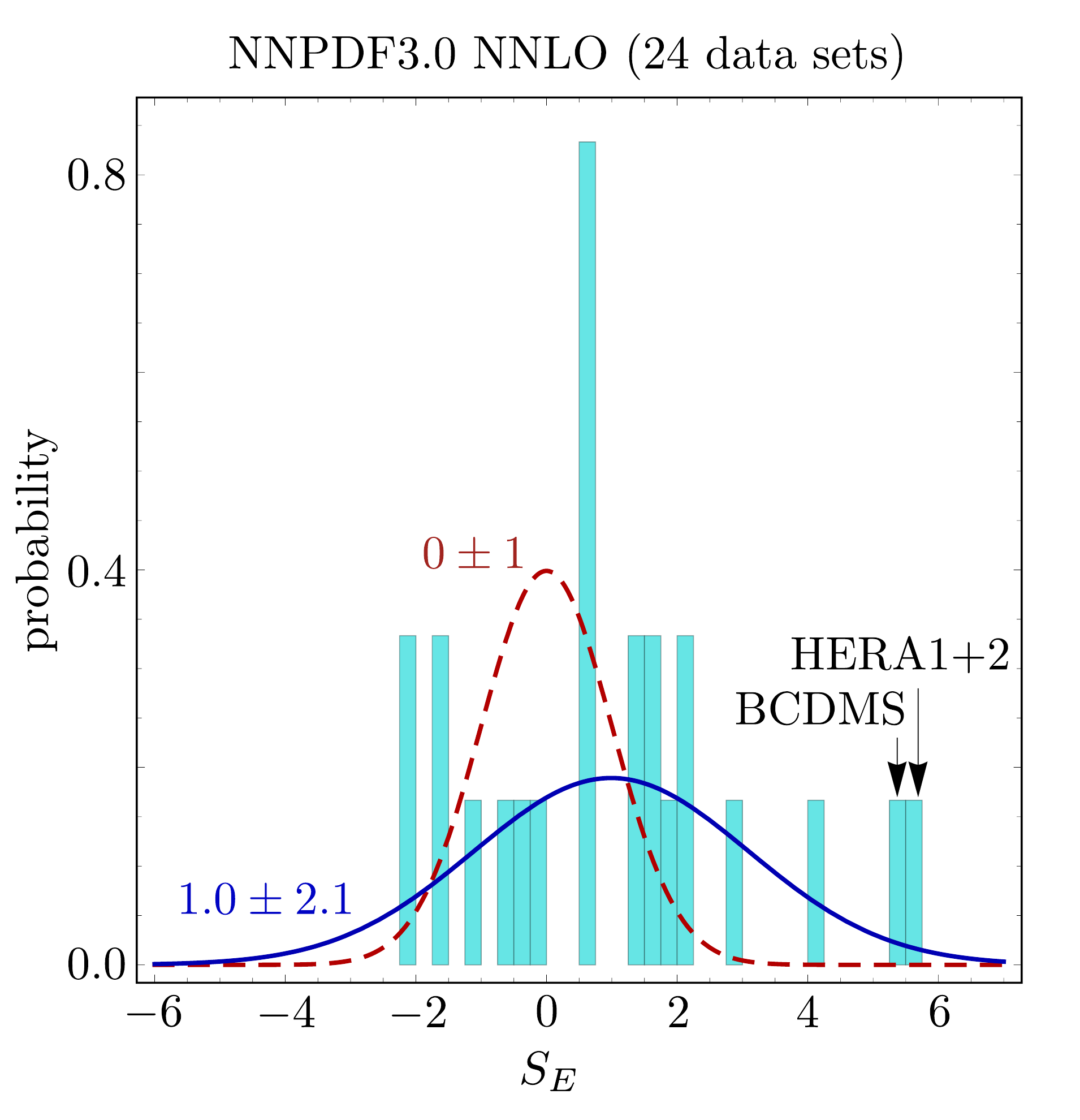}\quad 
\includegraphics[width=0.48\textwidth]{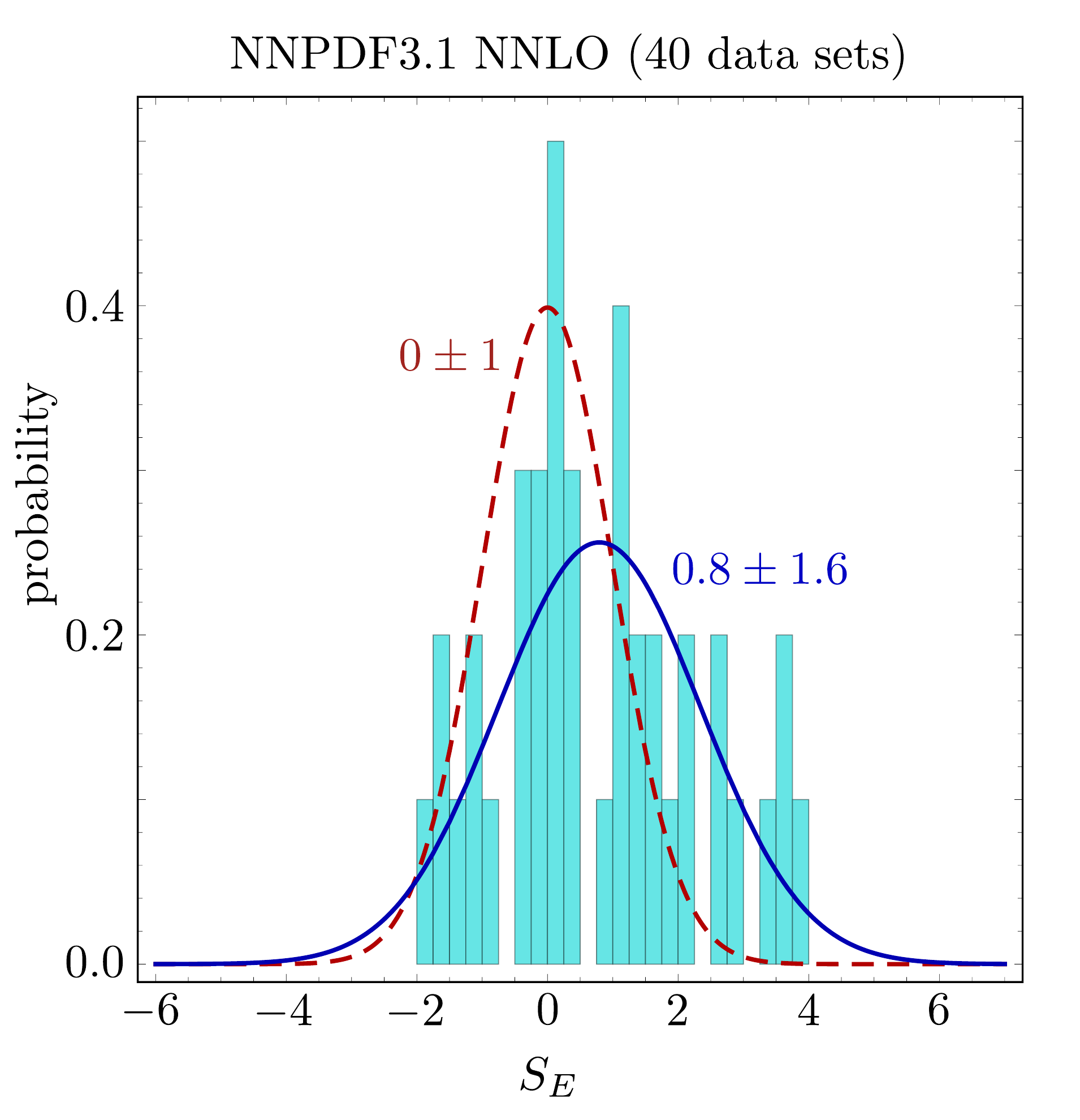}
\caption{Probability distributions in the effective Gaussian variable $S_E$ for $\chi^2$ values of the fitted data sets from the NNLO fits CT14HERA2, MMHT'2014, NNPDF3.0, and NNPDF3.1.}
\label{fig:SE} 
\end{figure*}

Now, let us turn to the distributions of $S_E$ from recent NNLO global
analyses shown in Fig.~\ref{fig:SE}. The parameters $S_E$ in the histograms are computed using Eq.~(\ref{eq:Sdef}) from $\chi^2$ and $N_E$ values listed in Table 5 of \cite{Harland-lang:2014zoa} for MMHT 2014 NNLO, Tables 2.1-2.3 and 3.1 of \cite{Ball:2017nwa} for NNPDF 3.1 and 3.0, and internal tables of $\chi^2$ for CT14HERA2 NNLO \cite{Hou:2016nqm}.
It is obvious that variations in $S_E$ are
broader than the standard normal distribution expected in an ideal fit
to all experiments, both in the positive and negative directions. We
can estimate the differences by the mean and standard deviation for
each observed distribution. For CT14HERA2 and MMHT2014 fits, the means
are close to zero, indicating that, while some experiments are not fit
well, the other experiments are fit too well. On the other hand, for
the NNPDF3.0 and NNPDF3.1 analyses, the observed mean is of order 0.8 to 1
-- more experiments are not fitted well than fitted too well. For the
four fits, the probability values for matching the expected ${\cal N} (0,1)$ distribution according to the Anderson-Darling test are
\begin{equation}
\begin{split}
P_\textrm{A-D} ={}& 6.4\times 10^{-3},\hskip 1 cm \textrm{CT14HERA2 NNLO}
\;,
\\
P_\textrm{A-D} ={}& 6.4\times 10^{-3},\hskip 1 cm \textrm{MMHT2014 NNLO}
\;,
\\
P_\textrm{A-D} ={}& 2.6\times 10^{-5},\hskip 1 cm \textrm{NNPDF3.0 NNLO}
\;,
\\
P_\textrm{A-D} ={}& 1.6\times 10^{-5},\hskip 1 cm \textrm{NNPDF3.1 NNLO}
\;.
\end{split}
\label{eq:adproton}
\end{equation}
In all four cases, it is very unlikely that the observed distribution came from the expected Gaussian distribution.

We emphasize that none of the four PDF fits described above is a good fit according to the $P_\textrm{A-D}$ values obtained by breaking the data into smaller data sets, even though each fit is acceptable according to its total $\chi^2$ value. 

The expectation that the $S_E$ distribution should match an $\cN(0,1)$ distribution is based in part on the assumption that the parameters $R_k$ representing imperfections in the theory are negligible in Eq.~(\ref{eq:chiEexpectation}). The evident failure of the distributions in Fig.~\ref{fig:SE} to match $\cN(0,1)$ may indicate that the theory is not precise enough to match very precise experiments. In fact, some  elevated $S_E$ values are contributed by the most precise experiments, such as the combined HERA 1+2 DIS data \cite{Abramowicz:2015mha} and some LHC measurements. These experiments test QCD at unprecedented (NNLO) precision and thus may reveal evidence for new dynamical mechanisms. For instance, $S_E\approx 5.5$ for HERA 1+2 DIS data can be reduced to $S_E \approx 3$ by including small-$x$ resummation in DIS or by evaluating NNLO DIS cross sections with an $x$-dependent factorization scale (see the discussion in Sec.~\ref{sec:specialregions}). Similarly, the description of HERA 1+2 DIS and fixed-target DIS data such as BCDMS \cite{Benvenuti:1989fm} is improved in the NNPDF3.1 analysis as compared to NNPDF3.0 in part by introducing the ``fitted charm'', an independent and possibly process-dependent nonperturbative function that has similarities to power-suppressed (``higher-twist'') terms in DIS. In Sec.~\ref{sec:heavyquarks}, we briefly reviewed the rationale for optionally including the ``fitted charm'' in some PDF fits and the current limitations to its theoretical understanding. 

A different class of concerns arises when too many data sets have negative $S_E$, indicating that the the data sets are systematically described better than would be expected in a good fit in which the data fluctuations around the best-fit theory are random. When too many experiments have negative $S_E$, the reduced global $\chi^2$ may hide some problems, such as overestimated experimental uncertainties or overfitting the statistical fluctuations because of using too flexible theory. Sometimes, a very low global $\chi^2 \ll N_D-N_P$ is taken as evidence for perfect theory that justifies aggressive estimates for PDF uncertainties based exclusively on the $\Delta \chi^2=1$ criterion. In fact, the resulting small PDF uncertainties would be wrong: not only it is improbable that the very low $\chi^2$ is caused by random fluctuations consistently with the experimental errors, the parameterization uncertainty also needs to be estimated by trying other PDF parameterization forms that render $\chi^2$ of up to about $N_D - N_P$.

%--------------------------

\subsection{Test of consistency between experiments}
\label{sec:consistency}

We can carry this analysis further by asking whether different
experiments imposed consistent constraints on PDF parameters $a$.
To this end, we consider an observable $\sigma(a)$ that depends on the
parton parameters, as in Sec.~\ref{sec:crosssection}. The observable
could be a cross section, as suggested by the notation, or, extending
the notion of ``observable'' a bit, it could be the value of the
PDF $f_{a/p}(x,\mu^2)$ for a particular flavor at a
particular momentum fraction $x$ and scale $\mu$.  

As in Sec.~\ref{sec:crosssection}, as long as we consider parameters $a$ that are not far from the best fit parameters $a_\textrm{fit}$, we can apply a linear approximation for the evaluation of $\sigma(a)$, 
\begin{equation}
\label{eq:Deltasigma1}
\sigma(a) = \sigma(a_\textrm{fit}) + (a_\alpha - a^\textrm{fit}_\alpha)\,\sigma'_\alpha
\;,
\end{equation}
with $\sigma'_\alpha = \partial \sigma(a)/\partial a_\alpha$. Furthermore, if we define a special vector $e(\sigma)$ according to Eq.~(\ref{eq:e1forsigma}), 
\begin{equation}
\label{eq:e1forsigmabis}
e(\sigma)_\alpha = 
\frac{H^{-1}_{\alpha \beta}\,\sigma'_\beta}
{\sqrt{\sigma'_\gamma H^{-1}_{\gamma \delta}\sigma'_\delta}}
\;,
\end{equation}
then we found in Sec.~\ref{sec:crosssection} that we can evaluate $\sigma(a) - \sigma(a_\textrm{fit})$ by setting
\begin{equation}
\label{eq:avariatione}
a = a_\textrm{fit} + t\, e(\sigma)
\end{equation}
in Eq.~(\ref{eq:Deltasigma1}). Variations of $a - a_\textrm{fit}$ in orthogonal directions $\{e(\sigma)^{(2)}, e(\sigma)^{(3)}, \dots \} $ do not contribute to $\sigma(a)$. That is
\begin{equation}
\begin{split}
\label{eq:Deltasigmavalue}
&\sigma\!\left(a_\textrm{fit} + t\, e(\sigma) + \sum_{n\ge 2} t_n e(\sigma)^{(n)}\right)
\\&\qquad\qquad 
= 
\sigma(a_\textrm{fit})
+ t\,e(\sigma)_\alpha\sigma'_\alpha
\\&\qquad\qquad =
\sigma(a_\textrm{fit})
+
\sqrt{\sigma'_\gamma H^{-1}_{\gamma \delta}\sigma'_\delta}\
t
\;.
\end{split}
\end{equation}
The result is independent of the parameters $t_n$. That is, $t$ directly measures $\sigma(a)$. Let us denote the corresponding cross section by $\sigma_0(t)$.

The parameters $a_\textrm{fit}$ correspond to the minimum of the global $\chi^2$, so that, according to Eq.~(\ref{eq:chisqshifteda}),
\begin{equation}
\label{eq:chisqshifteda1}
\chi^2 (
D,a_\textrm{fit} + t\, e(\sigma) ) 
= \chi^2(D,a_\textrm{fit}) + t^2
\;.
\end{equation}
Furthermore, if we evaluate Eq.~(\ref{eq:chisqshifteda}) at a general point $a_\textrm{fit} + t\, e(\sigma) + {\scriptstyle \sum_n } t_n e(\sigma)^{(n)}$, we get
\begin{equation}
\begin{split}
\label{eq:chisqshiftedabis}
\chi^2 &\!\left(
D,a_\textrm{fit} + t\, e(\sigma) 
+ \sum_{n=2}^{N_P} t_n e(\sigma)^{(n)} \right) 
\\ &\qquad = \chi^2(D,a_\textrm{fit}) + t^2
+ \sum_{n=2}^{N_P} t_n^2
\;.
\end{split}
\end{equation}
If we regard parameter points $a$ as distributed at random according to a probability density proportional to $\exp(-(\chi^2- \chi^2_\textrm{min})/2)$, then, to find the probability $\rho$ for $a$ to lie in a plane of constant $\sigma(a) = \sigma_0(t)$, we simply integrate over the other variables $t_n$:
\begin{equation}
\begin{split}
\rho ={}& (2\pi)^{-N_P/2} \int\!dt_2 \cdots dt_{N_P}\
\exp\!\big[-(t^2
+ \sum_{n\geq 2} t_n^2)/2\big]
\\
={}&
(2\pi)^{-1/2}\,\exp[-t^2/2]
\\
={}& (2\pi)^{-1/2}\,
\exp\!\left[-\left(\chi^2(D,a_\textrm{fit} + t\, e) - \chi^2(D,a_\textrm{fit}\right)/2)\right]
.
\end{split}
\end{equation}
That is, $\chi^2(D,a_\textrm{fit} + t\, e(\sigma))$ gives both the probability for $a$ to lie at position $t$ along the line $a_\textrm{fit} + t\, e(\sigma)$ and the probability for $a$ to lie in the plane $\sigma(a) = \sigma_0(t)$ that intersects this line at position $t$.

It may be useful to note that one can find the direction of $e(\sigma)$ quite  simply. Up to its normalization, $e(\sigma)$ is the vector from $a_\textrm{fit}$ to the point on the surface $\sigma(a) = \sigma_0(t)$ that minimizes $\chi^2$ on this surface. The standard Lagrange multiplier method \cite{Stump:2001gu} produces this vector.

After this introduction, let us explore the role of a single
experiment, $E$, in the fit. Consider $\chi^2(D,a)$ for $a$ that
varies along the line $a = a_\textrm{fit} + t\, e(\sigma)$. The
parameter $t$ labels distance along this line. We use one of three
sets of data $D$: either all of the data, $D(\textrm{all})$, or all of
the data except for the data from experiment $E$, $D({\textrm{no
  }E})$, or the data from experiment $E$ alone, $D(E)$. We are
interested in how the function $\chi^2(D,a_\textrm{fit} + t\,
e(\sigma))$ depends on $t$ when we make these different choices for
what data set $D$ we use in computing $\chi^2$.\footnote{The
  dependence of $\chi^2$ on the position of the parameters $a$ along
  the line $a_\textrm{fit} + t\, e(\sigma)$ near $a = a_\textrm{fit}$ is easily determined. The dependence on $a$ over the whole plane $\sigma(a) = \sigma_0$ would require knowing $\chi^2$ in the entire parameter space. This is not so meaningful if we use just a small subset of the data, $D(E)$.}

\begin{figure*}%--------------------------
\begin{center} 
\includegraphics[width=0.48\textwidth]{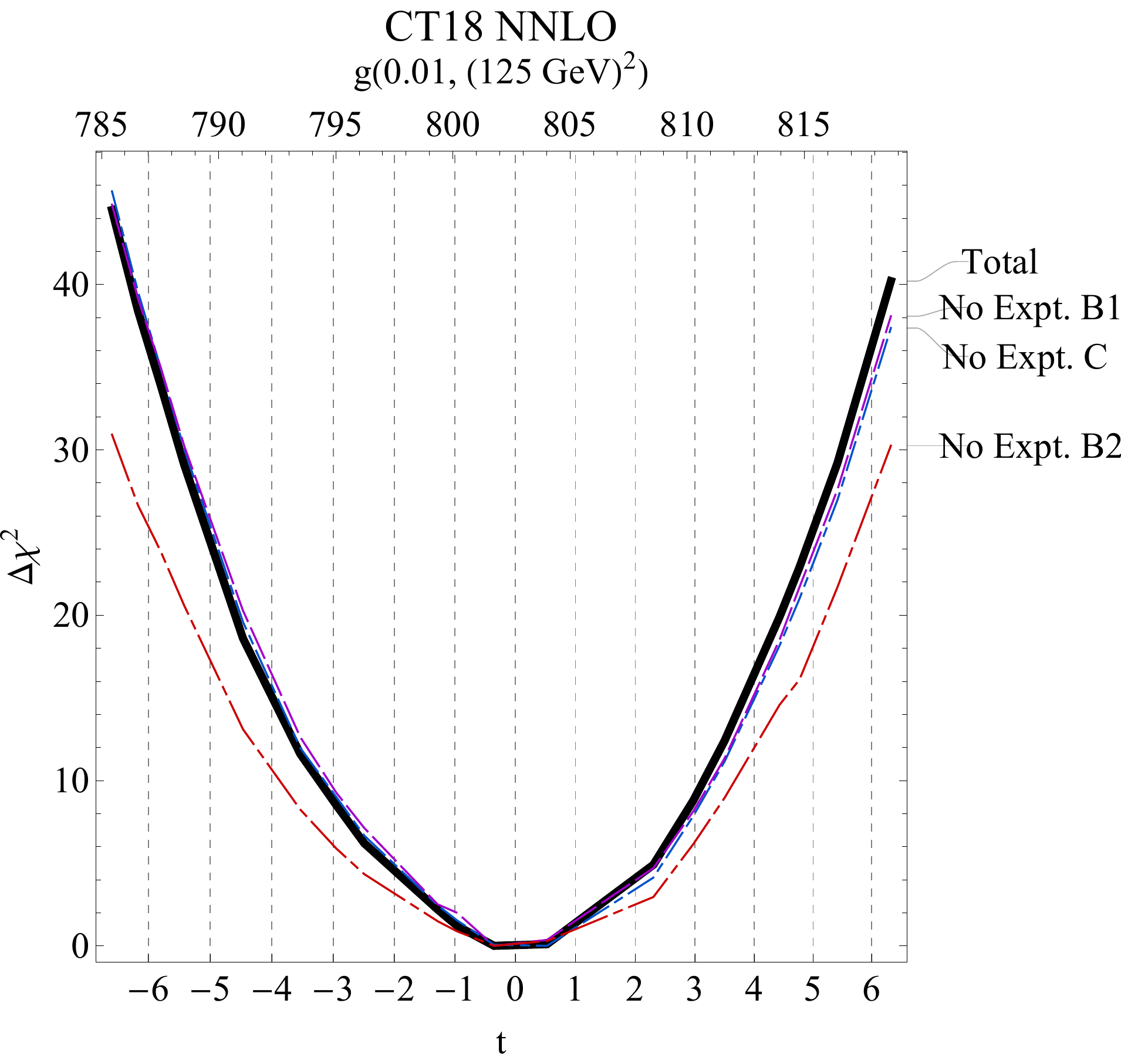}\quad
\includegraphics[width=0.48\textwidth]{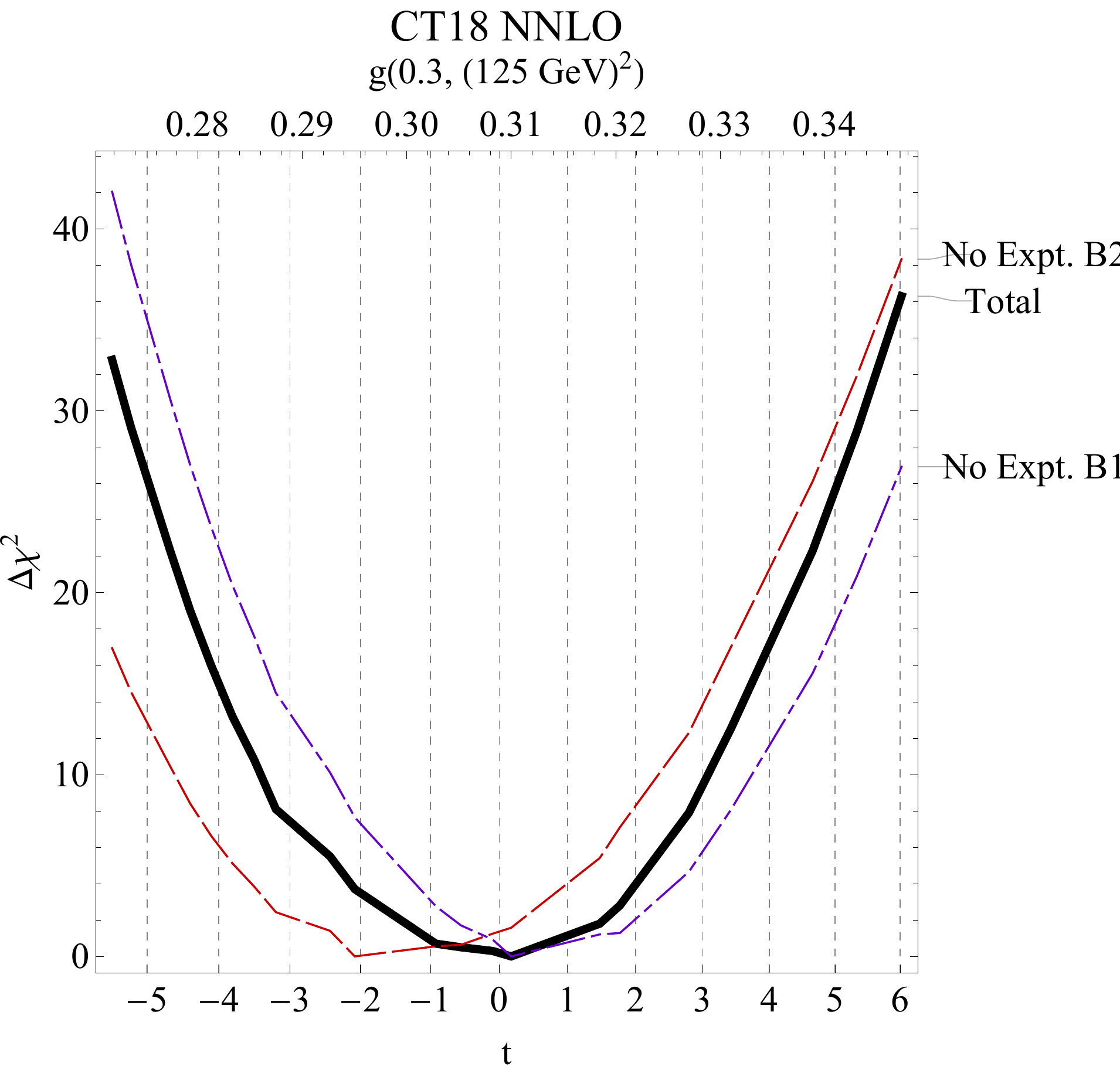}
\\ (a)\hspace{3.4in}(b)
\end{center}
\caption{(a) Dependence of $\chi^2$ in a CT18 NNLO fit as a function of distance $t$ in parameter space corresponding to changes in $g(0.01,(125\mbox{ GeV})^2)$. The black solid curve shows the total $\chi^2$, while the remaining three curves show $\chi^2$ as a function of $t$ with particular experiments removed from the data set. 
(b) $\chi^2$ as before, but along a line corresponding to changes in $g(0.3,(125\mbox{ GeV})^2)$.
\label{fig:EMakesDifference}}
\end{figure*}%--------------------------

We will ask two questions concerning the role of experiment $E$ in determining $t$. 

The first question is ``Does experiment $E$ exert a substantial pull on the value of $t$?'' To answer this question, we ask what would happen if we omitted the data from experiment $E$ from the evaluation of $\chi^2$. Then the minimum value of $\chi^2\left(D({\textrm{no }E}), a_\textrm{fit} + t\, e(\sigma)\right)$ will occur at a value $t({\textrm{no }E})$ that will typically be different from the value $t({\textrm{all}}) = 0$ that we get using all of the data. The $1\sigma$ uncertainty in $t({\textrm{all}})$ is $\Delta t({\textrm{all}}) = 1$. If the difference between $t(\textrm{no }E)$ and $t(\textrm{all}) = 0$ is smaller than this uncertainty, then we may conclude that experiment $E$ does not exert a substantial pull on $t$. That is, for experiment $E$ to substantially pull $t$, we need
\begin{equation}
\label{eq:Ematters}
|t(\textrm{no }E)| > f
\;,
\end{equation}
where $f$ is a parameter we could pick, perhaps $f = 1$.

This is illustrated in Fig.~\ref{fig:EMakesDifference}, which is based on the CT18 NNLO fit \cite{Hou:2019efy}. For this illustration, we choose a very conservative value of $f$, $f = 0.5$.
 
In panel (a), we choose the gluon distribution at $x  = 0.01$ and $\mu = 125 \mbox{ GeV}$ as our observable $\sigma$. The heavy black curve is the difference of $\chi^2(D(\textrm{all}),a_\textrm{fit} + t\, e(\sigma))$ and its minimum value as a function of the parameter $t$. The corresponding values of $g(0.01,125\mbox{ GeV})$ are shown along the top of the plot. We also show curves for the differences of $\chi^2(D(\textrm{no }E),a_\textrm{fit} + t\, e(\sigma))$ and their minimum values for three choices of data sets $E$, labeled experiments B1, B2, and C. [The experiments are taken from an actual global fit.]  We see that the minima of all of these curves lie in the range $-0.5 < t < 0.5$, indicating that none of these data sets exerts a substantial pull on $t$ in the sense of Eq.~(\ref{eq:Ematters}) with $f = 0.5$.

In panel (b), we choose the gluon distribution at $x  = 0.3$ and $\mu
= 125 \mbox{ GeV}$ as our observable $\sigma$. Again, the heavy black
curve is constructed from $\chi^2(D(\textrm{all}),a_\textrm{fit} + t\,
e(\sigma))$.  We also show curves for $\chi^2(D(\textrm{no
}E),a_\textrm{fit} + t\, e(\sigma))$ for two choices of data sets $E$:
experiments B1 and B2. These correspond to two data sets obtained from
the same experiment B for two different collision energies
$\sqrt{s}$. We see that the minima of these curves lie outside the
range $-0.5 < t < 0.5$, indicating that both of these data sets  exerts a substantial pull on $t$ in the sense of Eq.~(\ref{eq:Ematters}) with $f = 0.5$. 

\begin{figure}%--------------------------
\begin{center} 
\includegraphics[width=0.48\textwidth]{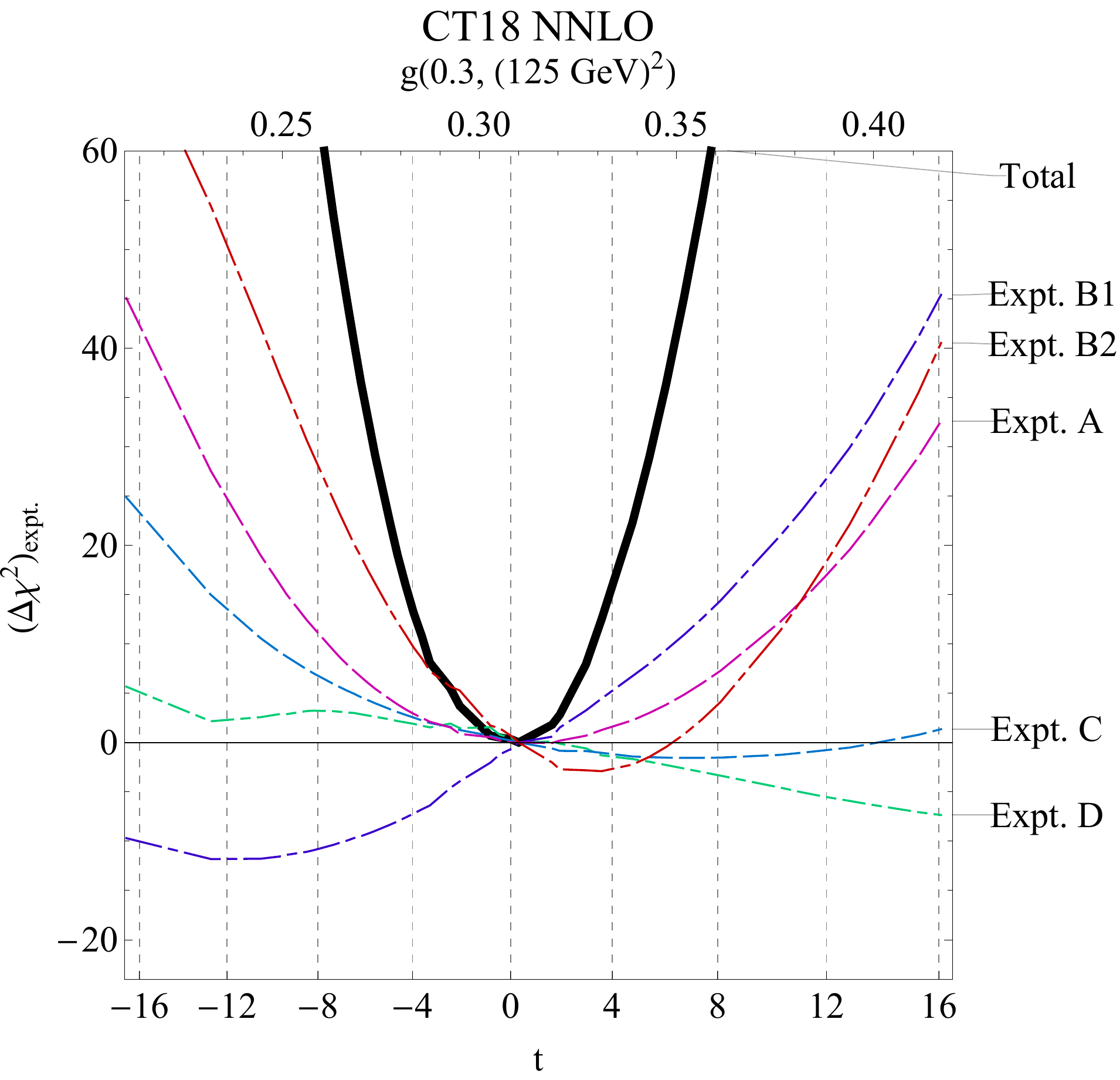}
\end{center}
\caption{Dependence of $\chi^2$ in a CT18 NNLO fit as a function of distance $t$ in parameter space corresponding to changes in $g(0.3,(125\mbox{ GeV})^2)$. The black solid curve shows the total $\chi^2$, while the remaining curves show $\chi^2$ as a function of $t$ with some particular experiments $E$ alone.
\label{fig:Einconsistent}}
\end{figure}%--------------------------

Now suppose that experiment $E$ {\em does} exert a substantial pull on $t$. Then we need to check whether the global fit solution, $t(\textrm{all})$, is consistent with what experiment $E$ says. Let us define
\begin{equation}
\Delta \chi^2_E(t) = 
\chi^2(D(E),a_\textrm{fit} + t\, e(\sigma)) - \chi^2(D(E),a_\textrm{fit})
\;.
\end{equation}
According to experiment $E$ alone, the best fit $t(E)$ is obtained by minimizing $\Delta \chi^2_E(t)$. The $1\sigma$ uncertainty range for $t(E)$ is given by $|\Delta \chi^2_E(t) - \Delta \chi^2_E(t(E)))| = 1$. Thus the result $t = 0$ from the full fit is consistent with the result from experiment $E$ alone if
\begin{equation}
\label{eq:Einconsistent}
|\Delta \chi^2_E(0) - \Delta \chi^2_E(t(E))|
< n^2
\;,
\end{equation}
where $n$ is a parameter that we could pick, perhaps $n = 2$ for consistency within a 95\% confidence interval.

This is illustrated in Fig.~\ref{fig:Einconsistent}, again based on
the CT18 NNLO fit and the observable $\sigma = g(0.3,
(125 \textrm{ GeV})^2)$. In Fig.~\ref{fig:EMakesDifference}(b), we saw
that two data sets, B1 and B2, make at least a marginal difference in the overall
fit. Now we plot $\Delta \chi^2_\textrm{Total}(t)$ for the overall fit as a heavy black line and also $\Delta \chi^2_E(t)$ for $E = $ B1 and $E = $ B2: the data sets obtained in the same experiment B that was repeated at two collider energies $\sqrt s$.  We also exhibit the $\Delta \chi^2_E(t)$ curves for three other data sets. We see that $\Delta \chi^2_\mathrm{B1}(t)$ is about 10 units higher at $t = 0$ than it is at its minimum. Thus the consistency condition (\ref{eq:Einconsistent}) with $n = 2$ is violated for experiment B1. On the other hand, $\Delta \chi^2_\mathrm{B2}(t)$ is only about 3 units higher at $t = 0$ than it is at its minimum. Thus the consistency condition (\ref{eq:Einconsistent}) with $n = 2$ is satisfied for experiment B2. 

\begin{figure*}%--------------------------
\begin{center}
\includegraphics[width=0.48\textwidth]{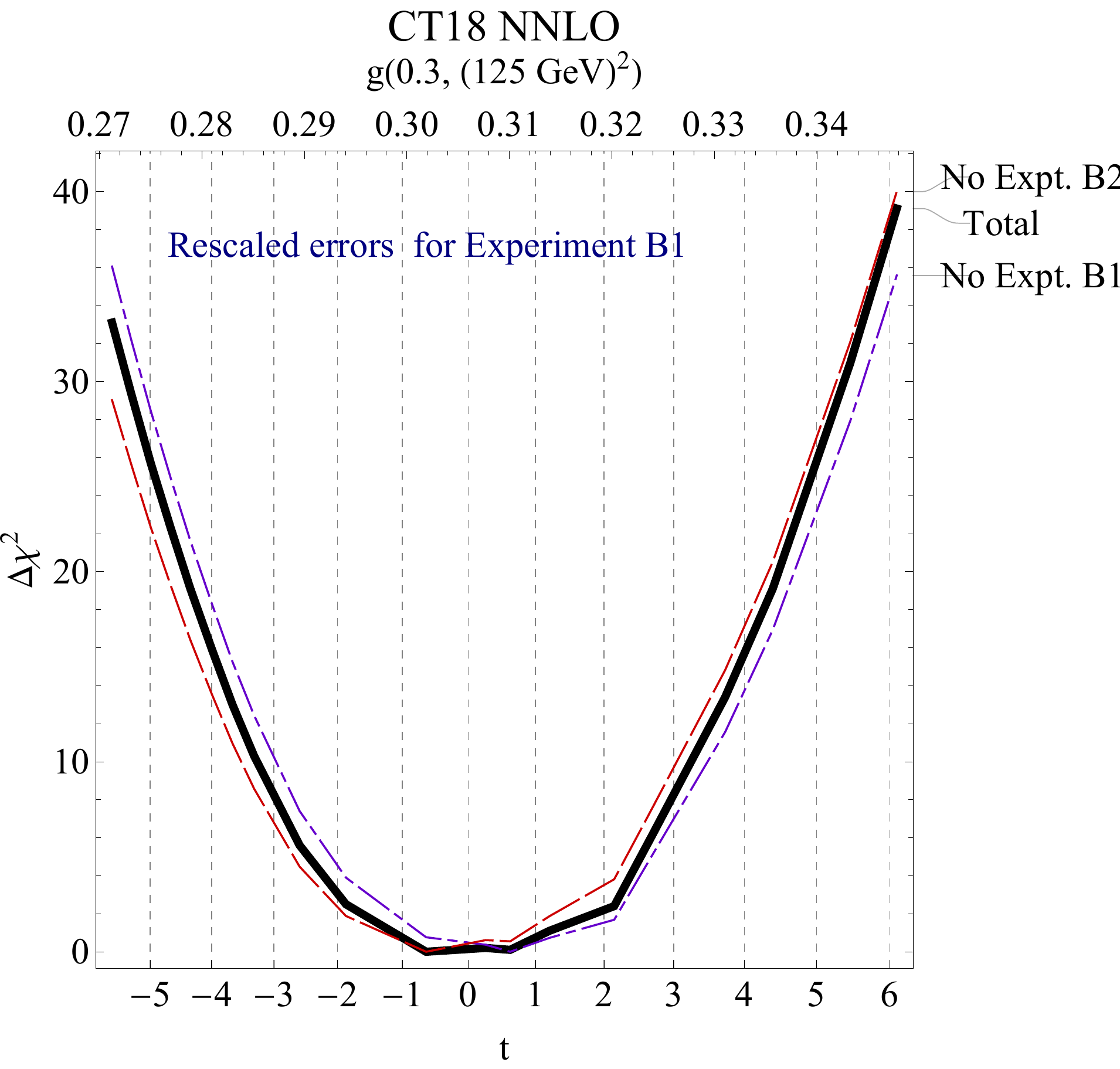}\quad
\includegraphics[width=0.48\textwidth]{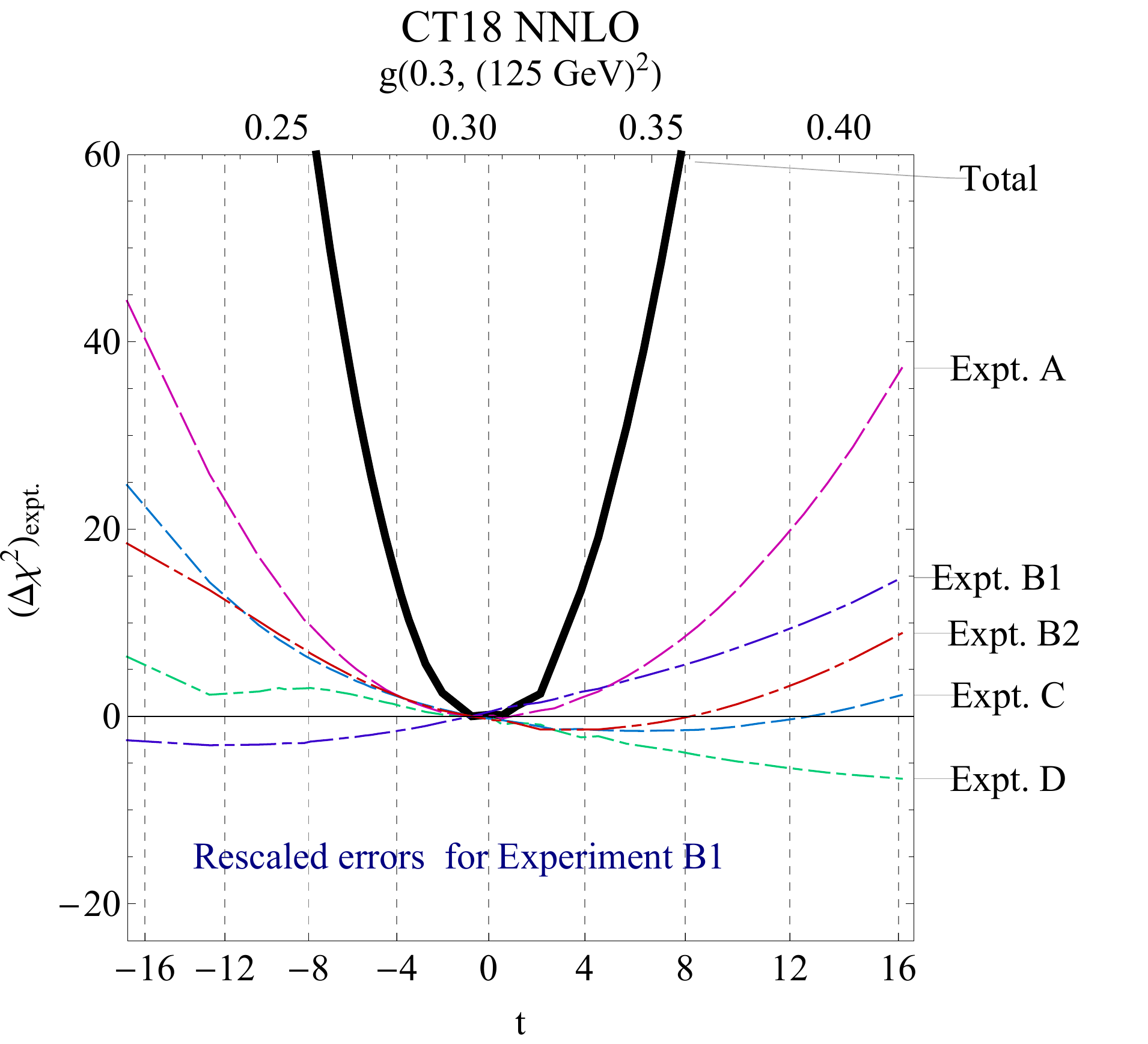}\\ (a)\hspace{3.4in}(b)
\end{center}
\caption{Same as
  Figs.~\ref{fig:EMakesDifference}(b) and \ref{fig:Einconsistent}, if
  the estimated errors for the experiment B1 are multiplied by $\sqrt{2}$.
\label{fig:Ebetter}}
\end{figure*}

At this level of inconsistency for experiment B1, it is not credible that we are simply looking at statistical fluctuations. One simple but crude way to remove the inconsistency would be to increase the error estimates for the discrepant data set(s). To illustrate how this might work, in Fig.~\ref{fig:Ebetter}, we have refitted the PDFs by assuming increased quoted errors for the data set B1. Namely, we refitted after multiplying the quoted errors of B1 by a constant factor $\sqrt{2}$. That is, in Eq.~(\ref{eq:chisqDa}), we multiply $C_{ij}$ for $i,j \in B1$ by a common factor $1/2$.

Fig.~\ref{fig:Ebetter}(a) shows differences of $\chi^2$ from their minimum values as functions of $t$ for the the full data set $D(\textrm{all})$ and then for $D(\textrm{no }E)$ with $E = \mathrm{B1}$ (with the rescaled errors) and B2, as in Fig.~\ref{fig:EMakesDifference}(b). In general, we would expect that increasing the estimated errors from certain data sets would change the position of the best fit for the observable $\sigma$ and increase the estimated error on the prediction for $\sigma$. In this case, for $g(0.3, (125 \textrm{ GeV})^2)$, neither the position of the minimum nor the estimated error changes by much.

In Fig.~\ref{fig:Ebetter}(a), we see that the minima of the two curves occur well within the region $-f < t < f$, even for $f = 0.5$. Since the criterion (\ref{eq:Ematters}) no longer indicates that these two data sets exert substantial pulls on $t$, we need not examine the criterion (\ref{eq:Einconsistent}) for a discrepancy between a data set and the overall fit. 

If we do examine the criterion (\ref{eq:Einconsistent}), we obtain Fig.~\ref{fig:Ebetter}(b), where we show $\Delta \chi^2_E(t)$ for the the full data set $D(\textrm{all})$ and then for $D(E)$ with $E = \mathrm{B1}$ (rescaled errors) and B2. We see that $\Delta \chi^2_E(t)$ is less than four units higher at $t = 0$ than it is at its minimum both for B1 and B2. These represent less than $2\,\sigma$ discrepancies, which are not nearly as alarming as the discrepancy for experiment B1 that we saw in Fig.~\ref{fig:Einconsistent}. 

In summary, this analysis gives us criteria for checking whether there is a problem associated with the data from experiment $E$ in determining $t$. There is a problem if experiment $E$ exerts a substantial pull on the value of $t$, Eq.~(\ref{eq:Ematters}), and if the fit based on just experiment $E$ is inconsistent with the global fit, Eq.~(\ref{eq:Einconsistent}). There is one set of criteria for each independent direction $e(\sigma)$ corresponding to an observable $\sigma$ and for each experiment $E$.  

We explored how one can make the results from a data set $E$ more consistent with the rest of the data by simply rescaling the errors for this data set. This is a very crude method. We do not recommend using it for finding the best fit. In the following section, we explore a more subtle method.

A less precise alternative is to leave the disagreeing experiment(s) and best fit based on these experiment(s) unchanged, but increase the PDF uncertainty to reflect the incompatibility in the experimental constraints. This possibility is discussed in Sec.~\ref{sec:tolerance}.

%--------------------------

\subsection{A more conservative way to adjust the errors}
\label{sec:conservativeErrorAdjustment}

Suppose that the experiment $E$ exerts a substantial pull on the fit value of the observable $\sigma$ according to the criterion (\ref{eq:Ematters}),  and $\chi^2$ for data set $E$ along the line $a = a_\textrm{fit} + t\, e(\sigma)$ is  not consistent according to the criterion (\ref{eq:Einconsistent})
with its value at $t = 0$, the best-fit value of $t$ according to the
fit to all data. Then it may be helpful to increase the experimental
errors for experiment $E$. We stated that simply increasing the total
error estimate for experiment $E$ by dividing $C_{ij}$ for $i,j \in E$
by a common factor is a rather crude strategy. A more focused strategy
would be to add a systematic error of the form~(\ref{eq:Dkdef}) such
that it resolves the inconsistency of the constraints on our
observable $\sigma$, but does not affect the fit in any other way.
We parameterize this systematic error as
\begin{equation}
\label{eq:newbeta}
\sigma_k\beta_{k,\textrm{new}} = \xi \bar \beta_k
\;.
\end{equation}
where $\xi$ is a constant that we can adjust, and
\begin{equation}
\label{eq:barbetadef}
\bar \beta_k = 
\theta(k\in E)\, 
\frac{T_{k \beta}\, e(\sigma)_\beta}
{[e(\sigma)_\alpha H(E)_{\alpha\beta} e(\sigma)_\beta]^{1/2}}
\;.
\end{equation}
In the normalization factor, $H(E)_{\alpha\beta}$ is the Hessian matrix, as in Eq.~(\ref{eq:Halphabeta}), but only including the data from data set $E$:
\begin{equation}
\label{eq:HE0}
H(E)_{\alpha\beta} = 
\sum_{i,j \in E}
T_{i\alpha}T_{j\beta} C_{ij}
\;.
\end{equation}
With this factor, $\bar \beta_k$ is independent of the normalization of the vector $e(\sigma)$. The vector $e(\sigma)$ does have a definite normalization (\ref{eq:e1forsigmabis}), but in Eq.~(\ref{eq:newbeta}), only the direction of $e(\sigma)$ matters. A simple relation for the normalization factor $e(\sigma)_\alpha H(E)_{\alpha\beta} e(\sigma)_\beta$ is given below in Eq.~(\ref{eq:betanormalization}).

Let us examine whether adding a new systematic error of this form can repair the incompatibility between data set E and the rest of the data while not much affecting the remaining fit.

From Eq.~(\ref{eq:Cinversedef}) with the systematic error (\ref{eq:newbeta}) added, the covariance matrix becomes
\begin{equation}
\label{eq:Cinversedefbis}
C(\xi^2)^{-1}_{ij} = \sigma_i\sigma_j \delta_{ij} 
+ \sum_J \sigma_i\beta_{i J}\, \sigma_j\beta_{j J} 
+ \xi^2 \bar \beta_i \bar \beta_j
\;.
\end{equation}
Thus
\begin{equation}
\frac{d}{d\xi^2}\,C(\xi^2)^{-1}_{ij}
=
\bar \beta_i \bar \beta_j
\;.
\end{equation}
Using $dC/d\xi^2 = - C\, [dC^{-1}/d\xi^2]\, C$, this is
\begin{equation}
\frac{d}{d\xi^2}\,C(\xi^2)_{ij}
=
- \sum_{k}C(\xi^2)_{ik}\bar \beta_k
\sum_{l }C(\xi^2)_{jl}\bar \beta_l
\;.
\end{equation}
It is straightforward to solve this differential equation to obtain
\begin{equation}
\label{eq:Cxiresult0}
C(\xi^2)_{ij} = C_{ij}
- \frac{\xi^2 \sum_{k}C_{ik}\bar \beta_k\ 
\sum_{l }C_{jl}\bar \beta_l}
{1 + \xi^2 \sum_{kl} \bar \beta_k C_{kl}  \bar\beta_l}
\;.
\end{equation}
Here $C_{ij} = C(0)_{ij}$ is the covariance matrix without the added systematic error. With the definition Eq.~(\ref{eq:barbetadef}) of $\bar\beta_k$, we have
\begin{equation}
\sum_{kl} \bar \beta_k C_{kl}  \bar\beta_l = 1
\;,
\end{equation}
so
\begin{equation}
\label{eq:Cxiresult}
C(\xi^2)_{ij} = C_{ij}
- \frac{\xi^2}{1 + \xi^2}\,
\sum_{k}C_{ik}\bar \beta_k\ 
\sum_{l }C_{jl}\bar \beta_l
\;.
\end{equation}

What is the effect on $\chi^2$ for experiment $E$ of adding this systematic error? Consider $\chi^2$ for experiment $E$, for parameters
\begin{equation}
a_\alpha = a^\textrm{fit}_\alpha + t\,e_\alpha
\;,
\end{equation}
where the $a^\textrm{fit}_\alpha$ are the parameters from the global fit before adding the extra systematic error, and the vector $e$ could be the special vector $e(\sigma)$ for observable $\sigma$, but could also be any other vector in the many dimensional space of parameters. We have, from Eqs.~(\ref{eq:chisqDa}) and (\ref{eq:theoryT1}),
\begin{equation}
\begin{split}
\label{eq:chisqDEa}
\chi^2(D&(E),a_\textrm{fit} + t e,\xi) 
\\& = 
\sum_{i,j \in E}
[D_i - T_i(a_\textrm{fit})- t\, T_{i\alpha} e_\alpha]
\\&\qquad\quad\ \times
[D_j - T_j(a_\textrm{fit}) - t\, T_{j\beta} e_\beta]\, 
C_{ij}(\xi)
\;.
\end{split}
\end{equation}
We can write this as
\begin{equation}
\begin{split}
\label{eq:chisqDEaexpanded}
\chi^2(D(E)&,a_\textrm{fit} + t e,\xi) 
\\={}&
\chi^2(D(E),a_\textrm{fit},\xi)
- 2 t B(E,\xi)_\beta e_\beta
\\&
+ t^2 e_\alpha  H(E,\xi)_{\alpha\beta} e_\beta
\;,
\end{split}
\end{equation}
where
\begin{equation}
B(E,\xi)_\beta = 
\sum_{i,j \in E}
[D_i - T_i(a_\textrm{fit})]
C_{ij}(\xi) T_{j\beta}
\end{equation}
gives the contribution linear in $t$ and
\begin{equation}
H(E,\xi)_{\alpha\beta} = 
\sum_{i,j \in E}
T_{i\alpha}T_{j\beta} C_{ij}(\xi)
\end{equation}
is the Hessian matrix including the added systematic error, but just
for the data from experiment $E$.
Following the notation from
Eq.~(\ref{eq:HE0}),
we  define
\begin{equation}
H(E)_{\alpha\beta} \equiv H(E,0)_{\alpha\beta}, \quad B(E)_{\alpha\beta} \equiv B(E,0)_{\alpha\beta}.
\end{equation}

Without the added systematic error, we have
\begin{equation}
\begin{split}
\label{eq:chisqDEaexpanded0}
\chi^2(D(E)&,a_\textrm{fit} + t e,0) 
\\={}&
\chi^2(D(E),a_\textrm{fit},0)
- 2 t B(E)_\gamma e_\gamma
\\&
+ t^2\, e_\alpha H(E)_{\alpha\beta} e_\beta
\;.
\end{split}
\end{equation}

When we add the new systematic error, the result changes to
\begin{align}
\label{eq:chisqDEaexpanded1}
\chi^2(D(E)&,a_\textrm{fit} + t e,\xi) 
\notag
\\={}& \chi^2(D(E),a_\textrm{fit} + t e,0)
\\&
- \frac{\xi^2}{1 + \xi^2}\,
\frac{\left[B(E)_\beta\, e(\sigma)_\beta\right]^2}
{e(\sigma)_\alpha H(E)_{\alpha\beta} e(\sigma)_\beta}
\notag
\\&
+ 2t\,  \frac{\xi^2}{1 + \xi^2}
\frac{[B(E)_{\gamma} e(\sigma)_\gamma] 
[e(\sigma)_\alpha H(E)_{\alpha\beta} e_\beta]}
{e(\sigma)_\alpha H(E)_{\alpha\beta} e(\sigma)_\beta}
\notag
\\&
- t^2\,\frac{\xi^2 }{1 + \xi^2}\,
\frac{[e(\sigma)_\alpha H(E)_{\alpha\beta} e_\beta]^2}
{e(\sigma)_\alpha H(E)_{\alpha\beta} e(\sigma)_\beta}
\;.
\notag
\end{align}
Here we used Eq.~(\ref{eq:Cxiresult}) to separate from
$\chi^2(D(E),a,0)$ the extra terms resulting from the new systematic
error (proportional to $\xi^2$). 

We will now show that the new systematic error reduces an apparent
tension between data set $E$ and the rest of the data along the line
associated with $\sigma$ (our observable of interest). It does not
modify constraints in the other directions. In the last two terms of
Eq.~(\ref{eq:chisqDEaexpanded1}), the numerators contain
$[e(\sigma)_\alpha H(E)_{\alpha\beta} e_\beta]$, the  inner product
between the unit vector $e(\sigma)$ defining the direction associated with $\sigma$, and another (possibly
orthogonal) unit vector $e$ that defines the line $a_{fit}+t e$ along
which we choose to scan $\chi^2(D(E),a_\textrm{fit} + t e,\xi)$. When
we scan along the direction $e = e(\sigma)$, the $\chi^2(D(E),a_\textrm{fit} + t e,\xi)$ function
changes as
\begin{equation}
\begin{split}
\label{eq:chisqDEaexpanded2}
\chi^2(D(E)&,a_\textrm{fit} + t e(\sigma),\xi) 
\\={}& \chi^2(D(E),a_\textrm{fit},0)
\\&
- \frac{\xi^2}{1 + \xi^2}\,
\frac{\left[B(E)_\beta\, e(\sigma)_\beta\right]^2}
{e(\sigma)_\alpha H(E)_{\alpha\beta} e(\sigma)_\beta}
\\&
- 2t\,  \frac{1}{1 + \xi^2}
[B(E)_{\gamma} e(\sigma)_\gamma] 
\\&
+ t^2\,\frac{1}{1 + \xi^2}\,
[e(\sigma)_\alpha H(E)_{\alpha\beta} e(\sigma)_\beta]
\;.
\end{split}
\end{equation}
If we turn off the systematic error (set $\xi=0$) for a moment,
from Eq.~(\ref{eq:chisqDEaexpanded2}) we can numerically find the factor
$e(\sigma)_\alpha H(E)_{\alpha\beta} e(\sigma)_\beta$ that appears
here and in the definition (\ref{eq:barbetadef}): 
\begin{equation}
\label{eq:betanormalization}
e(\sigma)_\alpha H(E)_{\alpha\beta} e(\sigma)_\beta
=
\frac{1}{2}\, \frac{d^2}{dt^2}\,
\chi^2(D(E),a_\textrm{fit} + t e(\sigma),0)
\;.
\end{equation}

If we turn the systematic error back on by choosing $\xi\neq 0$, 
the second term makes $\chi^2$ at $t = 0$ smaller as $\xi$ increases
without affecting the shape of $\chi^2$ as a function of $t$. In the
remaining terms, the coefficients of $t$ and $t^2$ are reduced by the
same factor as $\xi$ increases. The shape of $\chi^2$ versus $t$
simply becomes shallower as desired,
reducing the tension between data set $E$ and the rest of
the data along the line $a = a_\textrm{fit} + t e(\sigma)$. 

\begin{figure}%--------------------------
\begin{center}
\includegraphics[width=0.48\textwidth]{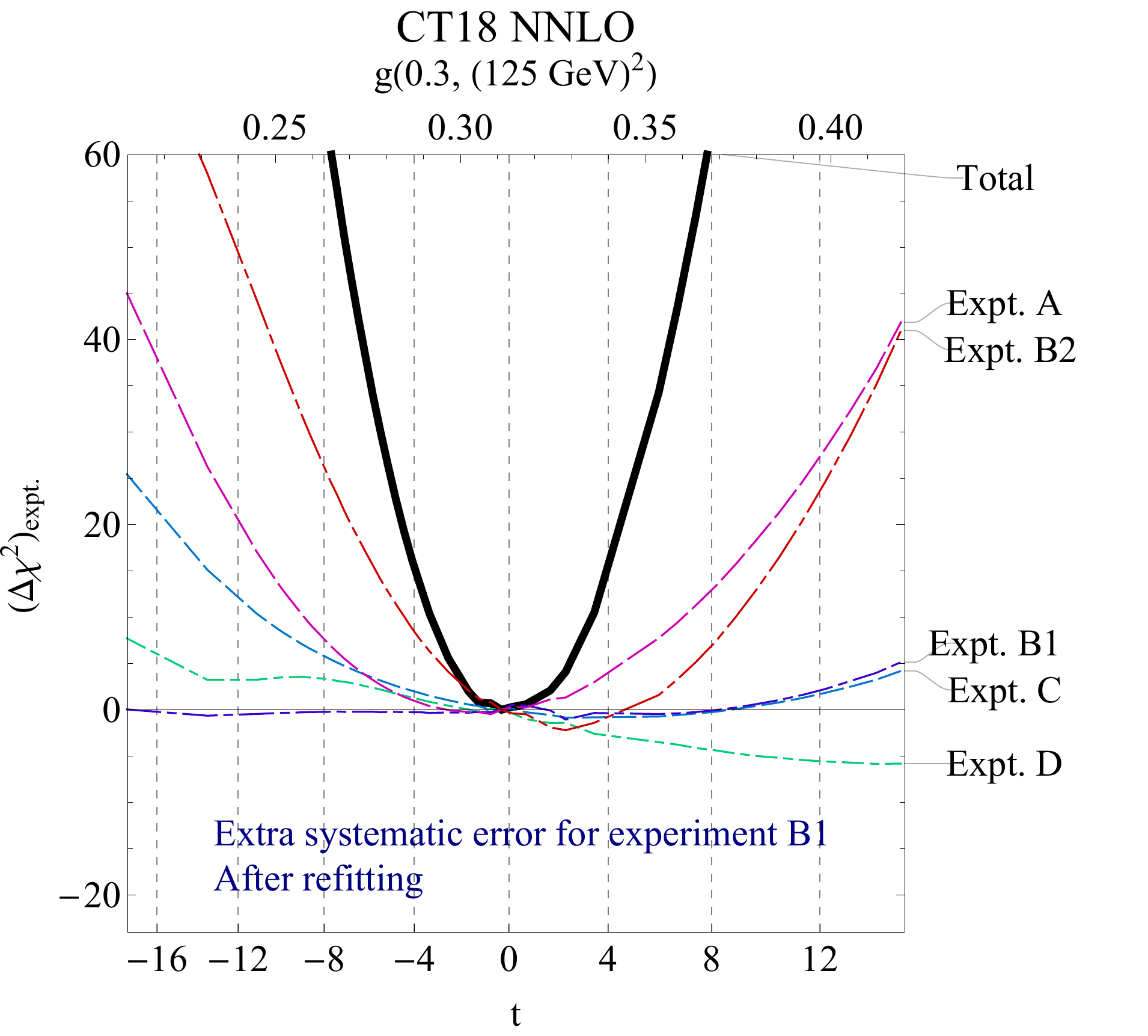}
\end{center}
\caption{$\chi^2$ curves for individual experiments as in
  Fig.~\ref{fig:Einconsistent} but with an extra systematic error
  added for experiment B1 according to Eqs.~(\ref{eq:newbeta}) and (\ref{eq:barbetadef}). The fit has been repeated with the new systematic error for experiment B1. The new fit gives a new best-fit choice $a_\mathrm{fit}$. Now the observable $g(0.3,(125 \GeV)^2)$ defines a new direction $e(\sigma)$ in parameter space. This plot uses the new $a_\mathrm{fit}$, $e(\sigma)$, and total $\chi^2$ values after the fit.
\label{fig:chisqnewfitB1}}
\end{figure}%--------------------------

On the other hand, in the directions that are orthogonal to the
direction of variation of $\sigma$, the unit vectors $e$ obey
\begin{equation}
e(\sigma)_\alpha H(E)_{\alpha\beta} e_\beta = 0
\;.
\end{equation}
If, for example, the parameter space is 26-dimensional, then there is a 25-dimensional vector space in which $e$ could lie and satisfy this condition. In this case, Eq.~(\ref{eq:chisqDEaexpanded1}) gives
\begin{equation}
\begin{split}
\label{eq:chisqDEaexpanded3}
\chi^2(D(E)&,a_\textrm{fit} + t e,\xi) 
\\={}& \chi^2(D(E),a_\textrm{fit} + t e,0)
\\&
- \frac{\xi^2}{1 + \xi^2}\,
\frac{\left[B(E)_\beta\, e(\sigma)_\beta\right]^2}
{e(\sigma)_\alpha H(E)_{\alpha\beta} e(\sigma)_\beta}
\;.
\end{split}
\end{equation}
That is, $\chi^2$ for data set $E$ at $t = 0$ is smaller when $\xi
>0$, but the shape of $\chi^2$ as a function of $t$ is not changed at
all.  Thus the new systematic error is a conservative choice in that
it alleviates the incompatibility problem while having a minimal
effect on the rest of the fit. 

Let us see how this prescription resolves the tension between
experiment B1 and other experiments that we observed in 
Fig.~\ref{fig:Einconsistent}. There we examined the parameter fit
along a direction $e(\sigma)$ corresponding to $\sigma = g(0.3,(125
\GeV)^2)$. We saw that the curve of 
\begin{equation}
\Delta \chi^2_E(t) 
=\chi^2(D(E),a_\textrm{fit} + t\, e(\sigma)) - \chi^2(D(E),a_\textrm{fit})
\end{equation}
for a certain experiment $E = \mathrm{B1}$ was not consistent with the
choice of $t=0$ that minimizes the total $\chi^2$. To alleviate this
problem, we add the new systematic error (\ref{eq:newbeta}) for
experiment B1 with $\xi$ defined by $1/(1+\xi^2) \approx 4/10$.
With this error added, the new $\Delta \chi^2_\textrm{B1}(t)$ for B1 now satisfies
\begin{equation}
\Delta \chi^2_\textrm{B1,new}(t) = 
\frac{\Delta \chi^2_\textrm{B1,old}(t)}{1+ \xi^2}
\
\approx \frac{4}{10}\,\Delta \chi^2_\textrm{B1,old}(t)
\;.
\end{equation}
The new $\Delta \chi^2_\textrm{B1}(t)$ curve becomes flat enough so
that consistency with the rest of the data along the line
$a_\textrm{fit} + t\, e(\sigma)$ is no longer a problem. 

We can now perform the global fit again, with the modified systematic
error for experiment B1. Then the best fit parameters $a_\mathrm{fit}$
change. The direction vector $e(\sigma)$ corresponding to the
observable $g(0.3, (125 \GeV)^2$ also changes. The new fit gives us a
new plot analogous to Fig.~\ref{fig:Einconsistent} in which all of the
$\chi^2$ curves have changed. The result is shown in
Fig.~\ref{fig:chisqnewfitB1}. In the new fit, $\chi^2$ for experiment
B1 is very flat, indicating that experiment B1 is now not
significantly affecting the determination of $g(0.3,(125
\GeV)^2)$. The best fit value of $g(0.3,(125 \GeV)^2)$ has changed
from 0.309 to 0.312. The estimated error on the fit value of
$g(0.3,(125 \GeV)^2)$, determined by the second derivative of the
total $\chi^2$ curve with respect to $g(0.3,(125 \GeV)^2)$, is about
2\% larger. 

\subsection{Summary of measures of goodness of fit}

The statistical analysis that we have presented in Sec.~III relies on several assumptions. One assumes that experimental systematic errors are adequately represented by a Gaussian distribution described by a covariance matrix given by the experiments. Typically, one also assumes that theory errors are small enough to be neglected. Additionally, one assumes that the theoretical prediction is, to a good approximation, a linear function of the PDF parameters in the region $a_\textrm{fit} \pm \delta a$ near the best fit parameter choice. Some of these assumptions could be wrong.

We have argued that we should not simply trust these assumptions, but should test them by using a strong set of goodness-of-fit criteria. Taken together, these tests are much more stringent than that obtained by simply noting the global $\chi^2$ value. If the fit passes all of these tests, we can have some confidence in the results and the errors on the results. If the fit does not pass all tests, then remediation is needed. We do not offer a fixed prescription, but we have pointed out some possibilities.

The choice that has usually been made is to leave the fit as it is but use larger error estimates on the final PDFs than those found with the 
parameter-fitting criterion that requires $\Delta \chi^2=4$ at the 95\%
probability level. Estimation of trustworthy PDF errors in such imperfect
situation can be difficult and sometimes controversial. Before
a new generation of PDFs is published, it may undergo many months of multifaceted PDF testing in order to establish the realistic estimates for PDF uncertainties. The
increase over the nominal PDF errors that results from this procedure
is often referred to as {\it applying tolerance to the PDF uncertainty}. 

\subsection{Global and dynamic tolerance}
\label{sec:tolerance}

Tolerance is relevant at the stage of determination of PDF uncertainties, after the best-fit PDF has been found. In its simplest realization, tolerance defines an allowed range for the variation 
\begin{eqnarray}
  \label{eq:variatione}
a = a_\textrm{fit} + t e
\end{eqnarray}
of the parameters $a$ along a direction $e$, at a probability level $v$. 
Often, $e$ is one of the eigenvectors of $H$, $H_{\alpha\beta} e_\beta = h e_\alpha$. However, any direction $e$ is a possible choice.  We define the normalization of $e$ using the Hessian matrix, as in Eq.~(\ref{eq:enormalization}): $e_\alpha H_{\alpha\beta} e_\beta = 1$. Then the dependence of $\chi^2$ on $t$ is given by Eq.~(\ref{eq:chisqshifteda}):
\begin{equation}
\label{eq:chisqshiftedbbis}
\chi^2 (
D,a_\textrm{fit} 
+ t e) 
= \chi^2(D,a_\textrm{fit})
+ t^2
\;.
\end{equation}
According to Eq.~(\ref{eq:vectdistribution}), if the experiments that
determine $a_\textrm{fit}$ were repeated many times, then the component
$t$ of $a_\textrm{fit} - \bar a$ in the direction $e$ would be
distributed according to ${\cal N}(0,1)$.
Thus, if we pick a probability $v$ and ask that 
\begin{equation}
\label{eq:trange}
-t_\textrm{lim}(v) < t < t_\textrm{lim}(v)
\end{equation}
with probability $v$, the limiting value $t_\textrm{lim}(v)$ is determined by
\begin{equation}
\label{eq:toleranceP}
\int_{-t_\textrm{lim}(v)}^{t_\textrm{lim}(v)} d\bar t\ p(\bar t) = v
\;,
\end{equation}
where $p(\bar t)$ is the Gaussian distribution ${\cal N}(0,1)$. 
Then $t_\textrm{lim}(0) \approx 0$, $t_\textrm{lim}(0.68) \approx 1$,  $t_\textrm{lim}(0.8) \approx 1.3$, and $t_\textrm{lim}(0.95) \approx 2$.

When we believe that the above procedure misestimates the true
uncertainty on $t$, we could try to find a better probability distribution
$p(\bar t)$ to use in Eq.~(\ref{eq:toleranceP}).
For example, we could use ${\cal N}(0,T)$ with $T^2 > 1$ as our
$p(\bar t)$,
\begin{equation}
  p(\bar t) = \frac{1}{\sqrt{2\pi}} \exp(-\bar t^2/(2 T^2))
  \;.
\end{equation}
Here we use the same value $T$ for every direction vector $e$ \cite{Pumplin:2001ct}. The value $T^2$ in this case is
referred to as the {\it global tolerance.} With $v = 0.68$, the allowed variation of
PDF parameters will be constrained to satisfy $-T < t < T$ along any vector direction with this prescription.  With $v = 0.95$, the allowed variation of
PDF parameters will be constrained to satisfy $-4T < t < 4T$.

The {\it dynamic tolerance} introduced by the MSTW group
\cite{Martin:2009iq} is determined by a
similar consideration, by constructing $p(\bar t)$ from $\chi^2$
distributions for individual experiments $E$. If $P_N(\chi^2)$ is the
$\chi^2$ distribution with $N$ degrees of freedom, we can define $\xi(N,v)$ by
\begin{equation}
\int_{-\infty}^{\xi(N,v)} d\chi^2\ P_N(\chi^2) = v
\;,
\end{equation}
so that $\chi^2 < \xi(N,v)$ with probability $v$. Note that we choose
a one-sided limit here. Since, according to Eq.~(\ref{eq:Sdef}), $S =
\sqrt{2\chi^2}-\sqrt{2N-1}$ closely obeys the ${\cal N}(0,1)$
distribution, we can relate $\xi(N,v)$ to $t_\textrm{lim}(v)$ to a good approximation:
\begin{equation}
\label{eq:xiNv}
\xi(N,v) \approx \frac{1}{2}\left[\sqrt{2N-1} + t_\textrm{lim}(2 v - 1)\right]^2
\;.
\end{equation}

With this information,
the ``dynamic tolerance'' prescription of \cite{Martin:2009iq} assigns
an allowed interval
\begin{equation}
T_\textrm{min} < t < T_\textrm{max}
\end{equation}
for some eigenvector direction in the following way.  

We define $\chi^2(D(E),a)$ to be the part of $\chi^2$ coming from only the data in data set $E$, as in Sec.~\ref{sec:consistency}. We use $\chi^2(D(E),a_\textrm{fit} + t e)$ to define limits $T_\textrm{min}(E)$ and $T_\textrm{max}(E)$ arising from data set $E$, as explained below. Then we set
\begin{equation}
\begin{split}
\label{eq:Tmaxandmin}
T_\textrm{min} ={}& \max_E T_\textrm{min}(E)
\;,
\\
T_\textrm{max} ={}& \min_E T_\textrm{max}(E)
\;.
\end{split}
\end{equation}
For every $E$, \cite{Martin:2009iq} define the range $T_\textrm{min}(E) < t < T_\textrm{max}(E)$ by the criterion
\begin{equation}
\label{eq:chisqErange}
\frac{\chi^2(D(E),a_\textrm{fit} + t e)}{\chi^2(D(E),a_\textrm{fit})} < 
\frac{\xi(N_E,v)}{\xi(N_E,1/2)}
\;,
\end{equation}
where $N_E$ is the number of data in data set $E$.
To understand the result of applying this criterion, it is helpful to use some approximations.

First, using Eq.~(\ref{eq:xiNv}) gives
\begin{equation}
\label{eq:chisqErange1}
\frac{\chi^2(D(E),a_\textrm{fit} + t e)}{\chi^2(D(E),a_\textrm{fit})} < 
\left[1 + 
\frac{t_\textrm{lim}(2 v - 1)}{\sqrt{2 N_E - 1}}\right]^2
\;,
\end{equation}
Noting that $t_\textrm{lim}(2 v - 1)$ is of order 1, we see that for $N_E \gg 1$, this is
\begin{equation}
\label{eq:chisqErange2}
\frac{\chi^2(D(E),a_\textrm{fit} + t e)}{\chi^2(D(E),a_\textrm{fit})} < 
1 + 
\frac{\sqrt{2}\, t_\textrm{lim}(2 v - 1)}{\sqrt{N_E}}
\;,
\end{equation}

Now we examine the left-hand side of Eq.~(\ref{eq:chisqErange1}). The $\chi^2(D(E),a_\textrm{fit} + t e)$ is a quadratic function of $t$,
\begin{equation}
\label{eq:chisqoft}
\chi^2(D(E),a_\textrm{fit} + t e) = 
\chi^2(D(E),a_\textrm{fit}) 
+ A_1(E)\, t + A_2(E)\, t^2
\;,
\end{equation}
with the coefficients $A_1(E)$ and $A_2(E)$ given in Eq.~(\ref{eq:chisqDEaexpanded0}). We note that $0 < A_2(E) < 1$:
\begin{equation}
\begin{split}
A_2(E) ={}& \sum_{i,j \in E} e_\alpha T_{i\alpha} C_{ij} T_{j\beta} e_\beta
\\ <{}&
\sum_{i,j} e_\alpha T_{i\alpha} C_{ij} T_{j\beta} e_\beta
= e_\alpha H_{\alpha\beta}  e_\beta
= 1
\;.
\end{split}
\end{equation}
The coefficient $A_1(E)$ could have either sign and could be large.

Inserting Eq.~(\ref{eq:chisqoft}) into Eq.~(\ref{eq:chisqErange2}) gives
\begin{equation}
\label{eq:chisqErange3}
1 +
\frac{A_1(E) t + A_2(E) t^2}{\chi^2(D(E),a_\textrm{fit})} < 
1 + 
\frac{\sqrt{2}\,t_\textrm{lim}(2 v - 1)}{\sqrt{N_E}}
\;,
\end{equation}
The large terms, 1, here cancel exactly. This gives
\begin{equation}
\label{eq:chisqErange4}
A_1(E)\, t + A_2(E)\, t^2 < 
\frac{\chi^2(D(E),a_\textrm{fit})}{\sqrt{N_E}}\,
\sqrt{2}\,t_\textrm{lim}(2 v - 1)
\;.
\end{equation}
To understand this, we can estimate $\chi^2(D(E),a_\textrm{fit})$ by its expectation value, which, according to Eq.~(\ref{eq:chisqexpectation}) is approximately $N_E$. This gives
\begin{equation}
\label{eq:chisqErange5}
A_1(E)\, t + A_2(E)\, t^2 < 
\sqrt{2 N_E}\,t_\textrm{lim}(2 v - 1)
\;.
\end{equation}
This gives upper and lower limits on $t$ for each experiment $E$. If,
for example, we take $v = 0.9$ then $t_\textrm{lim}(2 v - 1) =
t_\textrm{lim}(0.8) \approx 1.3$.  For simplicity, consider the case
that $A_1(E)$ is small. If $A_2(E)$ is also small, then this
inequality restricts $t$ only weakly. That is, $|T_\textrm{min}(E)|$
and $|T_\textrm{max}(E)|$ are large. We always have $A_2(E) < 1$. If
$A_2(E)$ is close to 1, then this inequality can provide a significant
restriction on $t$. However, the restriction is only significant if
$\sqrt{2 N_E}$ is not too large. For data sets with many data, the
restriction is always weak. Thus the most restrictive values of
$|T_\textrm{min}(E)|$ and $|T_\textrm{max}(E)|$, and thus the overall
values of $|T_\textrm{min}|$ and $|T_\textrm{max}|$, are likely to
come from data sets in which $\sqrt{2 N_E}$ is not too large, and
$A_2(E)$ is not too small. 
For most experiments, the values of $|T_\textrm{min}(E)|$ or
$|T_\textrm{max}(E)|$ tend to be substantially greater than 1.

We do not attempt to justify the definition (\ref{eq:chisqErange}) of the range for $t$ or its approximate version (\ref{eq:chisqErange5}). We do note, however, that the factor $\sqrt{2 N_E}$ in Eq.~(\ref{eq:chisqErange5}) is familiar: it is the standard deviation for the distribution of $\chi^2$, as in Eq.~(\ref{eq:chisqvariance}) for $N_D \gg N_P$.

We have described one approach to defining a tolerance factor. There are several other approaches. It is beyond our scope to explore these in detail.

%-------------------------------------------------------------------

% !TEX root = main.tex
\newcommand{\nN}{\mathtt{N}}
\newcommand{\nA}{\mathtt{A}}
\newcommand{\nZ}{\mathtt{Z}}

\section{Parton distributions for heavy ions}\label{sec:NuclearPDFs}
The concepts discussed in this article can be applied to  nuclear parton distribution functions (nPDFs), nonperturbative QCD functions that are increasingly employed to model the structure of heavy nuclei in high-energy scattering. The concept of collinear QCD factorization that is central for describing scattering of free hadrons is also relevant for the growing number of measurements in collisions of heavy nuclei. The experimental data available for constraining the nPDFs is still very limited in their span over $x$ and $\mu^2$. They are anticipated to grow quickly as the Large Hadron Collider and as especially the envisioned Electron-Ion Collider produce new results. We will review the key features of the nPDFs and will refer the reader to the original publications by nuclear PDF analysis groups (EPPS \cite{Eskola:2016oht}, nCTEQ \cite{Kovarik:2015cma}, DSSZ \cite{Deflorian:2011fp}, HKN \cite{Hirai:2007sx}), KA \cite{Khanpour:2016pph} and NNPDF \cite{AbdulKhalek:2019mzd}) for more details.
\subsection{Universality of nuclear PDFs}\label{sec:NuclearPDFsUniversality}
\begin{figure}%--------------------------
\includegraphics[width=0.33\textwidth]{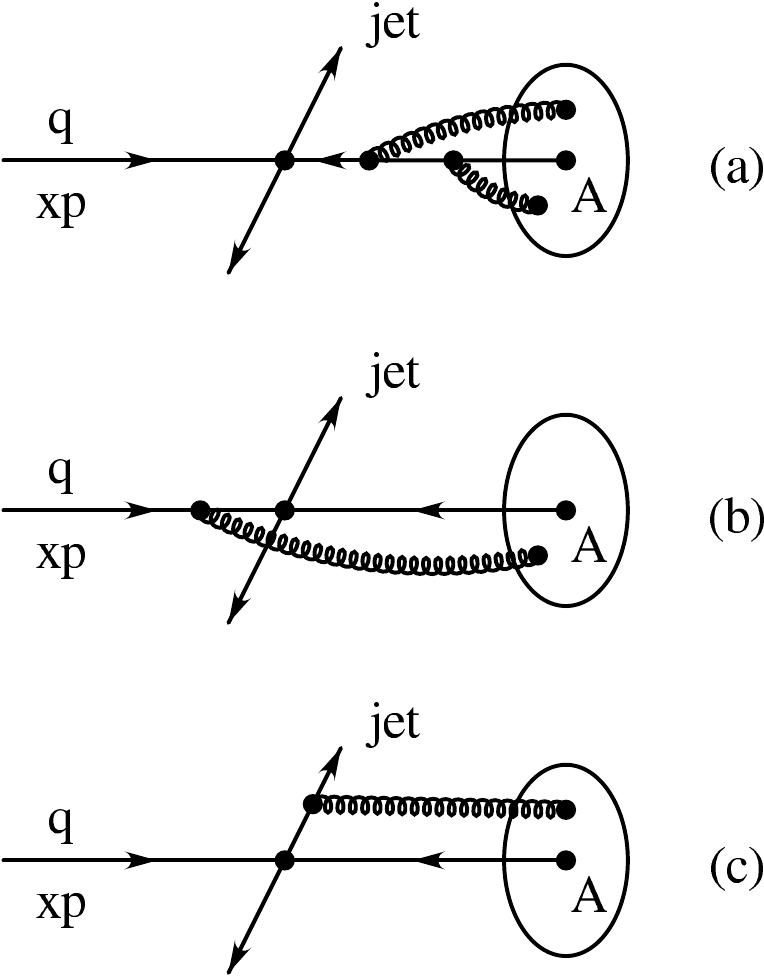}
\caption{Interactions of the nucleus with the initial and the final state partons. From \cite{Qiu:2003cg}.}
\label{fig:nuc}
\end{figure}%--------------------------
As in the standard case of protons, the structure functions and
cross-sections in collisions involving one or more nuclei are related
to nPDFs via perturbative QCD factorization \cite{Collins:1982wa,
  Collins:1985ue, Bodwin:1984hc, Collins:1988ig}.  
Even though  inclusive collinear factorization has been proven  only in
a few cases (e.g. deeply inelastic scattering on hadrons and 
lepton pair production in hadron collisions), it has formed the basis
for the analysis of proton PDFs.  Eqs.~\eqref{eq:factorization} or
\eqref{eq:DISfactorization} express QCD factorization for free-nucleon
QCD observables. 
We will now show how to extend these factorization
formulas to high-energy scattering of heavy nuclei. We will denote a nucleus by
$\nN$ and its atomic number by $\nA$.   

We assume that, at sufficiently large $\sqrt{s}$, the QCD observables
of interest are dominated by independent parton scatterings, in which
only one parton per initial-state nucleus contributes to the hard
scattering. The interactions between a parton and a nucleus can be
classified in three categories \cite{Accardi:2004be,Qiu:2003cg} shown
in Fig.~\ref{fig:nuc}. Fig.~\ref{fig:nuc}(c) illustrates jet quenching, the changes in a jet as it moves through a large nucleus. Fig.~\ref{fig:nuc}(b) illustrates two partons from the nucleus participating in the hard interaction. These effects contribute to the power suppressed corrections to the factorization formulas \eqref{eq:factorization} and \eqref{eq:DISfactorization} for nuclei, but they can be important if the nucleus is big enough and the scattering is not too hard.

The additional interactions between the nucleus
and the initial-state parton within the same nucleus, shown in
Fig.~\ref{fig:nuc}(a), change the parton
distributions of the nucleus and do not affect the hard scattering, 
thereby leaving intact the form of factorization given by
Eqs.~\eqref{eq:factorization} and \eqref{eq:DISfactorization}. The only change in the
prescription replaces the free-proton PDFs by the nPDFs,
which account for the additional initial-state
effects and can be defined as

\begin{align}
\label{eq:fquarkN}
f_{i/\nN}(\xi_\nN,\mu^2 ) & =
\frac{1}{4\pi} \int\!dy^-\, e^{-\mi \xi_\nN P_\nN^+ y^-}
\\ & \times 
\bra{\nN}
\bar\psi_i(0,y^-,\bm 0) \gamma^+ W(y^-,0)\, \psi_i(0)
\ket{\nN}
\notag
\;.
\end{align}

This definition is analogous to the one in Eq.~\eqref{eq:fquarkfinal},
but the proton matrix element of the number density operator is
replaced by the nuclear one. This nPDF is defined with respect to the
whole nucleus with +-momentum $P_\nN^+$. Accordingly, the parton
described by this parton distribution function carries the +-momentum
$p^+ = \xi_\nN\,P_\nN^+$.
The momentum fraction $\xi_\nN$ is defined as
\begin{equation}
	\xi_\nN = \frac{p^+}{P_\nN^+}, \mbox{ with } 0 \leq \xi_\nN < 1.
\end{equation}

The modified factorization prescription for a cross-section in
collisions of nuclei $\nN_1$ and $\nN_2$
can be written using the nPDFs (\ref{eq:fquarkN}) as
\begin{equation}
\begin{split}
\label{eq:NUCfactorization}
\sigma[F] ={}& \sum_{a,b}\int_0^1 \!d\xi_a \int_0^1 \!d\xi_b\
f_{a/\nN_1}(\xi_a,\mu_\LF^2)\, f_{b/\nN_2}(\xi_b,\mu_\LF^2)
\\&\quad\times
\hat \sigma_{a,b,\xi_a,\xi_b,\mu_\LF^2}[F]
+ {\cal O}\!\left({M}/{Q}\right)
\;.
\end{split}
\end{equation}
Even in collisions of nuclei, hard processes such as production of
muon pairs or sufficiently high-$p_T$ jets are dominated by the
leading-power contributions in Eq.~\eqref{eq:NUCfactorization}. Thus
these processes can be well described using $f_{i/\nN}(\xi_N,\mu^2)$,
where the dependence on the scale is still governed by DGLAP
equations. However, the environment of the nuclear collisions is much
different from the free-nucleon collisions. For example, at small values of $\xi_\nN$, the parton
momenta, as viewed in the nucleus rest frame, are very small, so that
the parton wave functions spread over the whole nucleus and
beyond. Then ``saturation''
\cite{Mueller:1999wm,GolecBiernat:1999qd,Bartels:2002cj,Hautmann:2007cx}
or 
``shadowing'' \cite{Armesto:2006ph}
can substantially modify the nPDFs. 
The nPDFs can 
incorporate these and other initial-state
nuclear effects, such as the ``EMC"
effect \cite{Aubert:1983xm,Geesaman:1995yd,Malace:2014uea},
and still be universal.

On the other hand, jets produced in a partonic scattering can be
altered by their passage through nuclear matter, as in
Fig.~\ref{fig:nuc}(c), unless the jet's transverse momentum is very
large. This ``jet quenching'' can affect jet cross sections beyond
what is predicted by Eq.~(\ref{eq:NUCfactorization}) \cite{Aad:2010bu,
  Chatrchyan:2011sx, Adam:2015ewa}. 
Jet-quenched contributions will not factorize in the same way.

The nPDFs depend on the number of protons, $\nZ$, and number of
neutrons, $\nA-\nZ$.  Highly
nontrivial $\nA$ dependence arises from strong interactions of partons
inside the nucleus. There is also trivial $\nA$ dependence that would
be present even if the nucleons were free. Consider a very simple
model, in which a nucleus with +-momentum $P_\nN^+$ is just a collection of
comoving independent protons and neutrons, in which each nucleon
carries the same fraction $\xi_{p,n}=1/\nA$ of the total momentum $P_\nN^+$.

In this model, one could write the nPDF of the whole nucleus as
\begin{align}\label{eq:freenuc}
f_{a/\nN}(\xi_\nN,\mu^2)&\,d\xi_\nN = \\ \nonumber &\Big[\nZ\, f_{a/p}(\xi_\nA,\mu^2) + \big(\nA-\nZ\big)\, f_{a/n}(\xi_\nA,\mu^2)\Big]\,d\xi_\nA\;,
\end{align}
where $f_{a/p}$ and $f_{a/n}$ are the parton distributions in the free
proton and neutron, and $\xi_\nA$ is the momentum fraction of the
+-momentum of the parton with respect to the +-momentum of the
nucleon. The +-momentum $p^+$ of parton $a$ is
\begin{equation}
  p^{+} = \xi_\nN P_\nN^+ = \xi_\nA \xi_{p,n} P_\nN^+,
\end{equation}  
so that the momentum fractions $\xi_\nA$ and $\xi_\nN$ are related via
\begin{equation}\label{eq:xiA}
	\xi_\nA = \frac{\xi_\nN}{\xi_{p,n}},\quad \mbox{ or } \xi_\nA
        = \nA \, \xi_\nN \mbox{ if } \xi_{p,n}=\frac{1}{\nA}. 
\end{equation}
We can use this relation to rewrite Eq.~\eqref{eq:freenuc}
in terms of momentum fraction $\xi_\nN$ as
\begin{align}
f_{a/\nN}(& \xi_\nN ,\mu^2)\,d\xi_\nN = \\ \nonumber &\Big[\nZ\, f_{a/p}(\nA\,\xi_\nN,\mu^2) + \big(\nA-\nZ\big)\, f_{a/n}(\nA\,\xi_\nN,\mu^2)\Big]\,d(\nA\,\xi_\nN)\,.
\end{align}
In this model, $\xi_\nN$ is constrained to be in the range $0 \leq \xi_\nN
\leq 1/\nA$, since free-nucleon PDFs vanish for $\xi_\nA > 1$,
and each nucleon carries exactly the fraction $1/\nA$ of the +-momentum
of nucleus.

In reality, one nucleon can carry any fraction of the
nucleus' +-momentum, since the nucleons participate in Fermi motion relative to
each other \cite{Bodek:1980ar,Saito:1985ct}. We still find it helpful
to use the momentum fraction $\xi_\nA\equiv A \xi_\nN$. We now define it to be
the fraction of the {\it average} +-momentum $P^+_\nN/\nA$ of a bound nucleon.
The variable $\xi_\nA$ now takes values in the interval $0 \leq \xi_\nA
\leq \nA$, with contributions at $1 <  \xi_\nA < \nA$ arising from
in-nucleus motion.

The PDFs of bound nucleons in the nucleus do not coincide with the
free-nucleon PDFs. However, if nuclear modifications are moderate, we
can start from Eq.~\eqref{eq:freenuc} to get a reasonable {\it ansatz} for
the parameterizations of nuclear PDFs.

We define a nuclear PDF of an average nucleon in a nucleus with atomic
number $\nA$, denoted by $f_{a}^\nA(\xi_\nA,\mu^2)$. This nPDF has the
form
\begin{equation}\label{eq:PDFnuc}
f_{a}^\nA(\xi_\nA,\mu^2) = \frac{\nZ}{\nA}\, f_{a/p}^{\nA}(\xi_\nA,\mu^2) + \frac{\big(\nA-\nZ\big)}{\nA}\, f_{a/n}^{\nA}(\xi_\nA,\mu^2)\;.
\end{equation}
In Eq.~\eqref{eq:PDFnuc},  $f_{a/p}^{\nA}(\xi_\nA,\mu^2)$ and
$f_{a/n}^{\nA}(\xi_\nA,\mu^2)$ are the PDFs in the bound proton and bound
 neutron. They are different from the free-nucleon PDFs
$f_{a/p,n}(\xi_\nA,\mu^2)$. They depend on the momentum fraction
$\xi_\nA$ defined above. We can relate the two types of nPDFs that we just
discussed:
\begin{equation}
	f_{a/\nN}(\xi_\nN,\mu^2)\,d\xi_\nN = \nA\,f_{a}^\nA(\xi_\nA,\mu^2)\,d\xi_\nA\,.
\end{equation}

Either the nPDFs $f_{a/\nN}(\xi_\nN,\mu^2)$ in the nucleus or the nPDFs
$f_{a}^\nA(\xi_\nA,\mu^2)$ for an average nucleon are acceptable for use in QCD
calculations. But, ``trivial'' $\nA$ dependence makes
it difficult to compare the nPDFs of the first kind,
$f_{a/\nN}(\xi_\nN,\mu^2)$,  for two different nuclei.

For example, consider the prominent feature of proton PDFs: the peaks
of the up- and down-quark distributions at $\xi\approx 1/3$.
Similar peaks are found in the respective
nPDFs $f_{u/\nN}(\xi_\nN)$ and $f_{d/\nN}(\xi_\nN)$
at $\xi_\nN\sim 1/(3\nA)$, i.e., the position of the peaks in these
nPDFs depends on the nucleus. In addition, the respective valence-quark distributions are
normalized by the sum rules in a nucleus-dependent way:
\begin{equation}
\begin{split}
\int_0^1 \big[ f_{u/\nN}(\xi_\nN,\mu^2)-f_{\bar{u}/\nN}(\xi_\nN,\mu^2)\big]\,d\xi_\nN =&\ \nA+\nZ
\;,
\\
\int_0^1 \big[ f_{d/\nN}(\xi_\nN,\mu^2)-f_{\bar{d}/\nN}(\xi_\nN,\mu^2)\big]\,d\xi_\nN =&\ 2\nA-\nZ
\;.
\end{split}
\end{equation}

In contrast, the nPDFs for an average nucleon not only take into
account the trivial $A$ dependence, they also correctly incorporate
the specific ratio of protons to neutrons. 
The nPDFs $f_{a/p}^{\nA}(\xi_\nA,\mu^2)$ of a bound proton satisfy the sum rules 
\begin{equation}
\begin{split}
\label{eq:NucSumRuleA}
\int_0^\nA \big[ f^{\nA}_{u/p}(\xi_\nA,\mu^2)-f^{\nA}_{\bar{u}/p}(\xi_\nA,\mu^2)\big]\,d\xi_\nA =&\ 2
\;,
\\
\int_0^\nA \big[ f^{\nA}_{d/p}(\xi_\nA,\mu^2)-f^{\nA}_{\bar{d}/p}(\xi_\nA,\mu^2)\big]\,d\xi_\nA =&\ 1
\;,
\end{split}
\end{equation}
which are much like the sum rules for the free proton.

Experimental analyses of nuclear DIS  account for the trivial $\nA$
dependence by presenting the cross-sections or DIS structure functions
not for the whole nucleus but rather per nucleon. Similarly, for
collisions between two nuclei with atomic numbers $\nA_1$ and $\nA_2$,
cross sections $\tilde \sigma[F] \equiv \sigma[F]/(\nA_1 \nA_2)$ per
nucleon are usually quoted. The cross-section $\tilde{\sigma}[F]$ can
be expressed using either type of nPDFs:
\begin{align}
\nonumber
\tilde{\sigma}[F] ={}& \frac{1}{\nA_1}\frac{1}{\nA_2}\sum_{a,b}\int_0^1\!d\xi_a \,d\xi_b\
f_{a/\nN_1}(\xi_a,\mu_\LF^2)\, f_{b/\nN_2}(\xi_b,\mu_\LF^2)
\\ \nonumber &\quad\times
\hat \sigma_{a,b,\xi_a,\xi_b,\mu_\LF^2}[F]
+ {\cal O}\!\left({M}/{Q}\right)
\\ \nonumber
={}& \sum_{a,b}\int_0^{\nA_1}\!d\xi'_a \int_0^{\nA_2}\!d\xi'_b\
f_{a}^{\nA_1}(\xi'_a,\mu_\LF^2)\, f_{b}^{\nA_2}(\xi'_b,\mu_\LF^2)
\\&\quad\times
\hat \sigma_{a,b,\xi'_a,\xi'_b,\mu_\LF^2}[F]
+ {\cal O}\!\left({M}/{Q}\right)
\;.\label{eq:NUCfactperN}
\end{align}
One should note that even in the case of non-interacting nucleons, the cross-section per nucleon $\tilde{\sigma}[F]$ would differ from the cross-section in proton-proton collisions due to the different flavor decomposition. 

To summarize, the trivial $\nA$ dependence reflecting the sheer number
of the nucleons can be captured by using the ansatz \eqref{eq:PDFnuc}
for the nPDF $f_a^\nA(\xi_\nA, \mu^2)$ per average bound nucleon.
On the right-hand side of Eq.~(\ref{eq:PDFnuc}), we introduced the PDFs
$f_{a/p}^\nA$ and $f_{a/n}^\nA$ for bound protons and neutrons that acquire
non-trivial $\nA$ dependence from a combination of nuclear
effects. Their parameterization at the input
scale $\mu_0$ is discussed in the next section.
   \subsection{Parameterizing the $\nA$-dependence}\label{sec:NuclearPDFsParameterization}
In principle one can extract the nPDFs $f_{a/p}^{\nA}(\xi_\nA,\mu^2)$
of a bound proton from experimental data for each nucleus separately,
without constructing a comprehensive model for initial-state nuclear effects. The
current nuclear scattering data, however, are insufficient to determine
the complete set of nPDFs for any single nucleus. The dependence of
nuclear effects on  $\xi$ (also denoted as $x$), $\nA$ and $\nZ$ is assumed to be unknown from
the first principles. Thus, it must be determined a global fit to experimental data.
To assemble all scattering data taken on various
nuclei within a common global analysis, a number of simplifying
assumptions need to be made.

First, given that the nuclear modifications in the bound-proton PDFs
$f_{a/p}^{\nA}$ are expected to be small, it makes sense to use the 
free-proton PDFs $f_{a/p}$ as the baseline for the parameterization of $f_{a/p}^{\nA}$.

Second, to use the available data, one makes an assumption that the
bulk of the nuclear corrections depends only on $\nA$, the total number of
nucleons of either isospin. 

Third,  the current data are not sufficient to constrain the nPDFs
for momentum fractions $\xi_\nA >1$, so all nPDF analyses assume that $0<\xi_\nA<1$.\footnote{In the following text we will denote the momentum fraction $\xi_\nA$ restricted to the interval $\xi_\nA\in (0,1)$ for simplicity by $x$.} However, future fits may be able to include the region $1<\xi_\nA$.

Fourth, we need to decide how to introduce the $\nA$ dependence in
$f_{a/p}^{\nA}$. In practice, one of two approaches is taken. 

The first approach introduces nuclear correction factors $R_a(x,\nA)$
at the input scale $\mu^2_0$:
\begin{equation}
  \begin{split}
\label{eq:nPDFfromPDF}         
	f_{a/p}^{\nA}(x,\nA,\mu_0^2) =&\, R_a(x,\nA)\,f_{a/p}(x,\mu_0^2)\;,\\[1mm] & \mbox{for }a = u_v,d_v,g,\bar{u}+\bar{d},s,\bar{s},\bar{d}/\bar{u}\;.
\end{split}
\end{equation}
In Eq.~\eqref{eq:nPDFfromPDF} $f_{a/p}(x,\mu_0^2)$, the corresponding
PDF for a free proton, is held fixed during any nPDF analysis. The PDF $f_{a/n}^{\nA}(x,\mu_0^2)$
of a bound neutron is related to $f_{a/p}^{\nA}(x,\mu_0^2)$
by charge symmetry. All free parameters
associated with the nuclear modification
are contained in $R_a$. For example,
the EPPS16 analysis \cite{Eskola:2016oht} uses the following piecewise expression:
\begin{equation}
	R_a(x,\nA) = \left\{\begin{array}{ll}
	a_0+a_1(x-x_a)^2 & 0\leq x\leq x_a\\[1mm] 
	b_0+b_1x^\alpha+b_2 x^{2\alpha}+b_3 x^{3\alpha} & x_a\leq x\leq x_e\\[1mm]
	c_0+(c_1-c_2 x)(1-x)^{-\beta} & x_e\leq x\leq 1
	\end{array}\right.\;,
\end{equation}
where $\alpha=10x_a$, and all parameters $a_k$, $b_k$ and $c_k$
implicitly depend on the atomic number $\nA$ and the PDF flavor $a$.
A similar approach that employs a nuclear correction factor
is followed by HKN07 \cite{Hirai:2007sx} and DSSZ
\cite{Deflorian:2011fp}. Each analysis uses a different
proton baseline, cf. \cite{Hirai:2007sx}, \cite{Deflorian:2011fp}, and \cite{Eskola:2016oht}.

    \begin{figure*}%--------------------------
    \includegraphics[width=0.48\textwidth]{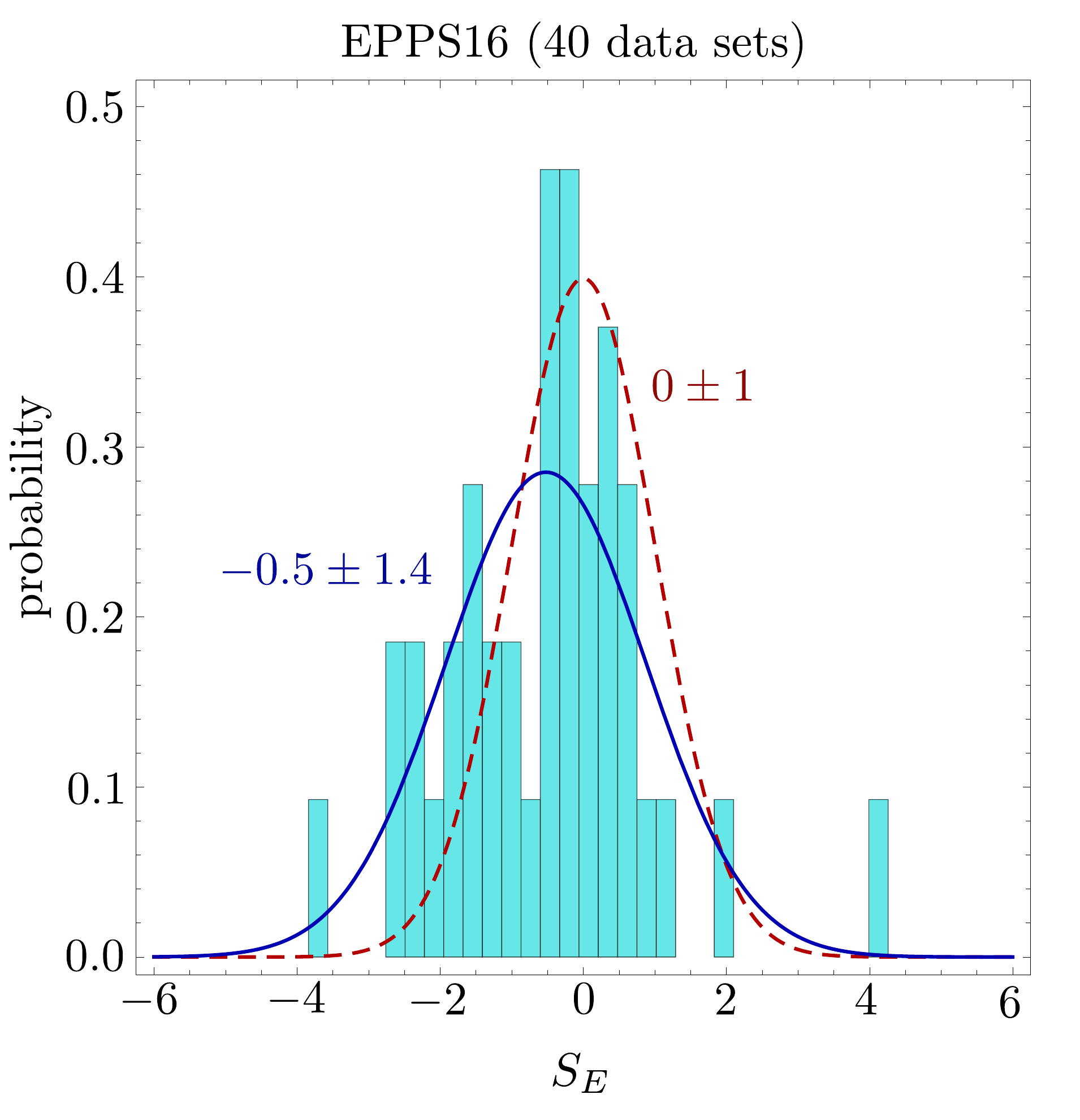}\quad 
    \includegraphics[width=0.48\textwidth]{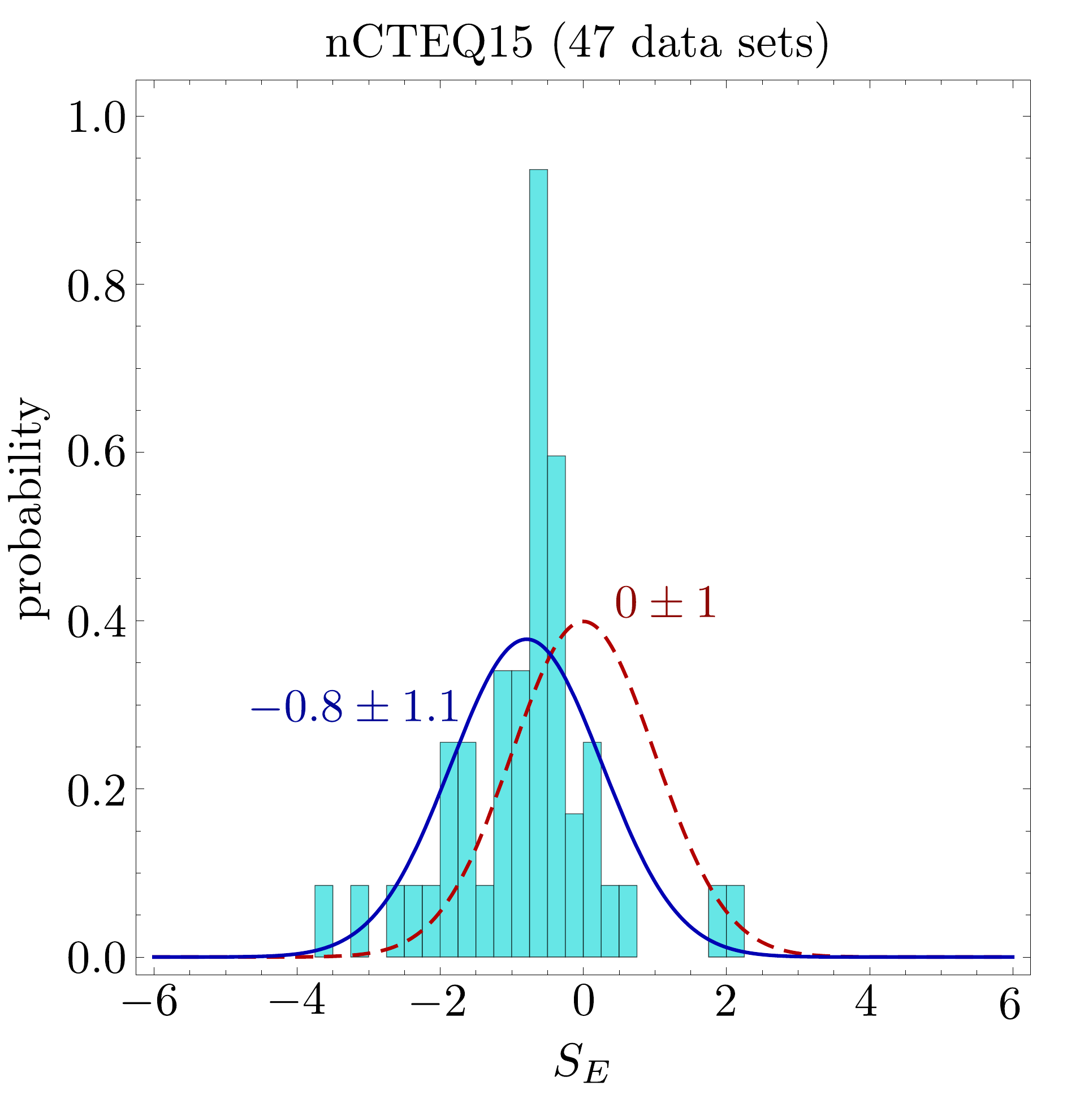}\\
    %\quad\\
    \includegraphics[width=0.48\textwidth]{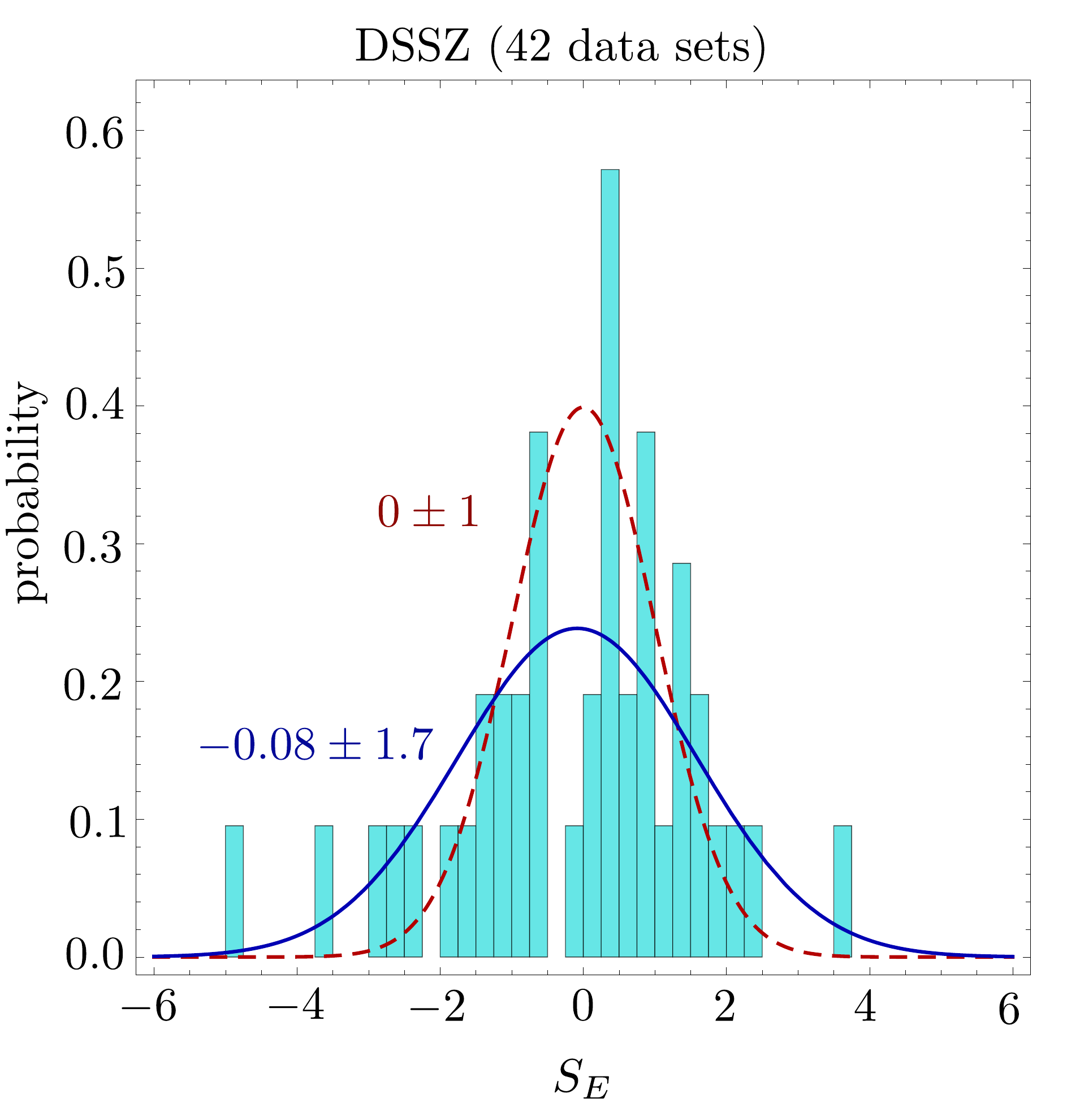}\quad 
    \includegraphics[width=0.48\textwidth]{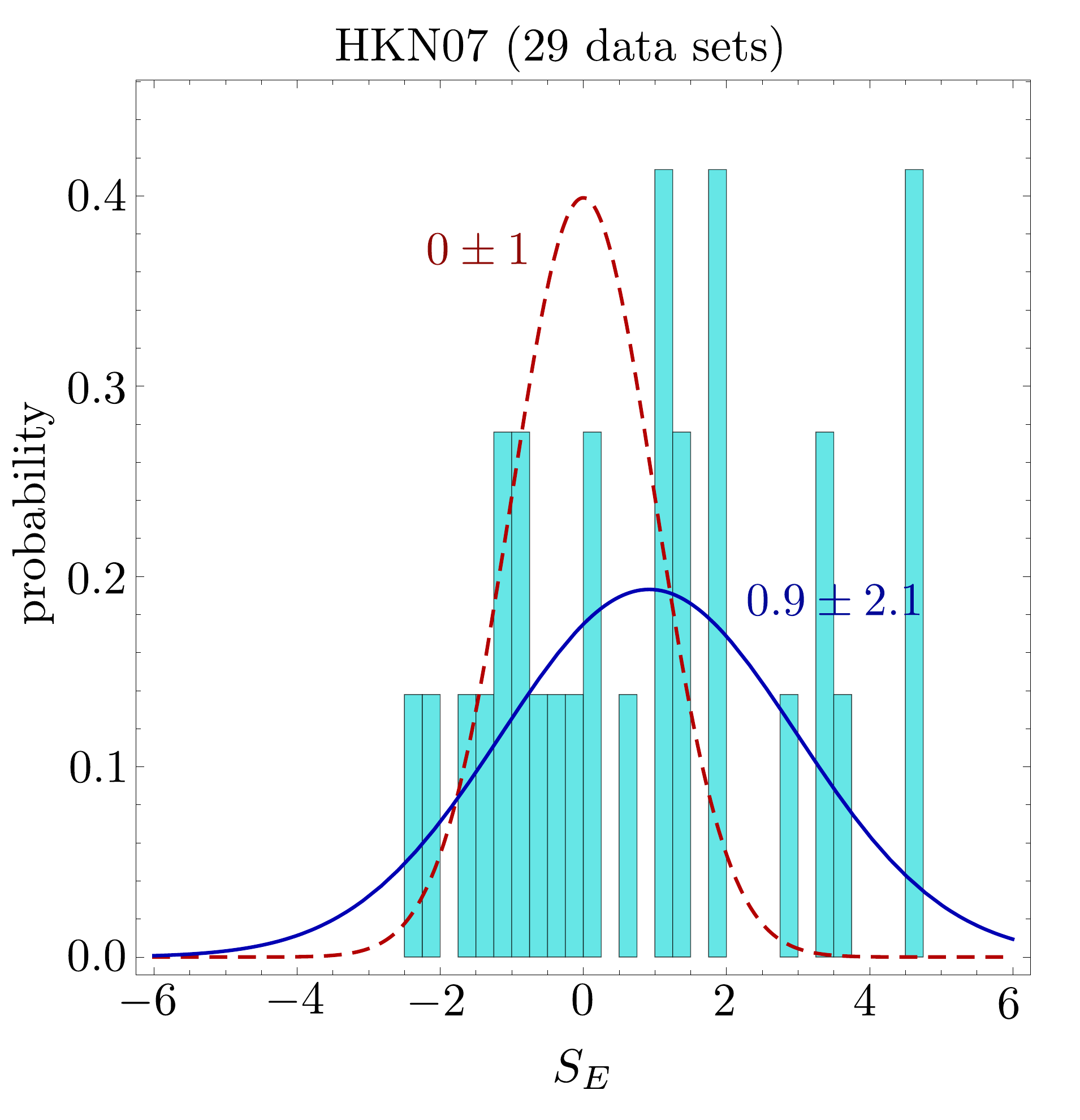}
    \caption{Probability distributions in the effective Gaussian variable $S_E$ for $\chi^2$ values of the fitted data sets from the NLO nuclear PDF fits 
    EPPS16, nCTEQ15, DSSZ and HKN07.}
    \label{fig:nucSE} 
    \end{figure*}%--------------------------

The second approach presented in \cite{Kovarik:2015cma} does not
operate with the nuclear correction factors $R_a$. It rather parameterizes
the whole nPDF $f_{a/p}^A(x,\mu_0^2)$ with a flexible functional form
used for the free-proton PDF $f_{a/p}(x,\mu_0^2)$, but with
$\nA$-dependent free parameters. As an example,
in the nCTEQ15 analysis, the explicit parameterization at the input scale is
\begin{equation}
\begin{split}
\label{eq:nCTEQ1}
	x f_{a/p}^{\nA}(x,\nA,\mu_0^2) =&\, c_0 x^{c_1}(1-x)^{c_2}e^{c_3 x}(1+e^{c_4}x)^{c_5}\;,\\[1mm] & 
	\mbox{for } a = u_v,d_v,g,\bar{u}+\bar{d},s,\bar{s}\;;
\end{split}
\end{equation}
and, similarly to the underlying CTEQ6 parameterization \cite{Pumplin:2002vw},
the parton combination $\bar{d}/\bar{u}$ is given by a different form:
\begin{equation}
\begin{split}
\label{eq:nCTEQ2}
	f_{\bar d/p}^{\nA}(x,\nA,\mu_0^2&)/f_{\bar u/p}^{\nA}(x,\nA,\mu_0^2) = \\[1mm] & c_0 x^{c_1}(1-x)^{c_2} + (1+c_3 x)(1-x)^{c_4}\;.
\end{split}
\end{equation}
All free parameters $c_k$ depend on the atomic number,
\begin{equation}
	c_k(\nA) = c_{k,0} + c_{k,1} (1-\nA^{-c_{k,2}})\;,\qquad k=1,\ldots,5.
\end{equation}
The coefficient $c_{k,0} = c_k(\nA=1)$ is the underlying proton
coefficient. It is held constant during the nCTEQ analysis. In the case of the nCTEQ15 analysis, the underlying proton coefficients $c_{k,0}$ are set to the coefficients from the proton analysis performed in \cite{Owens:2007kp}.

Over the years, there have been other approaches to including nuclear effects in analyses of parton distribution functions. For example, the first next-to-leading order analysis of nuclear parton distribution functions performed in \cite{deFlorian:2003qf} used a convolution approach to relate the nuclear PDF of an average nucleon to the one of a free nucleon as
\begin{equation}
    f_{a/p}^{\nA}(x,\mu_0^2) = \int_{x}^A \frac{dy}{y}W_i(y,A,Z) f_{a/p}(\frac{x}{y},\mu_0^2)\,.
\end{equation}
More recently the analysis of proton PDFs using the neural networks was also extended to the analysis of nuclear PDFs \cite{AbdulKhalek:2019mzd}. In that case the dependence on the number of nucleons is introduced as an additional parameter to the neural network. We will not go into details of any of these approaches as they have not (yet) been used in a global analysis of all nuclear scattering data.
\subsection{Comparisons of nuclear PDFs}\label{sec:NuclearPDFsComp}

The nPDFs $f_{a/p}^{\nA}(x,\nA,\mu_0^2)$ of the bound proton  are
determined from experimental data sets taken on many different
nuclei. Most of the data are still coming from  deeply inelastic
scattering and are provided in the form of nuclear correction factors
\begin{equation}
	R_{\rm DIS}(x,\mu^2) = \frac{F_2^{\nA_1}(x,\mu^2)}{F_2^{\nA_2}(x,\mu^2)}\,.
\end{equation}
The more recent data from neutrino DIS are provided as double
differential cross-sections $d^2\sigma/(dx\,dQ^2)$. The collider data
from Fermilab, RHIC and the LHC are also provided as differential
cross-sections (per nucleon). The coverage of the relevant nuclear
world data is nowhere close to that of the data available for
free-nucleon PDFs. Many features of nPDFs are still poorly known,
especially outside of the interval $0.01 \lesssim \xi_\nA \lesssim
0.5$. Most notably, no data constrain the nuclear gluon PDF at low momentum fractions.

The comparison of different nuclear PDF ensembles is a little trickier than comparing
free-proton PDFs. The deficit of precise data introduces strong
sensitivity to the prior and methodological assumptions, such as the
kinematic cuts, nPDF parameterization form, or the
choice of the baseline free-proton PDFs. 

The methods introduced in Sec.~\ref{sec:goodness} can illustrate the
differences between the various nPDF analyses. First, in
Fig.~\ref{fig:nucSE} we show the distributions of $S_E$, defined in
Sec.~\ref{sec:experimentchisqall}, from four recent NLO global nPDF
analyses. As in the case of the proton analyses shown in
Fig.~\ref{fig:SE}, the distributions of $S_E$ for the nPDF analyses
are broader than the standard normal distribution ${\cal N}(0,1)$
expected from an ideal fit. Looking at the means and standard
deviations of the distributions of $S_E$ shown in
Fig.~\ref{fig:nucSE}, we see that, except for the HKN07 analysis, all
means are negative, indicating that more experiments were fitted too
well. This indicates possibly overestimated uncertainties in multiple nuclear experiments. The prior assumptions made in the HKN07
analysis do not allow for a good description of many Drell-Yan total
cross-section measurements by E772 and E866 experiments at
Fermilab. Consequently, the $S_E$ distribution for HKN07 has its mean shifted to the right, and it is wider. Some caution is needed when comparing the $S_E$ distributions between the analyses in
detail. For example, one entry with high $S_E$ in the EPPS16 analysis
is the double-differential neutrino DIS cross-section from the CHORUS
collaboration. This experiment is not included in the nCTEQ15 and
HKN07 analyses. In the DSSZ, it is included only in the form of the
structure functions $F_2$. 

We can quantify the observation that the $S_E$ distributions are far from
the ideal ${\cal N}(0,1)$ distribution using the Anderson-Darling
test. The probability values that the distributions for the four nPDF analyses were drawn from
${\cal N}(0,1)$ 
are
\begin{equation}
\begin{split}
P_\textrm{A-D} ={}& 6.8\times 10^{-4},\hskip 1 cm \textrm{EPPS16}
\;,
\\
P_\textrm{A-D} ={}& 1.3\times 10^{-5},\hskip 1 cm \textrm{nCTEQ15}
\;,
\\
P_\textrm{A-D} ={}& 1.4\times 10^{-2},\hskip 1 cm \textrm{DSSZ}
\;,
\\
P_\textrm{A-D} ={}& 2.1\times 10^{-5},\hskip 1 cm \textrm{HKN07}
\;.
\end{split}
\end{equation}
With the possible exception of the DSSZ distribution, the
Anderson-Darling test confirms that it is very unlikely that the
distributions in question come from the expected Gaussian
distribution. This is reminiscent of what we found in the proton case
in Eq.~\eqref{eq:adproton}; however, in three cases out of four, the nuclear data are
fitted too well, rather than too poorly. 
    \begin{figure*}%--------------------------
    \includegraphics[width=0.82\textwidth]{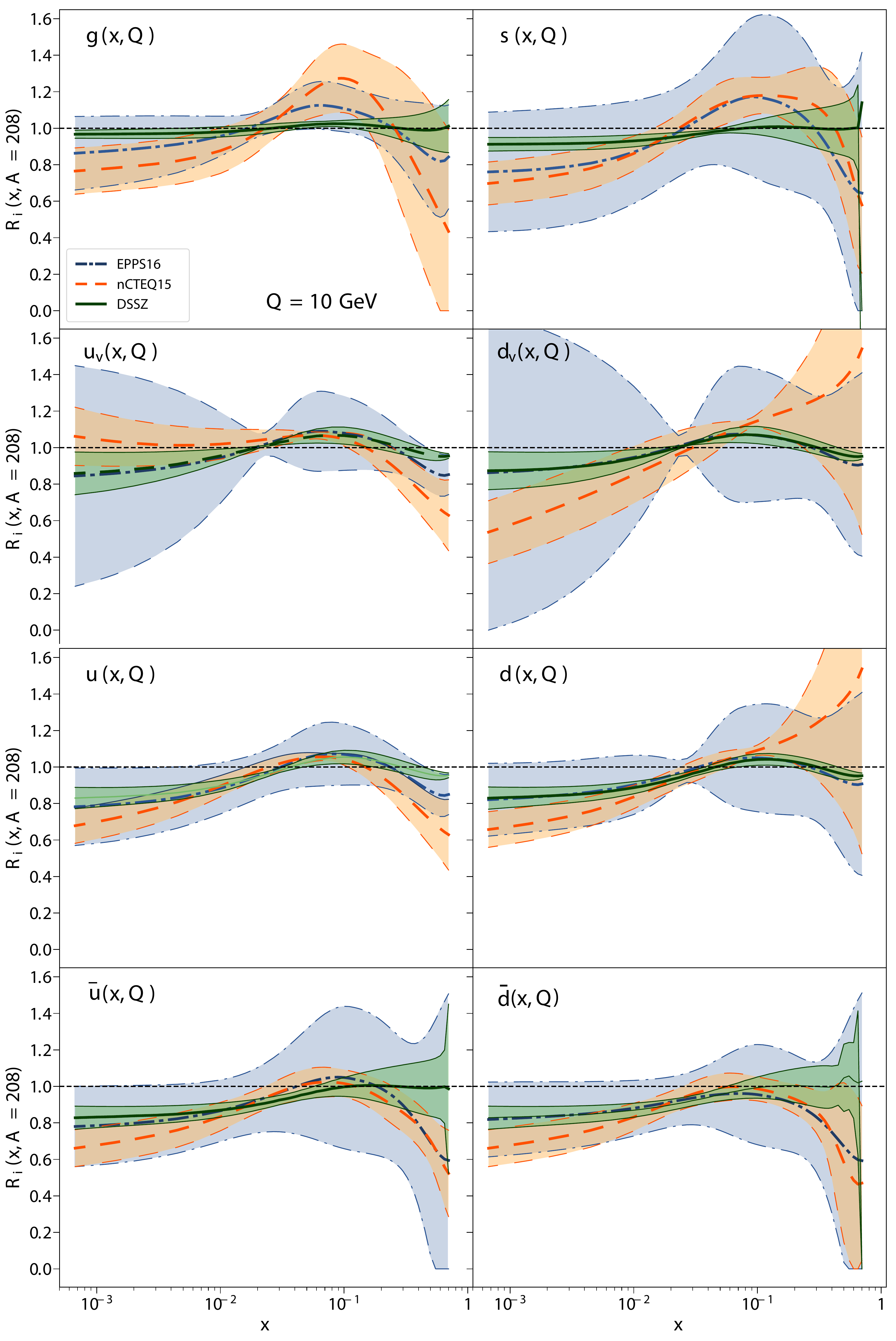}
    \caption{Nuclear correction factor $R_i(x,\nA) =
      f_{i/p}^{\nA}(x,\nA,Q^2)/f_{i/p}(x,Q^2)$  for lead ($\nA=208$)
      and the partons $i = g,s,u_v,d_v,\bar{u},\bar{d}$ and at the momentum transfer $Q=10$ GeV. }
    \label{fig:nPDFcomp}
    \end{figure*}%--------------------------

The momentum fraction dependence of nPDFs is often examined by plotting  scale-dependent nuclear correction factors, 
\begin{equation}\label{eq:nuccorr}
	R_{i}(x,\mu^2,\nA) = \frac{f_{i/p}^{\nA}(x,\nA,\mu^2)}{f_{i/p}(x,\mu^2)}\,,
\end{equation}
where $f_{i/p}(x,\mu^2)$ is the baseline free-proton PDF.
In Fig.~\ref{fig:nPDFcomp}, we turn to a comparison of the EPPS16,
nCTEQ15 and DSSZ nuclear PDFs presented as these  nuclear-correction factors. We show 
$R_{i}(x,Q^2,\nA)$  at $Q=10$ GeV for lead ($\nA=208$), for which the nuclear effects are the
largest. Broadly speaking, we can conclude that all three nPDF
families are consistent with each other within the indicated
uncertainties. Upon a closer inspection, we see that the central values of $R_i$ differ substantially among the three nPDF sets,
for a large part due to the strong dependence on the abovementioned
methodological assumptions, and most prominently due to the choice of the 
parameterization form. Furthermore,
even though it is conventional to compare the ratios $R_i$ rather than nPDFs themselves, this
quantity artificially introduces a dependence on the proton
baseline. Much of the dependence on the baseline is absent when one compares the
bound-proton PDFs $f_{a/p}^{\nA}(x,\nA,\mu^2)$ directly.

The other notable difference among the results in
Fig.~\ref{fig:nPDFcomp} is their strikingly different
uncertainties. One source of the differences are the various
definitions of the uncertainties. All nPDF analyses employ some version
of the global tolerance criterion that is based solely on the global $\chi^2$,
cf. Sec.~\ref{sec:tolerance}. The DSSZ
analysis uses the simplest version of the tolerance: their
uncertainties correspond to varying the underlying parameters along
the eigenvector directions (see Eq.~\eqref{eq:variatione}) by
$t=T=\sqrt{30}$. Both nCTEQ15 and EPPS16 analyses first examine a version of the
dynamical tolerance, as described in Sec.~\ref{sec:tolerance}, to
estimate proper global tolerances for their final nPDF
uncertainties. They determine the limits $T^i_\textrm{min}$ and
$T^i_\textrm{max}$ according to Eq.~\eqref{eq:Tmaxandmin}  using the
probability $v=0.90$ for each eigenvector direction $e_i$. Then, a global tolerance is constructed by averaging the changes in $\chi^2$ over all eigenvector directions as
\begin{align}\nonumber
    T^2 = &
    \sum_{i=1}^{N_P}\frac{\chi^2(a_{\rm fit}+T^i_\textrm{max}e_i)+\chi^2(a_{\rm fit}+T^i_\textrm{min}e_i)-2\chi^2_0}{2n}
    \\ \label{eq:tolnuc}
    = &\sum_{i=1}^{N_P}\frac{(T^i_\textrm{max})^2+(T^i_\textrm{min})^2}{2n}\,.
\end{align}
The nCTEQ15 analysis has 16 free parameters ($N_P=16$) and generates the error PDFs in a standard manner for the global tolerance of $T^2=35$. 

The EPPS16 analysis was the first to include the LHC data from
proton-lead collisions. It uses 20 free parameters, their prescription
given by Eq.~\eqref{eq:tolnuc} yields the global tolerance of $T^2=52$. 

    If all nuclear PDF analyses were to use the same nuclear data
    in a specific range of momentum fractions, and all analyses had a
    flexible parameterization form, the uncertainties would be very
    similar. At present, the compared nPDF analyses do not fit the same
    data. Furthermore, as the four nPDF analyses rely on the
    traditional minimization of global
    $\chi^2$, introducing more free nPDF parameters that can be
    constrained by the nuclear data would lead to unstable global
    fits. In Secs.~\ref{sec:differentdata} and \ref{sec:numberofparameters}, 
    we showed how one can find the optimal number $N_P$ of free parameters
    needed to obtain a stable fit to a given set of hadronic data. For the
    current nPDF analyses, the optimal number of free parameters
    appears to be no more than 15-20.
    Adding new data, for example the LHC data that are included
    in the EPPS16 analysis, allows one to expand the constraints to
    a wider range of momentum fractions or new parton flavor
    combinations. With more LHC data expected in the near future, it
    will be possible to open up additional free
    parameters in the initial nPDF parameterizations, leading to a more
    realistic estimate of uncertainties on nuclear PDFs. 

\section{Conclusions}\label{sec:Conclusions}

We have reviewed certain aspects of the fitting of collinear parton distribution functions (PDFs) to data. This is a very large field. We have concentrated on just a few areas that could be of interest for the readers who use the PDFs or are interested in the rich subject of the global QCD analysis. 

First, we have described the basic definition of what parton distribution functions are, and how they relate to the description of data. We have also provided definitions and a brief description for parton distributions in nuclei instead of in just protons and neutrons. 

Second, we have described the basic statistical treatment needed to fit the PDFs using what is often called the Hessian method. Our description is simplified compared to what is actually used in current PDF fits. Most importantly, we have assumed that, in  a small enough neighborhood of the best fit, the theory predictions $T_k(a)$ are approximately linear functions of the parameters $a$. This is not exactly the case, but this approximation is reasonably good when the PDF uncertainties are small and allows us to derive results in a closed form. Working within this framework, we have explored the statistical reasoning behind the fitting procedure and have derived analytic expressions for the key results of a PDF fit, such as expectation values and uncertainties. 

We have then provided a battery of tests to critically examine whether the statistical assumptions are consistent with certain statistical measures that result from the fit. With these tests, one can identify specific features or data subsets in the multidimensional QCD fits that may indicate discrepancies between experimental measurements and theoretical predictions. Without insisting on a concrete recipe, we present some ideas of what one can do in the case of inconsistency. 

\subsection*{Acknowledgments}

We thank the Kavli Institute for Theoretical Physics at University of California in Santa Barbara, supported by National Science Foundation grant PHY11-25915, for hospitality during the start of the work on this review. We also thank the Aspen Center for Physics, supported by National Science Foundation grant PHY-1607611, for its hospitality during a later stage of our work. We are grateful to CTEQ colleagues for insightful discussions. Work at UO was supported by the United States Department of Energy under grant DE-SC0011640. Work at SMU was supported by the U.S. Department of Energy under grant DE-SC0010129. 

% !TEX root = main.tex
\appendix
\section{Transformation for $\chi^2(D,a,\lambda)$}
\label{sec:appendix}
 
In this appendix, we relate the form (\ref{eq:chisqDalambda}) for $\chi^2(D,a,\lambda)$ to the form Eq.~(\ref{eq:altchisqDalam}), in which it is apparent that the minimum of  $\chi^2(D,a,\lambda)$ with respect to the variables $\lambda$ is $\chi^2(D,a)$. We begin with $\chi^2(D,a,\lambda)$ as given in Eq.~(\ref{eq:chisqDalambda}),
\begin{align}
\label{eq:exponentstart}
\chi^2(D,a,\lambda)
={}& 
\sum_k
\left[
\frac{D_k - T_k(a)}{\sigma_k}
- \sum_I \beta_{kI}\lambda_I
\right]^2
+\sum_J \lambda_J^2
\notag
\\
={}& 
\sum_k 
\frac{(D_k - T_k(a))^2}{\sigma_k^2}
- 2 \sum_J \rho_J\lambda_J 
\\&
+ \sum_{I J} \lambda_I\lambda_J 
B_{IJ}
\;,
\notag
\end{align}
where $B_{IJ}$ was defined in Eq.~(\ref{eq:BIJbis2}),
\begin{equation}
\label{eq:BIJbis}
B_{IJ} = \delta_{IJ} 
+\sum_k \beta_{kI}\beta_{kJ}
\;,
\end{equation}
and where
\begin{equation}
\label{eq:rhoJ}
\rho_{J} \equiv \sum_k \frac{(D_k - T_k(a))}{\sigma_k}\,\beta_{kJ}
\;.
\end{equation}
Completing the square in the variables $\lambda$ gives
\begin{equation}
\begin{split}
\chi^2(D,a,\lambda)
={}& 
\sum_k 
\frac{(D_k - T_k(a))^2}{\sigma_k^2}
- \sum_{IJ} \rho_I \rho_J B^{-1}_{IJ}
\\&
+ \sum_{I J} 
\left[\lambda_I - \sum_K \rho_K B^{-1}_{KI}\right]
B_{IJ}
\\&\quad\ \ \times
\left[\lambda_J - \sum_L B^{-1}_{JL} \rho_L \right]
\;.
\end{split}
\end{equation}

Define shifted variables $\lambda$,
\begin{equation}
\label{eq:lambdashift}
\lambda'_I = \lambda_I - \sum_K B^{-1}_{IK} \rho_K 
\end{equation}
and the matrix
\begin{equation}
\label{eq:Cdefbis}
\widetilde C_{ij} = 
\frac{1}{\sigma_i \sigma_j}
\left\{
\delta_{ij}
- 
\sum_{IJ}  \beta_{iI} B^{-1}_{IJ} \beta_{jJ} 
\right\}
\;.
\end{equation}
This gives
\begin{equation}
\begin{split}
\label{eq:exponentresult1}
\chi^2(D,a,\lambda)
={}& 
\sum_{ij}
(D_i - T_k(a))(D_j - T_k(a)) \widetilde C_{ij}
\\&
+ \sum_{I J} 
\lambda'_I \lambda'_J B_{IJ}
\;.
\end{split}
\end{equation}
The matrix $\widetilde C_{ij}$ is, in fact, the covariance matrix $C_{ij}$. To prove this, use the definition (\ref{eq:Cinversedef}) of $C^{-1}_{ij}$, calculate $\sum_j \widetilde C_{ij} C^{-1}_{jk}$, and simplify the product using $\sum_j \beta_{jJ}\beta_{jL} = B_{JL} - \delta_{JL}$. The calculation gives
\begin{equation}
\sum_j \widetilde C_{ij} C^{-1}_{jk} = \delta_{ik}
\;,
\end{equation}
so that $\widetilde C_{ij} = C_{ij}$.

We arrive at the form of $\chi^2(D,a,\lambda)$ given in Eq.~(\ref{eq:altchisqDalam}):
\begin{equation}
\begin{split}
\label{eq:exponentresult2}
\chi^2(D,a,\lambda)
={}& 
\sum_{ij}
(D_i - T_k(a))(D_j - T_k(a)) C_{ij}
\\&
+ \sum_{I J} 
\lambda'_I \lambda'_J B_{IJ}
\;.
\end{split}
\end{equation}
It is clear that minimizing $\chi^2(D,a,\lambda)$ with respect to $\lambda$, which is equivalent to setting $\lambda_{I,J}'=0$, leaves only the first term in Eq.~(\ref{eq:exponentresult2}), which is $\chi^2(D,a)$ according to Eq.~(\ref{eq:chisqDa}).

%-------------------------------------------------------------------

%\bibliography{rmp2018}
%merlin.mbs apsrmp4-1.bst 2010-07-25 4.21a (PWD, AO, DPC) hacked
%Control: author (3) reversed first dotless

\end{document}